 \documentclass[final,3p,times]{elsarticle}

\usepackage{amssymb}
\usepackage{amsmath}
\usepackage{booktabs}

\usepackage{amsfonts}
\usepackage{algorithmic}
\usepackage{algorithm}
\usepackage{array}
\usepackage{subfig}
\usepackage{textcomp}
\usepackage{stfloats}
\usepackage{url}
\usepackage{graphicx}
\usepackage[utf8]{inputenc} 
\usepackage{multirow}

\usepackage{lineno}

\usepackage{ulem}
\usepackage{xcolor}

\begin{document}

\begin{frontmatter}





\title{Integrating prior knowledge in equation discovery: Interpretable symmetry-informed neural networks and symbolic regression via characteristic curves}


\author[label1]{Federico J. Gonzalez}
\ead{fgonzalez@ifir-conicet.gov.ar. ORCID: 0000-0003-2026-4129}
\affiliation[label1]{organization={Physics Institute of Rosario (IFIR), CONICET-UNR},
             addressline={Blvd. 27 de Febrero 210 Bis},
             city={Rosario},
             postcode={S2000EZP},
             country={Argentina}}


\begin{abstract}
Data-driven equation discovery aims to reconstruct governing equations directly from empirical observations. A fundamental challenge in this domain is the ill-posed nature of the inverse problem, where multiple distinct mathematical models may yield similar errors, thus complicating  model selection and failing to guarantee a unique representation of the true underlying mechanisms. Consequently, robust identification requires frameworks capable of incorporating strong inductive biases to constrain the search space and discard the undesirable models.  
The characteristic curves-based (CCs) framework  offers a modular approach ideally suited to address these challenges. This approach is grounded in the specification of structural families that (unlike generic black-box models) possess provable identifiability properties. Crucially, the CC-based formalism allows practitioners to explicitly embed domain expertise, such as geometric symmetries or parity constraints, directly into the learning process.  Furthermore, this modular design facilitates the integration of diverse post-processing tools. 
In this work, we build upon the recent neural network implementation of this formalism (NN-CC), which benefits from the universal approximation capabilities of NNs. Specifically, we extend this methodology by introducing two inductive biases: (i) symmetry constraints and (ii) post-processing with symbolic regression.  
Using a chaotic Duffing oscillator and a discontinuous stick-slip model under varying Gaussian noise levels, we show how these extensions effectively reduce the hypothesis space, systematically improving the discovery process.  
We also analyze the integration of sparse and symbolic regression (using SINDy and PySR) into the CC-based formalism. These extensions (SINDy-CC and SR-CC) consistently show improvements as prior information is incorporated. 
By enabling the integration of prior or hypothesized knowledge into the learning and post-processing stages, the CC-based formalism emerges as a promising candidate to address identifiability issues in purely data-driven methods, advancing the goal of interpretable and reliable  system identification. 
\end{abstract}




\end{frontmatter}


\section{Introduction}

\subsection{Background}

The data-driven equation discovery is a subfield of system identification that has emerged as a cornerstone of modern research across physics, engineering, and the life sciences\cite{Ljung2019,Brunton_Kutz_2022}. 
The increasing availability of high-fidelity observational data necessitates the creation of models that are simultaneously accurate, generalizable, parsimonious, and physically interpretable. Satisfying these four criteria simultaneously presents a non-trivial challenge, as they are often competing; a prime example is the widely discussed trade-off between expressivity (the capacity to capture complex, high-dimensional, and nonlinear behaviors) and interpretability\cite{Tulleken1993,Rudin2019,Atrey2025,Cappi2025}. 

Historically, nonlinear system identification relied on frameworks such as nonlinear auto-regressive moving-average with exogenous inputs (NARMAX) \cite{Ljung1999,Billings_2013}. While foundational, these approaches rely on discrete-time lagged variables, which makes the extraction of the underlying continuous-time ordinary differential equation (ODE) mathematically non-trivial\cite{Billings_2013}. 
Consequently, over the past two decades, the focus has progressively shifted toward equation discovery methods, whose objective is to reconstruct the governing ODEs or partial differential equations (PDEs) directly from time-series data. In parallel, the field has seen a growing adoption of highly expressive machine-learning-based techniques \cite{Brunton_Kutz_2022}.


Different paradigms have emerged: 
(a) sparse regression techniques, such as the sparse identification of nonlinear dynamics code (SINDy)\cite{Brunton_Kutz_2022,brunton2016},  which is based on a library of candidate functions and employs least absolute shrinkage and selection operator (LASSO)-type optimization\cite{Tibshirani2018} to select parsimonious models; 
(b) symbolic regression,  such as the python symbolic regression package (PySR)\cite{Cranmer2023PySR}, which utilizes genetic algorithms to explore combinations of basis functions using predefined operations and operators; 
(c) neural ODEs\cite{chen2019neuralODE,Both2021,Lai2021,Kim2021,Goyal2023,Bradley2024,Jeong2025}, which are highly expressive but often difficult to interpret physically; 
(d) physics informed neural networks (PINNs)\cite{raissi2017physicsI,raissi2017physicsII,raissi2019physics,Lu2021DeepXDE}, which embed governing equations directly into the loss function during training; 
(e) Hamiltonian neural networks (HNNs)\cite{Greydanus2019,Varghese2025}, which enforce a Hamiltonian structure to guarantee energy conservation;
and (f) Lagrangian neural networks (LNNs)\cite{Cranmer2020LNN}, which are designed to preserve variational symmetries.

The use of machine learning (ML) approaches exploits the intrinsic ability of neural networks (NNs) to represent arbitrarily complex functions (see Refs.~\cite{Cybenko1989,Pinkus1999,Hornik1989}; see also Ref.~\cite{Goodfellow2016} for a more recent approach). However, black-box ML approaches\cite{Patra2002,Wang2006,Rahman2022,Arridu2025,Prokop2025DatadrivenDO}, including neural ODEs\cite{chen2019neuralODE,Both2021,Kim2021,Goyal2023,Bradley2024,Jeong2025}, while flexible enough to learn complex nonlinear dynamics from data, often produce models that are difficult to convert into compact and analytical expressions for scientific discovery. In this context, 
PINNs\cite{raissi2017physicsI,raissi2017physicsII,raissi2019physics,Lu2021DeepXDE} have emerged to address these interpretability challenges by  
explicitly incorporating the model equations into the loss function, also taking advantage of automatic differentiation techniques\cite{Paszke2017,Baydin2018,Both2021}.
Although PINNs are widely applied to simulation and identification of PDEs\cite{Pakravan2021,Luo2025,Guo2025,Zhou2024,Yi2025,Gao2025}, they are also applied to ODEs\cite{Roy2023,Zou2024,Matthews2025,Saha2025,Nguyen2025,Zhao2025}. However, the standard PINNs technique requires full knowledge of the governing physical equations in explicit parametric form, as these must be incorporated directly into the loss function.  
Also, PINNs can become computationally demanding due to over-parameterization, impeding its application for high-dimensional systems\cite{Hu2024}. Some recent works have focused on implementing multiple NNs to overcome these problems\cite{He2024} and multi-level decompositions\cite{Dolean2024,Mahar2025}.

However, despite these advances, two persistent challenges remain in the mentioned techniques:
\begin{itemize}
\vspace{-0.2 cm}    \item[i)] \textit{Ambiguity in global model selection.} 
The challenge is the fundamental lack of structural uniqueness in purely data-driven discovery.  Because the inverse problem of inferring dynamics from finite data is often ill-posed, the search space typically contains multiple distinct mathematical models that fit the data with comparable accuracy. This ambiguity leads to identifying a set of equally plausible models, typically visualized as a Pareto frontier, where the horizontal axis represents model complexity (or some hyperparameter) and the vertical axis corresponds to the identification error \cite{Schmidt2009,Tanevski2020,Naozuka2022,Cranmer2023PySR,Egan2024}. Consequently, selecting a single, physically `true' model becomes dependent on arbitrary user-defined thresholds rather than intrinsic properties of the system. 
Different paths have been proposed to find a candidate among these competitors. Standard statistical approaches rely on information-theoretic metrics, such as the Akaike information criterion (AIC) or the Bayesian information criterion (BIC), which explicitly penalize model complexity to counterbalance goodness-of-fit and prevent overfitting \cite{Burnham2004,Mangan2017}. Alternative strategies focus on generalization capability, utilizing cross-validation or stability selection to identify structures that remain persistent across varying data subsets or noise realizations \cite{Meinshausen2010,Heinze2018}. Furthermore, selection can be guided by domain-specific requirements, such as enforcing physical constraints (e.g., conservation laws) or utilizing the focused information criterion (FIC) to optimize the model for a specific quantity of interest rather than global accuracy \cite{Udrescu2020,Claeskens2003}.

\vspace{-0.2 cm} \item[ii)] \textit{Limitations in exploratory component discovery.} The second challenge concerns the difficulty of performing exploratory modeling—where a practitioner possesses partial domain knowledge (e.g., symmetries or conservation laws) but seeks to identify specific unknown constitutive relations (e.g., friction or stiffness) from data.
Existing approaches struggle to balance structural flexibility with physical interpretability. Purely physics-informed methods (like standard PINNs) are often too rigid, as they typically require the specific functional forms of all constitutive laws to be fixed a priori. Conversely, geometric approaches (e.g., HNNs, LNNs) enforces global conservation laws but adding partial knowledge is not straightforward.  
Recent frameworks, such as physics-guided neural networks (PGNNs)\cite{Robinson2022} and universal differential equations (UDEs)\cite{Rackauckas2021}, attempt to address this by using neural networks to approximate only the unknown terms. 
Additionally, other physics-guided methods add strong inductive biases (such as dimensional homogeneity and symmetry groups) to restrict the hypothesis space to physically valid functional forms \cite{Udrescu2020,Cranmer2020bias,Yang2026}. However, a critical identifiability issue remains. Without strict constraints on the admissible function class, the inverse problem for the unknown term remains ill-posed; the learned function may compensate for modeling errors elsewhere, leading to a solution that reproduces the global trajectory accurately but fails to capture the true underlying physical mechanism \cite{Djeumou2022}. 
\end{itemize}

\vspace{-0.2 cm} 
Building on these advances, a particularly promising strategy to address both challenges is to enforce a structural `skeleton', by defining a family of admissible physical models that ensures identifiability by design, while using universal approximators to flexibly learn the specific constitutive relations. This approach enables the exploratory search by computing the loss across a set of hypothetical model families. Furthermore, the structural skeleton facilitates a more thorough assessment of consistency, interpretability, and physical plausibility, ultimately supporting more informed model selection. 

In this context, the recently introduced framework based on characteristic curves (CCs) \cite{Gonzalez2023,Gonzalez2024,Gonzalez2025} offers a distinct and modular alternative for this type of exploratory discovery. 
The CC-based approach is inspired by current–voltage (I–V) characteristics in electronic circuits, where each circuit element is described by its own constitutive relation. By analogy, CC-based approach decomposes complex dynamical systems into interpretable components, each governed by a learned CC, while the global structure of the model remains fixed by a chosen template. 
Hence, rather than enforcing a single, fully specified global equation, CC-based framework relies on predefined structural templates in which the overall form of the governing dynamics is prescribed, while the CCs are treated as unknown and inferred from data.  

This framework was initially developed for first-order dynamical systems using Fourier \cite{Gonzalez2023} and polynomial \cite{Gonzalez2024} representations, for which analytical relationships were derived linking Fourier-domain quantities to the coefficients of the polynomial expansions defining the CCs. Although these analytical formulations are restricted to first-order systems, they established the general CC-based modeling paradigm, which was extended to higher-order systems\cite{Gonzalez2025} using different representations or basis functions for the CCs. Specifically, the CCs were represented using polynomial basis functions (Poly-CC), sparse regression (SINDy-CC), and neural networks (NN-CC). 

In the NN-CC approach, each CC of the proposed structure is represented by a dedicated NN, taking advantage of the universality properties of the NNs. 
These universality properties can be used along with the prescription of specific templates to demonstrate uniqueness properties. In Ref.~\cite{Gonzalez2025}, these uniqueness properties were shown for two class of second-order families of functions, giving a step forward to address the structural ambiguity mentioned in the first challenge mentioned above.


Furthermore, the modularity of CC-based formalism facilitates the addition of structural and symmetry constraints motivated by prior physical knowledge, thereby being a candidate formalism to perform exploratory search by adding prior domain knowledge or model hypotheses, thereby addressing the second challenge mentioned above.

\subsection{Our contribution}

The primary goal of this work is to establish the CC-based methodology as a formal mechanism for embedding prior physical knowledge and practitioner hypotheses, thereby bridging the gap toward experimental applications. 
Specifically, we aim to demonstrate how the CC-based framework, particularly the NN-CC method, serves as a structural interface for injecting hypothesized constraints and symmetries directly into the discovery loop,  while also benefiting from post-processing tools. Although we validate these capabilities using NN-CC variants, the underlying CC-based formalism is designed as a solver-agnostic architecture. Consequently, this framework offers a clear pathway for established methods, such as sparse identification (e.g., SINDy) and symbolic regression (e.g, pySR), to benefit from the CC-based formalism, opening a promising avenue for future extensions of these techniques.  

As attention shifts toward real-world scenarios, measurement noise and data imperfections become critical factors that can substantially compromise model reliability. In experimental scenarios, data complexities extend beyond simple measurement errors to include colored noise, outliers, and missing observations (challenges particularly relevant in fault diagnostics and system monitoring \cite{Van2016,Bounemeur2018,Bounemeur2023,Abdelhamid2024,Chen2025}). While addressing these issues is essential for the eventual development of the formalism, additive Gaussian white noise remains the standard for initial methodological evaluation. 
Therefore, as a first step to handling complex experimental issues, this study assesses the performance of the proposed framework under Gaussian noise conditions.

At this point, it is worth mentioning that SINDy and SR techniques have been explored and extended in multiple directions to address robustness and applicability issues, including, for instance, Weak SINDy \cite{Messenger2021} and other approaches that take advantage on automatic differentiation \cite{Kaheman2022};  also, probabilistic and ensemble-based strategies, including Ensemble-SINDy \cite{Fasel2022}, Monte Carlo Markov chain (MCMC-SINDy) \cite{Zhu2024sparse}, and Gaussian-process-based variants such as GPSINDy \cite{Hsin2024} and multi-fidelity GP (MFGP-SINDy) \cite{Meng2025}. 
SINDy has also been extended to accommodate implicit representations and more complex structural constraints. Notable contributions include parallel implicit SINDy-PI \cite{Kaheman2020}, extended Lagrangian-SINDy \cite{Purnomo2023}, and implicit Runge-Kutta IRK-SINDy \cite{Anvari2025}, as well as more recent integrations with advanced learning paradigms, such as automatic regression (ARGOS) \cite{Egan2024}, reinforcement learning (SINDy-RL) \cite{Zolman2025}, and deep neural networks (iNeural-SINDy)\cite{Forootani2025}.  
For SR techniques, multiple directions were also investigated to address robustness and applicability (see Refs.~\cite{Dong2025,Chen2023,Cohen2024,Sun2025}). 

Furthermore, it is important to mention recent related research that highlights the importance of incorporating structural and symmetry constraints into data-driven modeling. 
Physics-informed methods such as HNNs and LNNs enforce structural symmetries to ensure physical validity, for example through energy conservation. 
More broadly, advances in equivariant neural networks\cite{Cohen2016,Weiler2019,Finzi2020,Wang2021symm}, have led to symmetry-informed discovery frameworks\cite{Higgins2022,Arora2024,Kahlmeyer2025,Yang2025simmetry}.  
These developments suggest that structural and symmetry-based constraints play a crucial role in enhancing generalization and in ensuring that identified models respect fundamental properties of the underlying system\cite{Yang2025simmetry,Yang2025,Gkimisis2026}.

In this context, in addition to the noise studies, the present work aims to extend the NN-CC framework by explicitly incorporating symmetry constraints and symbolic regression within the CC-based formalism. 
Specifically, we consider the baseline NN-CC approach and three modified variants: NN-CC$_\text{+post-SR}$, which applies SR as a post-processing step to the learned CCs; NN-CC$_\text{+sym}$, which incorporates known structural or physical symmetries directly into the learning process; and NN-CC$_\text{+sym+post-SR}$, which combines both approaches. These methods are evaluated on two representative second-order dynamical systems: a chaotic Duffing oscillator, which serves as a test example for continuous polynomial nonlinearities, and a discontinuous stick–slip system, which probes the ability of the framework to handle non-smooth constitutive relations. 


This work is structured as follows. Section~\ref{sec:methodology} presents the CC-based framework and details the implementation of symmetry constraints and post-SR, along with other techniques used for comparison.  Section~\ref{sec:results} evaluates the performance of the model on the two benchmark systems.
Section~\ref{sec:discussion}  discusses the results and perspectives for future work, and the main findings are presented in Sec~\ref{sec:conclusions}.


\section{Methodology}\label{sec:methodology}
\subsection{Workflow} \label{sec:methodology_workflow}

\begin{figure*}[!htpb]
\centering
\includegraphics[width=16 cm,trim={0.5cm 1.6cm 0.5cm 0.5cm}]{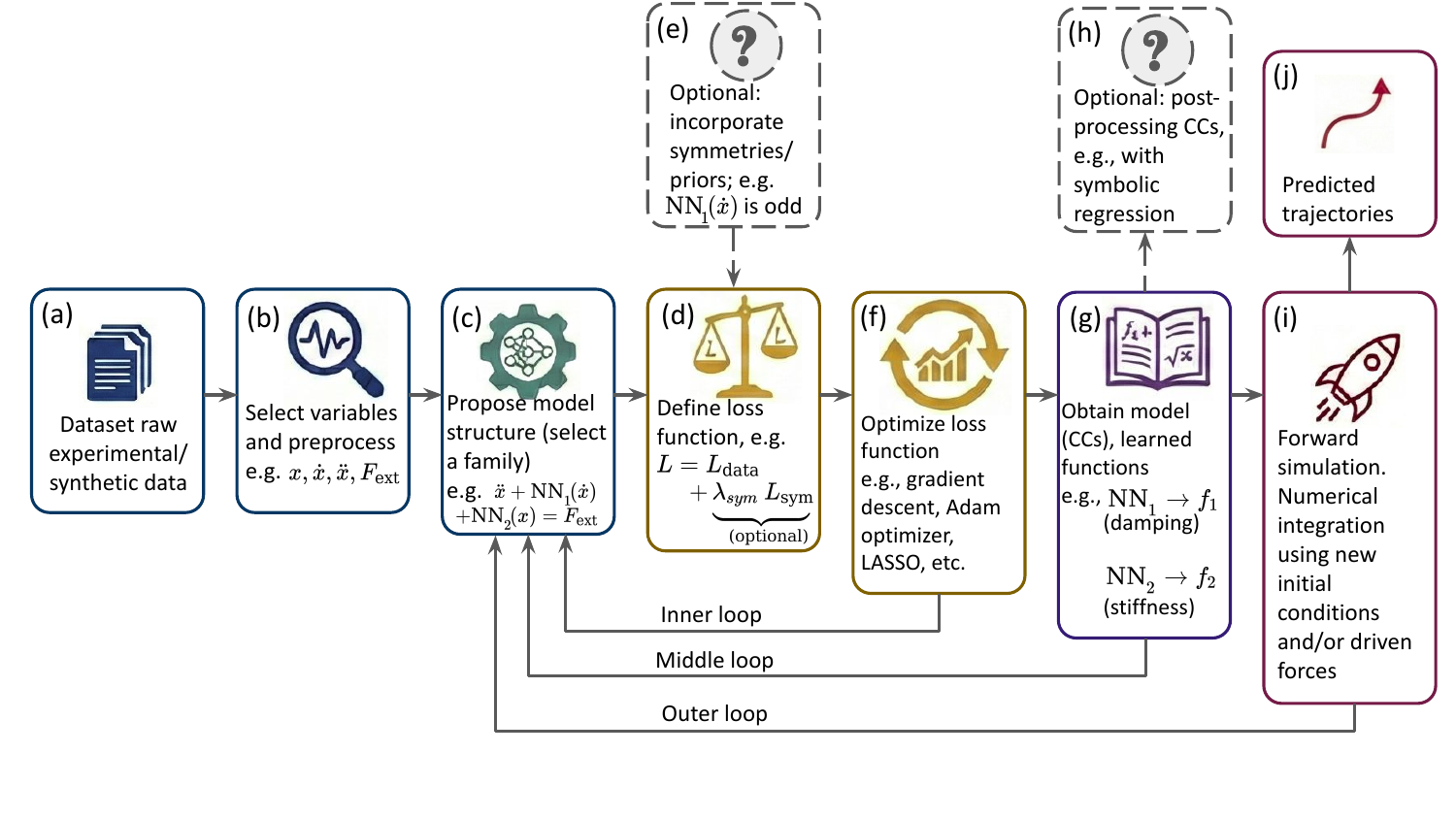}
\caption{Schematic workflow of the proposed CC-based formalism. Starting from raw data (a), the practitioner selects the variables that will intervene in the model (a), they define a equation structure (c) where unknown physical mechanisms CCs are approximated by learnable functions (e.g., Neural Networks, polynomials, or sparse regression terms). The optimization process (f) minimizes a loss function (d) that balances data reconstruction with physical priors and symmetries (e). The result (g) is a set of interpretable functions (e.g., damping $f_1$ and stiffness $f_2$) that accurately predict system dynamics via numerical integration (i) and (j). An optional stage of post-processing of the obtained CCs can be also performed (h). 
The workflow is exemplified with NNs, but other basis functions for the CCs are also analyzed.}
\label{fig:schematic}
\end{figure*}

Figure~\ref{fig:schematic} illustrates a systematic workflow for the CC-based formalism. The process transforms raw observations or measurements into interpretable physical models through the following stages:

\begin{itemize}
    \item[i)] \textit{Data acquisition and preprocessing.}
    Block (a): The pipeline begins with time-dependent raw datasets consisting of experimental or simulated trajectories. Block (b): We select the variables to be included in the models (e.g., position $x$ and external forces, $F_{ext}$). This step may require computing and preprocessing some variables such as time derivatives ($\dot{x}, \ddot{x}$) if they are not measured. 

    \item[ii)] \textit{Physics-informed model selection.} Block (c):
    Instead of learning a generic mapping as in black-box methods, a governing equation structure (family of models) is hypothesized. Within this structure, NN$_i$ functions (referred to as the CCs) are unknowns to be identified. 
    Hence, in the NN-CC method, each unknown function is represented with an individual NN (e.g. NN$_1$ and NN$_2$). 
    Here, the CCs are parametrized with NNs, but they can also be parametrized with polynomial or trigonometric basis functions.      It is advantageous to propose mathematical structures that share uniqueness properties, as detailed in \ref{app:uniquiness_CC}.

    \item[iii)] \textit{Symmetries/constraints and the construction of the loss function.} Block (d): We define a loss function ($L$ or $L_\text{tot}$) composed of a data-fidelity term ($L_\text{data}$) and optional symmetry terms ($L_\text{sym}$). Block (e): Prior physical knowledge, such as symmetry requirements (e.g., damping being an odd function of velocity), can be explicitly enforced and incorporated into the $L_\text{sym}$ definition, aiming to improve physical consistency.

    \item[iv)] \textit{Model optimization and error evaluation.} Block (f): Models are optimized using minimization algorithms appropriate for the chosen methodology (e.g., gradient descent or Adam for NNs, LASSO for sparse regression). If the obtained errors for the loss function are considerable after minimization (a quantity that can depend on the specific system), this suggests that the proposed structure or family of models is incorrect. In this case, we return to model selection (inner loop). See further details about model selection in \ref{app:model_selection}.

    \item[v)] \textit{Model extraction and interpretation.} Block (g): The learned CCs are visualized explicitly. This step enables the direct inspection of the learned constitutive relations, allowing the practitioner to verify whether the obtained CCs (e.g., NN$_1$ and NN$_2$) adhere to expected physical properties, also enabling them to hypothesize symmetries, saturation limits, or monotonic behavior to be incorporated in new training instances.  
    Critically, if the proposed structure accurately captures the system dynamics, the CCs must depend strictly on their assigned state variables (e.g., $x$ or $\dot{x}$) and, for instance, should not exhibit explicit time dependence. Under these conditions, the CCs themselves function as time invariants of the system, serving as a robust metric for assessing model precision. Consequently, this property, along with the uniqueness of the proposed structure, provides a rigorous criterion for model rejection.  If the extracted CCs vary significantly across datasets generated under different experimental conditions [such as varying initial conditions (ICs) or driving forces], it indicates that the hypothesized governing equation is incorrect. In such cases, the candidate model is rejected, and the procedure returns to the model selection phase (middle loop). See \ref{app:model_selection} for further details about model selection.   

    \item[vi)] \textit{Post-processing.} Block (h): The CC-based formalism enables an optional post-processing stage where the learned CCs are converted into analytical expressions using, e.g., symbolic regression tools. This approach simplifies the parametrization of the CCs, and for the case of the NN-CC method, it gives a low dimensional analytic representation of each constitutive relation. 
    
    \item[vii)] \textit{Forward simulations.} 
    Block (i): The discovered model is validated through forward integration, utilizing the original and new ICs and driving forces. This process allows for the definition of performance metrics that quantify the precision of the identified models. For instance, the root mean square error (RMSE) can be calculated by comparing the model forward simulations against reference theoretical trajectories. This approach is used for the discontinuous stick-slip example. Alternatively, for chaotic systems such as the Duffing oscillator, we define a separation time metric. 
    If these metrics reveal significant inaccuracies in the obtained models, we should further refine the model (outer loop).   
    Block (j): Once validated, the models are employed in forward simulations to generate production-level trajectories or for further characterization. 
 
\end{itemize}

Specifically, this work evaluates and compares NN-CC using blocks (e) and (h) in the presence of noise, along with other CC-based methods and standard methodologies.

\subsection{Application example to a second-order ODE family}
\label{sec:methodology_second_order}

While the CC-based methodology is adaptable to a broad range of families of systems (see \ref{app:uniquiness_CC}), in this work we focus on its application to a second-order family of systems, specifically to velocity-dependent friction systems (or generalized Rayleigh-type nonlinear oscillators)\cite{Gonzalez2025,Warminski2019}.  
This family exhibits a sufficiently rich dynamical landscape, including chaotic and non-smooth behaviors, to test the framework, while remaining low-dimensional enough to clearly illustrate the interpretability and uniqueness mechanisms.

This family can be written as
\begin{equation}
\ddot{x}(t)+f_1(\dot{x}(t))+f_2(x(t))=F_{ext}(t), \label{eq:model:veloc}%
\end{equation}
where $x(t)$ is the dynamical variable, the dotted subscripts refer to derivatives with respect to time, $F_{ext}(t)$ is an external driving force, $f_1(\dot{x})$ represents the velocity-dependent friction force, and $f_2(x)$ denotes the position-dependent elastic restoring force. The functions $f_1$ and $f_2$ are referred to as CCs within the CC-based formalism because they define the constitutive relations of friction and restoring elements.

\begin{figure*}[!htpb]
\centering
\includegraphics[width=12 cm]{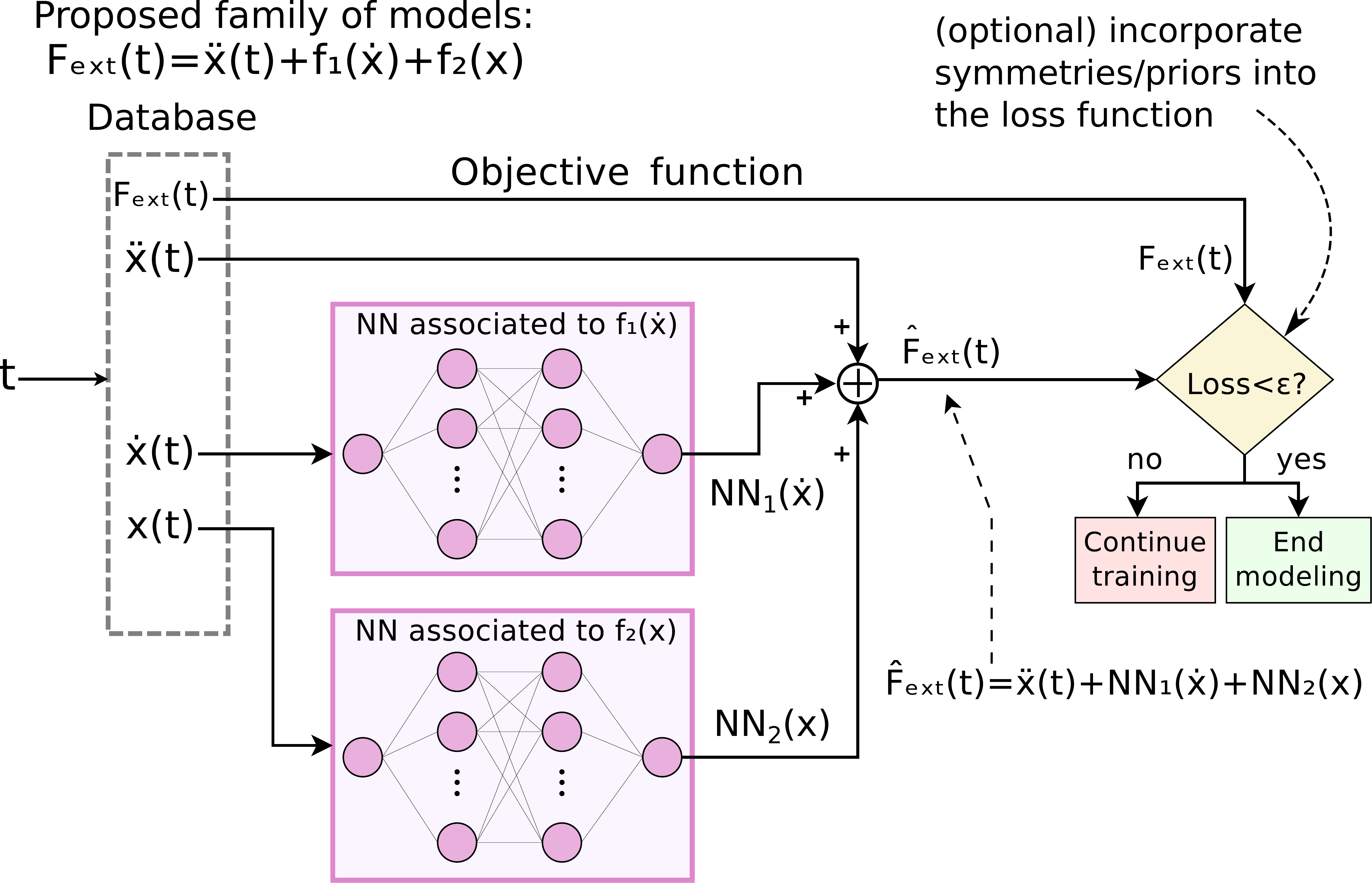}
\caption{Schematic diagram of the optimization process for the CC-based formalism applied to a family of second order systems. 
This schematic is parameterized using NNs (NN-CC method), but other basis functions for the CCs such as polynomials and sparse regression are also analyzed.}
\label{fig:schematic_second_order}
\end{figure*}

The practical implementation of the workflow for this family of models is illustrated in Fig.~\ref{fig:schematic_second_order}, exemplified for the NN-CC approach. As shown in the diagram: (i) each CC is represented by a dedicated neural network (NN$_1$ and NN$_2$); (ii) the outputs of these NNs are combined externally to satisfy the proposed family of models, providing an estimated driving force, $\hat{\text{F}}_{ext}$(t), which is compared against the objective function (i.e., the measured driving force, $\text{F}_{ext}$(t)); (iii) the training loop proceeds by minimizing this loss function (which may include additional symmetry terms or physical priors) to refine the parameterization of the CCs. 

It is important to emphasize again that while we have focused on the NN-CC approach in this explanation, the underlying logic of the workflow can be applied to other function approximators. Thus, an analogous scheme applies when expanding or expressing the CCs using other basis functions, such as polynomials, or within a sparse regression framework (e.g., LASSO or SINDy).

For example, within sparse regression CC-based frameworks, the architecture shown in Fig.~\ref{fig:schematic_second_order} remains conceptually identical, but: (i) instead of NNs, the CCs are represented as linear combination of library functions (e.g., 1, $x$,  $x^2$,  $\dot{x}$,  $\dot{x} \,|\dot{x}|$); (ii) the \textit{optimization} step is replaced by a sparse identification algorithm, such as LASSO, and the loss function is adapted to include the sparsity-promoting penalty (e.g., an $L_1$ norm) alongside optional symmetry terms.

This flexibility highlights that the CC-based formalism is a modular framework, allowing practitioners to choose the mathematical representation (e.g., NN representation, a set of basis functions, or using sparse regression techniques). 
This modularity provides an ideal platform to compare different methodologies under this unified formalism. In this work, we evaluate the following approaches:

\begin{itemize}

\item[i)] NN-CC: Each CC is represented by an independent NN, following the approach in Fig.~\ref{fig:schematic} but ignoring optional blocks (e) and (h).  This approach was previously discussed in Ref.~\cite{Gonzalez2025}; thus, here we only mention the main points.
In this approach, the unknown CCs $f_1(x)$ and $f_2(x)$ are represented by two independent feedforward NNs: $\text{NN}_1(x;\theta_1) \; \text{and}\;      \text{NN}_2(x;\theta_2)$, where $\theta_1$ and $\theta_2$ denote the respective weight and bias vector parameters. By selecting $x(t)$, $\dot{x}(t)$, $\ddot{x}(t)$ and $F_{ext}(t)$ as input variables for training (where derivatives may be directly measured or estimated from filtering procedures\cite{Kaptanoglu2021,Kaheman2022,Strebel2026}), we define a training dataset composed by $\{ x(t_i),\dot{x}(t_i),\ddot{x}(t_i),F_{ext}(t_i) \}_{i=0}^{N_{\text{data}}-1}$, where $N_{\text{data}}$ is the number of data points. We propose a family of models in agreement with Eq.~\ref{eq:model:veloc}. Then, the NN-predicted forcing, $\hat{F}_{ext}(t)$, is defined as: 
\begin{equation}
\hat{F}_{ext}(t)=\ddot{x}(t)+\text{NN}_1(\dot{x}(t)\,;\,\theta_1)+\text{NN}_2(x(t)\,;\,\theta_2)\; .
\label{eq:NN:predictedFext}
\end{equation}
To obtain $\theta_1$ and $\theta_2$, we minimize the mean-square error between the predicted and measured external forcings over the training dataset using the loss (or error) functional: 
\begin{equation}
    L_{\text{data}}(\theta_1,\theta_2)=\frac{1}{N_{\text{data}}} \sum_{i=0}^{N_{\text{data}}-1} \left[\hat{F}_{ext}(t_i)-F_{ext}(t_i)\right]^2  \; .
        \label{eq:loss}
\end{equation}
Optimization of $L_{\text{data}}$ with respect to both parameter sets $\{\theta_1,\theta_2\}$ yields the NN approximations of the CCs that best reproduce the dynamics under the given external forcing. By substituting the expression for $F_{ext}(t)$ [Eq.~\ref{eq:NN:predictedFext}] into Eq.~\ref{eq:loss}, we determine the optimal parameter vectors $\theta_1^*$ and $\theta_2^*$ that minimize the discrepancy between predicted and observed forcing. Formally, 
\begin{equation}
    \theta_1^*,\theta_2^*=\underset{\theta_1,\theta_2}{\text{argmin}} \; L_{\text{data}}(\theta_1,\theta_2) \; .
\end{equation}
This is performed by gradient optimization, $\nabla_{\theta_1}L$ and $\nabla_{\theta_2}L$, which are computed via backpropagation, and parameters are updated simultaneously in each iteration using a gradient-based optimizer (e.g., Adam or stochastic gradient descent):
\begin{equation}
\begin{cases}
\theta_1 &\leftarrow \theta_1 -\eta\; \nabla_{\theta_1}L_{\text{data}} \;\;\; \\ \theta_2 &\leftarrow \theta_2 -\eta\; \nabla_{\theta_2}L_{\text{data}} \;, 
    \end{cases}
\end{equation}
where $\eta$ is the learning rate. 
While the training process relies on standard gradient optimization techniques, this NN-CC-based methodology distinguishes itself from conventional PINNs strategies (which typically employ a single network for the entire problem) by utilizing one independent network for each CC, where all the NNs are optimized in tandem, meaning that the parameters of the all NNs are updated simultaneously to jointly satisfy the proposed equation. 

The importance of this decomposition is, for the case of the proposed equation, that the functions NN$_1$ and NN$_2$ are mathematically independent (except for an additive constant), thus there is a uniqueness in the representation (\ref{app:uniquiness_CC}). The ambiguity in the additive constant can be solved by adding a constraint (e.g. for the systems studied here, we used $f_2(0)=0$).  
In this NN-CC first approach, no symmetry constraints are imposed (except for the mentioned additive constant).

\item[ii)] NN-CC$_{\text{+post-SR}}$: A variant of NN-CC incorporating a post-processing step using SR. In this approach, the obtained CCs are evaluated at $N_{\mathrm{eval}}=1000$ points evenly spaced over the training data range, and SR is then applied to obtain analytical expressions for each CC. Then, forward simulations are performed using those analytical expressions.

\item[iii)] NN-CC$_{\text{+sym}}$: A variant of NN-CC incorporating symmetry constraints. Symmetry constraints are incorporated by adding the following term to the loss function:
\begin{equation}
L_{\mathrm{sym}}(\theta_1,\theta_2) 
=  \frac{1}{N_{\mathrm{sym}}} 
\sum_{j=1}^{N_{\mathrm{sym}}} 
\left[ \text{NN}_i(z_j;\theta_i) - \,s\, \text{NN}_i(-z_j;\theta_i) \right]^2,
\end{equation}

where $s=-1$ for odd symmetry and $s=+1$ for even symmetry, $i=\{1,2\}$, $\{z_i\}$ are points sampled from the domain of interest, and $N_{\mathrm{sym}} = 1000$ is the number of symmetry points, chosen equally spaced from the range of variations (minimum and maximum values) of the training data values. For the examples in this study, we enforce both NN$_1$ and NN$_2$ to be odd based on the observed CCs from the NN-CC method (see Sec.~\ref{sec:results}).  
 
The total loss is given by:
\begin{equation}
L_{\mathrm{tot}}(\theta_1,\theta_2) = L_{\mathrm{data}}(\theta_1,\theta_2) + \lambda_{\mathrm{sym}}\;L_{\mathrm{sym}}(\theta_1,\theta_2)\;,
\label{eq:loss_with_sym}
\end{equation}
where $L_{\mathrm{data}}(\theta_1,\theta_2)$ is the loss function of the NN-CC given by Eq.~\ref{eq:loss}, and $\lambda_\text{sym}$ is a regularization weight that balances the contribution of the symmetry constraint relative to the data-fitting loss. The selected value for $\lambda_{\mathrm{sym}}$, along with other hyperparameters of the NN-CC method are discussed in \ref{app:sensitivity_analysis}. Small values of $\lambda_{\mathrm{sym}}$ preserve the fidelity to the data while weakly enforcing symmetry, whereas excessively large values can over-prioritize symmetry and reduce the precision of the model to capture fine details present in the training data.

\item[iv)] NN-CC$_{\text{+sym+post-SR}}$: A variant of NN-CC incorporating both symmetries and post-SR procedures described above. 

\item[v)] Poly-CC: A method where the CCs are represented by a polynomial basis of degree $N$:
\begin{equation}
    f_i(z)\approx\sum_{j=0}^{N} c_{i,j} \;z^j\;,\;z=\{x,\dot{x}\}.
\end{equation}
We follow the same procedure as in Ref.~\cite{Gonzalez2025} by using $N=10$ and a domain rescaling strategy to improve numerical stability during the identification process. This approach has been shown to be effective for both first-order\cite{Gonzalez2023,Gonzalez2024} and second-order\cite{Gonzalez2025} systems, yielding errors comparable to those obtained with sparse regression techniques~\cite{Gonzalez2025}.

\item[vi)] SINDy: A method based on the SINDy framework\cite{brunton2016} with external forcing using the sequential thresholded least squares (STLSQ) optimizer with a sparsity threshold of $10^{-4}$. 
We enforce the additive model structure 
$\ddot{x}(t)+f_1(\dot{x}(t))+f_2(x(t))=k_0\,F_{ext}(t)$, where we utilize a generalized feature library composed of two distinct parts: a polynomial library for the state variables $(x,\dot{x})$ up to degree 10 with interaction terms disabled (i.e., excluding $x^n\dot{x}^m$ cross-terms), and an identity library for the external control input $F_{ext}(t)$. In this formulation, the coefficient $k_0$ is a free parameter identified by the algorithm.

\item[vii)] SINDy-CC: An extension of the previous SINDy method in which the model is rigorously constrained to match the physics of the external driving term, thereby enforcing $k_0=1$. This is implemented via the \texttt{ConstrainedSR3} (sparse relaxed regularized regression) optimizer~\cite{Kaptanoglu2021} using a linear equality constraint $C\,\mathbf{\xi}=\mathbf{d}$. The matrix $C$ contains a single non-zero entry corresponding to the index of the external force feature and $\mathbf{d}=[1]$ to force $k_0=1$.

\item[viii)] SINDy-CC$_\text{+sym}$:
A variant of SINDy-CC incorporating physical symmetry constraints into the candidate library. For the Duffing system, we apply a custom `odd library' restricted to polynomial terms with odd exponents ($x$, $x^3$, $\ldots$, $x^9$), excluding even powers and bias terms to strictly enforce $f(-x)=-f(x)$. 

\item[ix)] SINDy-CC$_\text{+sym+post-SR}$: 
A hybrid two-stage approach that incorporates post-SR to the structural identification of SINDy-CC$_\text{+sym}$. 

\item[x)] SR: a method based on the PySR package\cite{Cranmer2023PySR} where we use $x$ and $\dot{x}$ as input arguments and  $
F_{ext}-\ddot{x}$ as the objective function, thereby satisfying a model structure of the form $
F_{ext}-\ddot{x}=f(\,x,\dot{x}\,)$, where $f$ is identified via symbolic regression. 
We perform a post-processing analysis to determine whether the identified function $f(\,x,\dot{x}\,)$ admits a decomposition as $f_1(\dot{x})+f_2(x)$, consistent with the structure of the governing equation. 
For all SR models, the main hyperparameters used were:  
(1) number of iterations: $200$; 
(2) binary operators: \texttt{"+", "-", "*"};  
(3) unary operators: \texttt{"tanh"};  
(4) loss function: mean squared error;  
(5) number of populations: $10$;
(6) population size: $100$;  
(7) maximum equation size: $20$. 
We excluded the division operator \texttt{"/"} from the symbolic library to prevent potential singularities during forward integration. Furthermore, we selected \texttt{"tanh"} over \texttt{"sign"} to improve numerical stability, particularly for the stick-slip system.

\item[xi)] SR-CC: a structured variation of the SR method designed to strictly enforce a model structure of the form $
F_{ext}-\ddot{x}=f_1(\dot{x})+f_2(x)$. Unlike the standard SR approach, which searches the joint function space $f(x,\dot{x})$ and may produce crossed terms (e.g., $x\,\dot{x}$), SR-CC imposes the separation \textit{a priori} via an alternating optimization scheme (block coordinate descent). The algorithm iteratively identifies $f_1$ and $f_2$ in a sequential loop: (i) $f_2^\text{new}(x)$ is computed via SR using the residual target $y_2(x)=[F_{ext}-\ddot{x}]-f_1^\text{prev}(\dot{x})$, where $f_1^\text{prev}(\dot{x})$ is the estimate from the previous iteration (initialized as zero); (ii) $f_1^\text{new}(\dot{x})$ is computed via SR using the updated residual target $y_1(\dot{x})=[\ddot{x}-F_{ext}]-f_2^\text{new}(x)$. The computed functions are updated for the next iteration and  the cycle is repeated for a fixed number of outer iterations (set to 8 in this study, although stability was typically observed after 4 loops) to allow the functions to converge. The underlying engine and hyperparameters remain the same as in the standard SR method, though the number of generations per inner loop is reduced to 50 to maintain computational efficiency.

\item[xii)] Parametric: the CCs are defined explicitly using the full analytical expressions. For example, in the Duffing oscillator, we define $f_1(\dot{x})=a_1 \;\dot{x}$ and  $f_2(x)=a_2 \;x+a_3 \; x^3$, where $a_i$ ($i=1, \cdots ,3$) are identified by ordinary least-squares (OLS) regression.

\end{itemize}

It is important to emphasize that the methods (viii and ix) address a fundamentally more difficult identification problem than the CC-based approaches. While the CC framework utilizes physical priors to constrain the search to two univariate functions ($f_1(\dot{x})$ and $f_2(x)$), SINDy and SR must reconstruct the full bivariate function $f(x, \dot{x})$. This lack of structural constraints significantly expands the search space, increasing the complexity of the optimization and the risk of overfitting. 
Specifically, there is a potential issue in identifying spurious cross-coupling terms between position and velocity (e.g., $x\,\dot{x}$) that are not present in the ground truth system.

\section{Results}\label{sec:results}

Two representative examples are presented in this section: a chaotic Duffing system (Sec.~\ref{sec:duffing}), and a discontinuous stick-slip system (Sec.~\ref{sec:stickslip}). 
All numerical simulations in this study, including the theoretical simulations of the governing ODEs and the simulations for the different methods implemented, were performed using the LSODA solver (Livermore solver for ordinary differential equations with automatic method switching for stiff and non-stiff problems)\cite{Hindmarsh1983,Petzold1983}. This solver was chosen for its adaptive efficiency and numerical robustness, particularly for systems exhibiting stiffness, chaotic behavior, or rapid variations in the solution\cite{William2007numericalrecipes}.

For each simulation, we verified that the residuals of the governing equations at the integrated points remained small (below $10^{-8}$), confirming that the systems were correctly integrated. Simulations were carried out up to a maximum simulation time $t_{\rm max} = 40$~s, with the solutions interpolated to $N_{\rm data} = 1000$ uniformly spaced points. The only exception was the averaged separation time analysis for the Duffing system, where simulations were extended up to $t_{\rm max} = 80$~s to better capture the divergence between trajectories.

For the NN-CC approaches, we employed a NN architecture with three hidden layers, each comprising 100 neurons and rectified linear unit (ReLU) activation functions. 
This architecture and other hyperparameters are discussed in \ref{app:sensitivity_analysis}. 
Training was performed for up to 20.000 epochs or until the loss error reached $10^{-7}$, whichever was satisfied first.

\subsection{Example 1: Duffing oscillator in chaotic regime}\label{sec:duffing}

We consider a Duffing oscillator described by Eq.~\ref{eq:model:veloc}, where the CCs are defined as
\begin{equation}
\begin{cases}
    f_1(\dot{x})&=\delta\;\dot{x} \\f_2(x)&=\alpha \, x +\beta \, x^3 \;,
\end{cases}
\label{eq:cc:duffing}
\end{equation}

and the external forcing is given by $F_{ext}(t)=A\, \cos(\,\Omega\; t\,)$. 
For the definition of the CC parameters, we use $\alpha=-1.0$, $\beta=1.0$, and  $\delta=0.3$. The other coefficients will be varied, but for the first part of this section, we use $\Omega=1.2$, $A=0.5$, with ICs $x_0=0.5$ and $v_0=-0.5$. The quantities are expressed in nondimensional form for generality; thus, physical units are only relevant for a particular application\cite{Kovacic2011}.

To characterize the influence of noise during the identification stage, we use additive noise in $F_{ext}(t)$.
This approach serves as a controlled benchmark to evaluate the identification algorithms decoupled from preprocessing artifacts. While experimental noise typically affects the state variables ($x(t)$), introducing it at this stage of the formalism would conflate the model selection problem with the numerical differentiation problem. As noted in recent literature, the accuracy of identifying dynamics from noisy state measurements is inextricably linked to the specific choice of differentiation scheme and its hyperparameters\cite{VanBreugel2020,Egan2024}. To ensure a fair comparison of the core identification architectures without obscuring the results with filter-dependent performance, we focus the primary analysis on forcing noise. However, to validate the framework under state variable noise, we provide a supplementary analysis of noise on $x(t)$ utilizing a standard Savitzky-Golay differentiation scheme in \ref{app:noise_x}.

To establish a baseline for this analysis, a reference trajectory is first obtained by numerically integrating the Duffing equation with a noise-free external force, $F_{\rm ext}^{\rm clean}(t)$.  We then construct the noisy external force,  $F_{\rm ext}^{\rm noisy}(t)$, by introducing a stochastic component, $\varepsilon(t)$:

\begin{equation}
F_{\rm ext}^{\rm noisy}(t) = F_{\rm ext}^{\rm clean}(t) + \varepsilon(t),
\end{equation}
where $\varepsilon(t)$ is modeled as a zero-mean Gaussian white noise process, $\varepsilon(t)\sim \mathcal{N}(0,\sigma_{n}^2)$, where $\sigma_n$ is the standard deviation of the noise. 
The values of $\sigma_n$ are selected to satisfy a desired signal-to-noise ratio (SNR) in decibels (dB), based on this definition:
\begin{equation}
{\rm SNR(dB)} \equiv 10 \log_{10} \left(\frac{\sigma_{\rm clean}^2}{\sigma_{\rm n}^2}\right) \,,
\label{eq:snr_def}
\end{equation}
where $\sigma_{\rm clean}$ is the standard deviation computed for the clean external force as:
\begin{equation}
\sigma_{\rm clean}^2 = \frac{1}{N_{\text{data}}} \sum_{i=1}^{N} \big(F_{\rm ext}^{\rm clean}(t_i) - \bar{F}_{\rm ext}^{\rm clean}\big)^2 \,, 
\end{equation}
where $\bar{F}_{\rm ext}^{\rm clean}$ is the time-average of the signal. 
A noise fraction $\eta$ is often defined as:
\begin{equation}
\sigma_{\rm n} \equiv \eta \, \sigma_{\rm clean}\;.
\label{eq:sigma_noise}
\end{equation}
Using this definition, Eq.~\ref{eq:snr_def} 
can be written as
\begin{equation}
{\rm SNR(dB)}= 20 \log_{10} \left(\frac{1}{\eta}\right) \;.
\end{equation}

Thus, by setting different noise fractions, we adjust the SNR. For example, with noise fractions of $\eta$ = 0.01, 0.1, and 1.0 (i.e., percentages of noise of 1, 10 and 100 \%) we obtain SNRs of 40, 20, and 0 dB, respectively. This approach allows us to evaluate different models obtained under controlled noise levels.

\begin{figure*}[!htpb]
\centering
\includegraphics[width=16 cm]{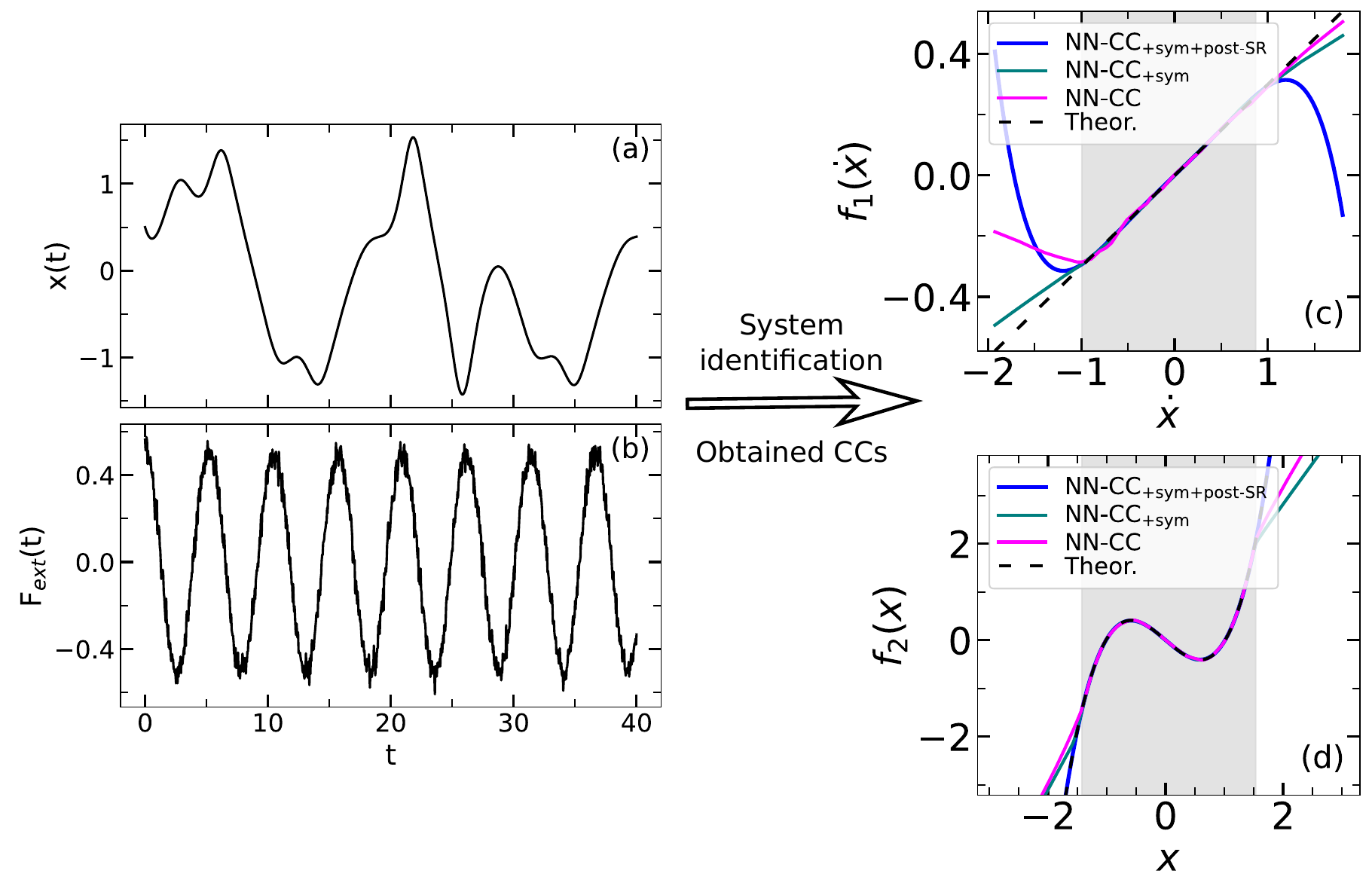}
\caption{System identification workflow based on the Duffing example. 
(a) and (b) show the input data: dynamical variable $x(t)$ and driving force $F_{ext}(t)$ with SNR~=~20~dB, respectively. (c) and (d) show the identified models obtained using the NN-CC$_{\text{+sym+post-SR}}$ method: $f_1(\dot{x})$ and $f_2(x)$, respectively. The gray zones indicate the range of training dataset values.}
\label{fig:duffing:trajectory}
\end{figure*}

Figure~\ref{fig:duffing:trajectory} illustrates the identification workflow based on an example with SNR~=~20~dB. Figures~\ref{fig:duffing:trajectory} (a) and (b) show the dynamical variable $x(t)$ and the external noisy force, $F_{\rm ext}^{\rm noisy}(t)$, respectively. For simplicity, the latter will be denoted simply as $F_{\rm ext}(t)$ hereafter.

Following the schematic procedure shown in Fig.~\ref{fig:schematic}, we select $x(t)$, $\dot{x}(t)$, $\ddot{x}(t)$, and $F_{\rm ext}(t)$ as the relevant variables and define the database [Fig.~\ref{fig:schematic}(b)]. We then propose the family of second-order velocity-dependent systems given by Eq~\ref{eq:model:veloc} [Fig.~\ref{fig:schematic}(c)]. This family is selected because we known beforehand the underlying mathematical structure. Nevertheless, other families could also be considered and subsequently rejected through the inner, middle, or outer loops of the identification workflow shown in Fig.~\ref{fig:schematic} (see \ref{app:model_selection} for further details). 

According to block (f) of Fig.~\ref{fig:schematic}, after the minimization process, we obtained total loss errors of $\approx1.1\times10^{-3}$ and $\approx1.2\times10^{-3}$ for NN-CC and NN-CC$_\text{+sym}$, respectively (see \ref{app:sensitivity_analysis} for more details).

Figures~\ref{fig:duffing:trajectory} (c) and (d) show the obtained CCs after the optimization stage for three identification methods: NN-CC, NN-CC$_\text{+sym}$, and NN-CC$_\text{+sym+post-SR}$ [corresponding to the step of Fig.~\ref{fig:schematic}(g)]. We also show the theoretical CCs with black dashed lines. 

The maximum and minimum values of $x(t)$ in the dataset are approximately 1.45 and -1.5, respectively (see the range of values in  Fig.~\ref{fig:duffing:trajectory} (a)). This defines a training data range shown by a gray zone in Fig.~\ref{fig:duffing:trajectory} (d) (analogously for $\dot{x}(t)$ and the corresponding gray zone of Fig.~\ref{fig:duffing:trajectory} (c)). 
Predictions beyond the gray regions correspond to extrapolations of the models since the models are evaluated outside of training data range. 
It can be seen that, within the training-data range, the obtained CCs from the three models are similar between them and also in good agreement with the theoretical curves, whereas noticeable differences between the obtained models emerge in the extrapolation regions.

A model that is able to recover CCs in better agreement with the theoretical CCs outside the training-data range can be regarded as having improved generalization properties, in the sense that it learns from data constrained to a limited range while producing reasonable behavior in unseen regions. We refer to this as `state-space extrapolation', to distinguish it from dynamics interpolation discussed below.

Extrapolation issues within the NN-CC framework have been previously investigated in Refs.~\cite{Gonzalez2024,Gonzalez2025}. These works considered linear extrapolation strategies, as well as polynomial basis expansions and sparse regression techniques for representing the CCs. The results showed that polynomial representations can exhibit poor extrapolation behavior, particularly when the true CCs asymptotically approach constant values outside the training domain.

In the present work, we do not further analyze generalization issues of kind 1, as we consider it a substantial topic that warrants separate investigation. Nevertheless, we consider important to mention that, the CC-based framework provides a convenient means to assess extrapolation issues, since the explicit visualization of the learned CCs offers a direct diagnostic tool to analyze whether the obtained models in extrapolated regions are physically plausible or potentially unreliable.

Extrapolation beyond the training domain requires a separate and careful analysis and is therefore not considered here. All forward simulations are restricted to the gray regions spanned by the training data, and state-space extrapolation is excluded.  
We therefore focus exclusively on dynamics interpolation, which probes the model ability to reproduce dynamics within the training domain under novel initial conditions and driving forces.

\begin{figure*}[!htpb]
    \centering
    \subfloat{\includegraphics[width=0.32\textwidth]{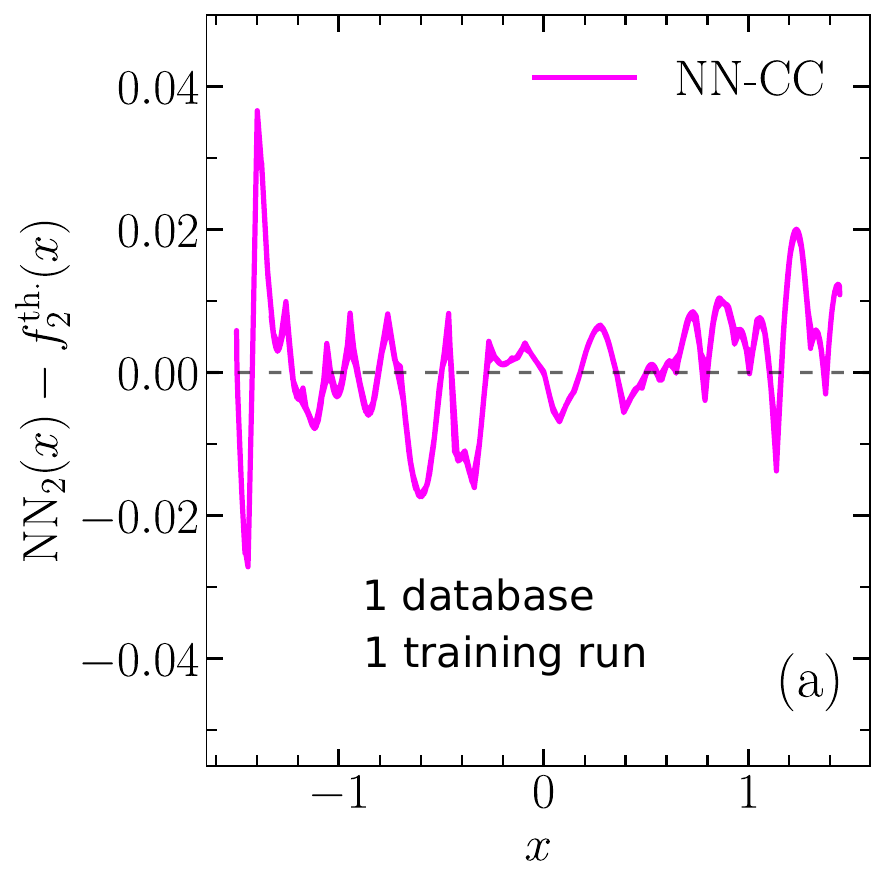}}
    \subfloat{\includegraphics[width=0.32\textwidth]{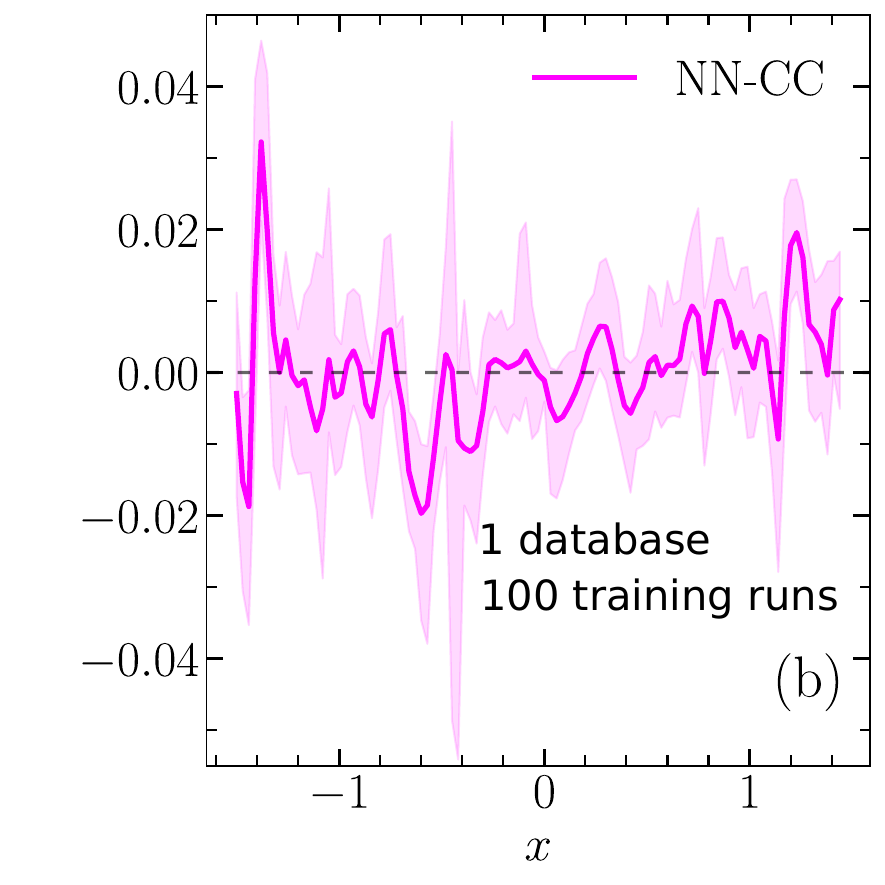}}    \subfloat{\includegraphics[width=0.32\textwidth]{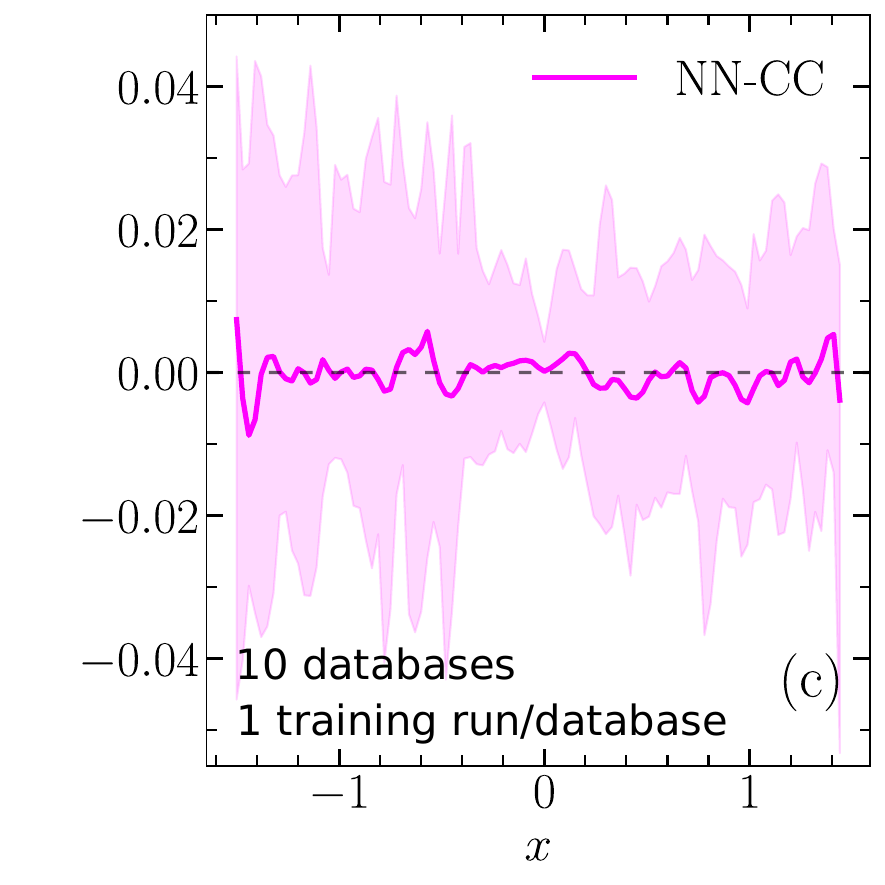}}\quad
    \subfloat{\includegraphics[width=0.32\textwidth]{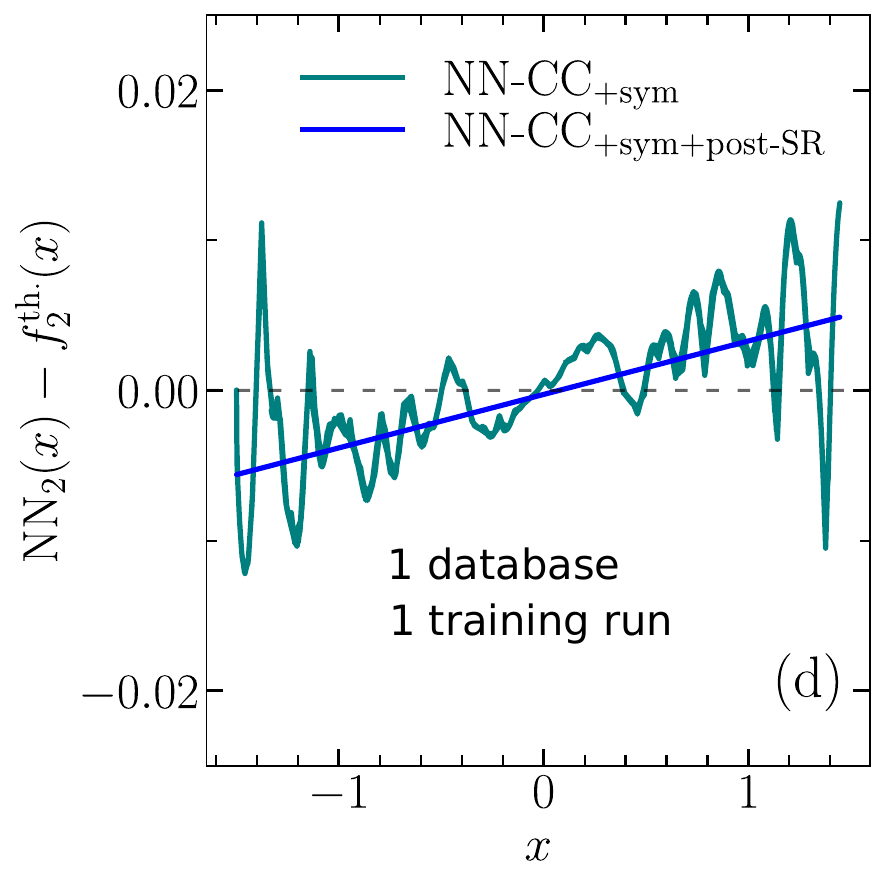}}
    \subfloat{\includegraphics[width=0.32\textwidth]{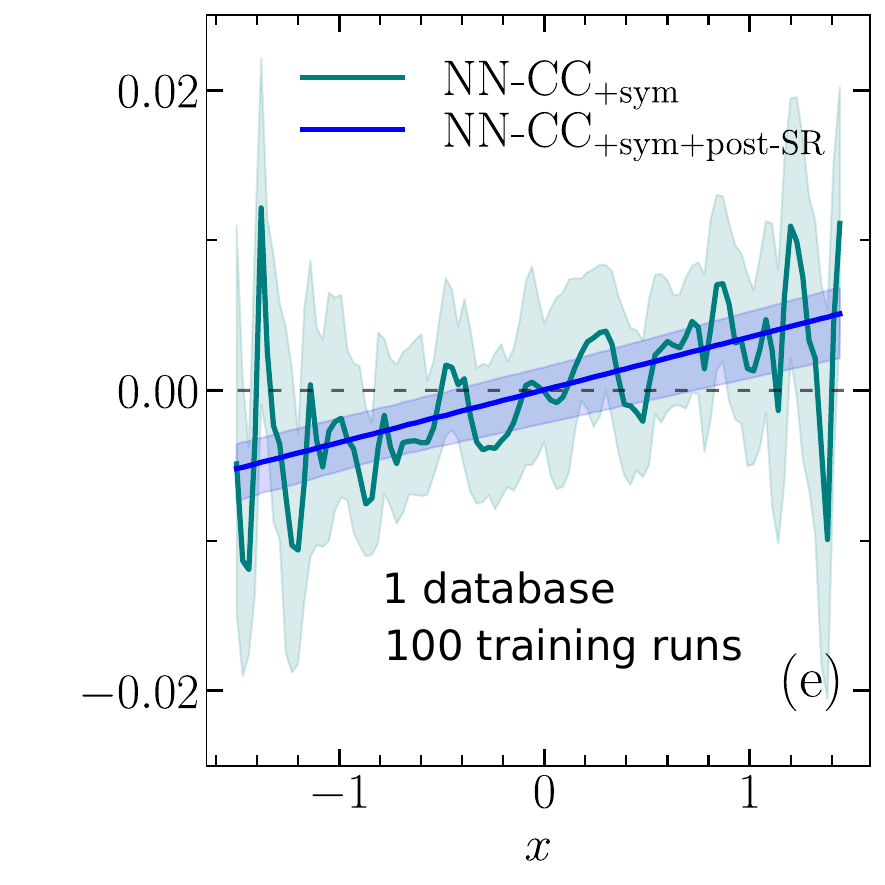}}
    \subfloat{\includegraphics[width=0.32\textwidth]{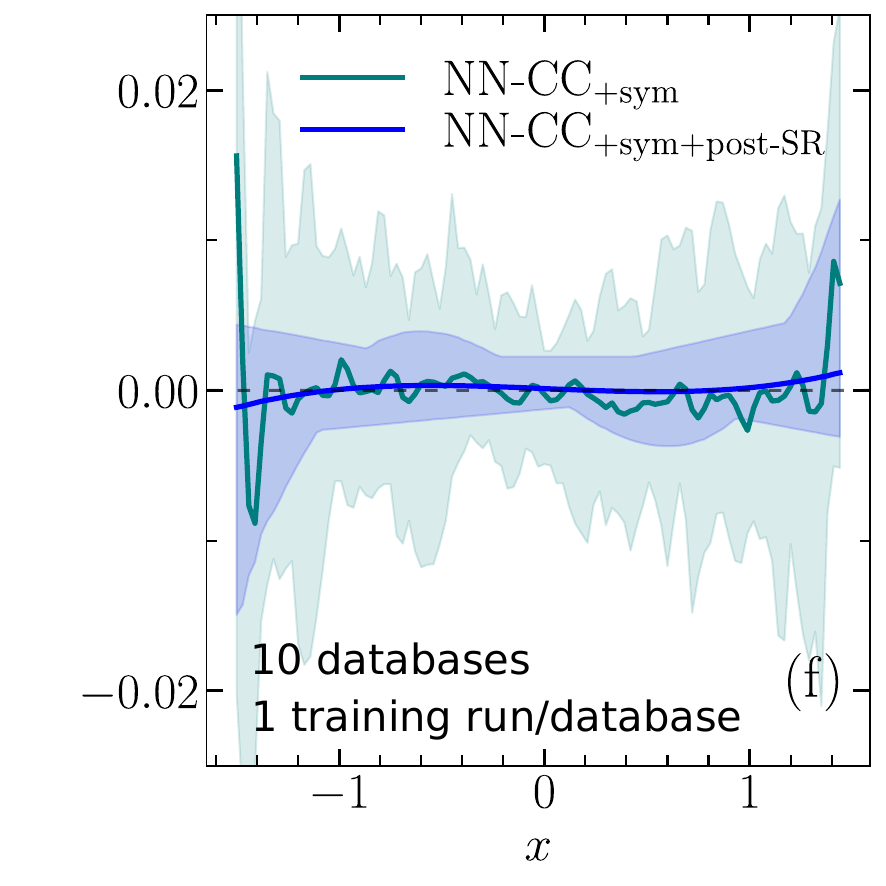}}
    \caption{
   Deviation of predicted CCs from theoretical values, NN$_2(x)-f_2^{\text{th}}(x)$. 
   Panels (a–c) correspond to NN-CC and panels (d–f) to NN-CC$_\text{+sym}$ and NN-CC$_\text{+sym+post-SR}$: (a,d) single training run; (b,e) 100 independent training instances; (c,f) 10 noise realizations with one training run per realization.
   In the 100-run cases, solid lines indicate the mean, while shaded regions show the [min, max] interval across runs.
    }
    \label{fig:duffing:overfitting}
\end{figure*}


Although the identified CCs in Fig.~\ref{fig:duffing:trajectory} are visually indistinguishable within the gray regions, Fig.~\ref{fig:duffing:overfitting} reveals some discrepancies. 
Figure~\ref{fig:duffing:overfitting}(a) shows the difference between identified and theoretical CCs for the $f_2$ function, defined as $\Delta f_2 = \text{NN}_2(x)-f_2^{th}(x)$, using the dataset with SNR $=20$ dB discussed above. It is evident that this residual error exhibits spurious oscillations around the ground truth. 

To investigate the origin of these oscillations, we performed a statistical analysis using 100 independent training runs on the same dataset. The results are shown in Fig.~\ref{fig:duffing:overfitting}(b), where the solid line corresponds to the mean prediction, and the shaded regions indicate the minimum and maximum bounds across all runs. Remarkably, the shaded regions are extremely narrow and tightly follow the mean, which itself retains the oscillatory pattern observed in Fig.~\ref{fig:duffing:overfitting}(a). 
To quantify the precision of the ensemble statistics, we define the mean interval width (MIW). Specifically $\text{MIW}_{\text{range}}$ corresponds to the average of the shaded regions in Fig.~\ref{fig:duffing:overfitting} and $\text{MIW}_{99\%}$ to the average range of the 99\% confidence interval. These values for each ensemble curve of Fig.~\ref{fig:duffing:overfitting} are shown in
Table~\ref{tab:miw_comparison}. Notably, for the baseline NN-CC method corresponding to Fig.~\ref{fig:duffing:overfitting}(b) (first row), the extremely small $\text{MIW}_{99\%}$ value of $<10^{-4}$ reinforces that all identified CCs are practically identical for this case of 100 training runs with one database. 
This indicates that the 100 independent models converge to practically the same solution, suggesting that the oscillations are not due to stochastic optimization noise (e.g., getting stuck in different local minima) but rather arising from a systematic overfitting bias to the specific noise realization present in the training data.

To validate this hypothesis, Fig.~\ref{fig:duffing:overfitting}(c) presents the results of an ensemble trained on 10 \textit{distinct} noise realizations (with one trained model per realization). In strong contrast to the previous case, the mean error (solid line) and the uncertainty bounds are now symmetric and centered around zero ($y=0$). This confirms that the bias observed in Fig.~\ref{fig:duffing:overfitting}(b) is indeed specific to the single noise instance used for training.  Thus, when averaged over multiple noise realizations, the spurious oscillations cancel out, and the method recovers the true physical curve without bias. 


\begin{table}[!htpb]
    \centering
    \caption{Comparison of mean interval width (MIW) metrics for the plots corresponding to Fig.~\ref{fig:duffing:overfitting}.}
    \label{tab:miw_comparison}
    \begin{tabular}{llcc}
        \toprule
        Fig.~\ref{fig:duffing:overfitting} & Method & $\text{MIW}_{\text{range}}$ & $\text{MIW}_{\text{99\% CI}}$ \\
        \midrule
        (b) & NN-CC & 0.022 & $<10^{-4}$ \\
        \multirow{2}{*}{(e)} & NN-CC$_{\text{+sym}}$ & 0.013 & $<10^{-4}$ \\
                            & NN-CC$_{\text{+sym+post-SR}}$ & 0.004 & $<10^{-4}$ \\        
        \addlinespace

        (c) & NN-CC & 0.037 & 0.003 \\
        \multirow{2}{*}{(f)} & NN-CC$_{\text{+sym}}$ & 0.016 & 0.001 \\
                            & NN-CC$_{\text{+sym+post-SR}}$ & 0.006 & 0.0006 \\
        \bottomrule
    \end{tabular}
\end{table}

It is important to note that this overfitting bias observed in Fig.~\ref{fig:duffing:overfitting}(b) is an expected consequence of the theoretical properties of the CC-based framework. As derived in \ref{app:uniquiness_CC}, the CC-based formalism ensures a unique mapping between the provided trajectory and the constitutive functions $f_1$ and $f_2$. Consequently, the inverse problem is injective: multiple NNs trained on the exact same noisy trajectory are mathematically constrained to converge to the same unique functions $f_1$ and $f_2$ that best fit that specific dataset, effectively `learning' the noise structure as part of the dynamics.

To analyze the impact of the ensemble strategies to the different NN-CC variants, Fig.~\ref{fig:duffing:overfitting}(d) presents the results for the same dataset used in Fig.~\ref{fig:duffing:overfitting}(a), but employing the NN-CC$_\text{+sym}$ and NN-CC$_\text{+sym+post-SR}$ models.
The inclusion of symmetry constraints (NN-CC$_\text{+sym}$) significantly attenuates the spurious oscillations around the ground truth compared to the baseline NN-CC; note that the $y$-axis scale has been reduced by half (notice also the reduction by approximately half on $\text{MIW}_{\text{range}}$ in Table~\ref{tab:miw_comparison}, from 0.022 to 0.013). 

This reduction by approximately half can be understood as follows. Let $\text{NN}_2(x): \mathbb{R} \to \mathbb{R}$ be a function such that both $f(x)$ and $f(-x)$ are well defined on the domain of interest. Then $\text{NN}_2(x)$ admits a unique decomposition into even and odd components, 
\begin{equation}
\text{NN}_2(x) = \text{NN}_2^{\mathrm{even}}(x) + \text{NN}_2^{\mathrm{odd}}(x),
\end{equation}
where
\begin{align}
\text{NN}_2^{\mathrm{even}}(x) &= \tfrac{1}{2}\bigl[\text{NN}_2(x)+\text{NN}_2(-x)\bigr],   \\
     \text{NN}_2^{\mathrm{odd}}(x) &=  \tfrac{1}{2}\bigl[\text{NN}_2(x)-\text{NN}_2(-x)\bigr] \,.
\end{align}
This decomposition follows from the fact that the only function that is both even and odd is the zero function. 

As a consequence, when odd symmetry is enforced in a NN, the hypothesis space is restricted to the odd subspace, thereby excluding all even-symmetric components from the representation. As a result, the model is prevented from fitting any even component present in the data.

For additive noise with zero mean and no preferred symmetry, the noise energy is, in expectation, equally distributed between the even and odd subspaces. Enforcing odd symmetry therefore removes the even-symmetric noise component, yielding an approximate factor-of-two reduction in the effective noise seen by the model. From this perspective, symmetry constraints act as a physics-informed inductive bias that reduces variance and stabilizes training.

The obtained $f_2(x)$ function with the NN-CC$_\text{+sym+post-SR}$ model shown in Fig.~\ref{fig:duffing:overfitting}(d) effectively filters out the remaining irregularities, correctly identifying the analytical structure $f_2(x)=-1.04\,x+1.02\,x^3$, which is in excellent agreement with the theoretical governing equation.

The ensemble statistics of the NN-CC variants are shown in Figs.~\ref{fig:duffing:overfitting}(e) and (f). Figure~\ref{fig:duffing:overfitting}(e) shows the ensemble statistics for 100 independent training runs with the same noise realization. While the NN-CC$_\text{+sym+post-SR}$ model suppresses spurious oscillations (evidenced by narrower shaded regions), it exhibits a systematic bias where the confidence bounds are not perfectly centered at $y=0$.   This confirms that while post-SR correctly identifies the functional form, the specific coefficients remain biased by the particular noise instance (i.e., $-1.04$ instead of $-1.05$). However, as shown in Fig.~\ref{fig:duffing:overfitting}(f), averaging over 10 different noise realizations eliminates this parametric bias, resulting in error bounds that are symmetric and centered at zero.

These findings suggest that symmetry constraints and post-SR work synergistically to minimize error and smooth the identified curves. The analysis demonstrates that while individual experiments are subject to realization-specific bias, ensemble averaging over multiple noise realizations recovers the unbiased system dynamics. In practice, this convergence can also be approached by extending data collection windows, effectively substituting ensemble averaging with time averaging.

To quantify the errors in the CCs, we define the ensemble root-mean-square error (RMSE) for each CC. This metric considers deviations across $N_\text{runs}$ independent training runs (corresponding to distinct noise realizations) and is defined as:
\begin{equation}
\mathrm{RMSE}[f_i(z)] = \left\{\frac{1}{N_{\mathrm{eval}}} 
\sum_{j=1}^{N_{\mathrm{eval}}} 
\frac{1}{N_{\text{runs}}} \sum_{k=1}^{N_{\text{runs}}}  
\left[ f_{i}^{(k)}(z_j) - f_{i}^{\mathrm{th}}(z_j) \right]^2 \right\}^{1/2}\; ,
\label{eq:rmse_fi}
\end{equation}

where $z$ represents the state variable corresponding to the function (i.e., $z=\dot{x}$ for $i=1$ and $z=x$ for $i=2$). Here, $f_i^{\mathrm{th}}$ denotes the theoretical ground truth, and $f_i^{(k)}$ represents the CC identified in the $k-th$ independent run. The error is computed over a grid of $N_{\mathrm{eval}}=1000$ points $z_j$, uniformly spaced between the minimum and maximum values of the train domain. For this analysis, we used $N_\text{runs}=10$ to ensure statistical robustness against specific noise realizations, as discussed above.

\begin{figure*}[!htpb]
    \centering
    \subfloat{
        \includegraphics[width=0.35\textwidth]{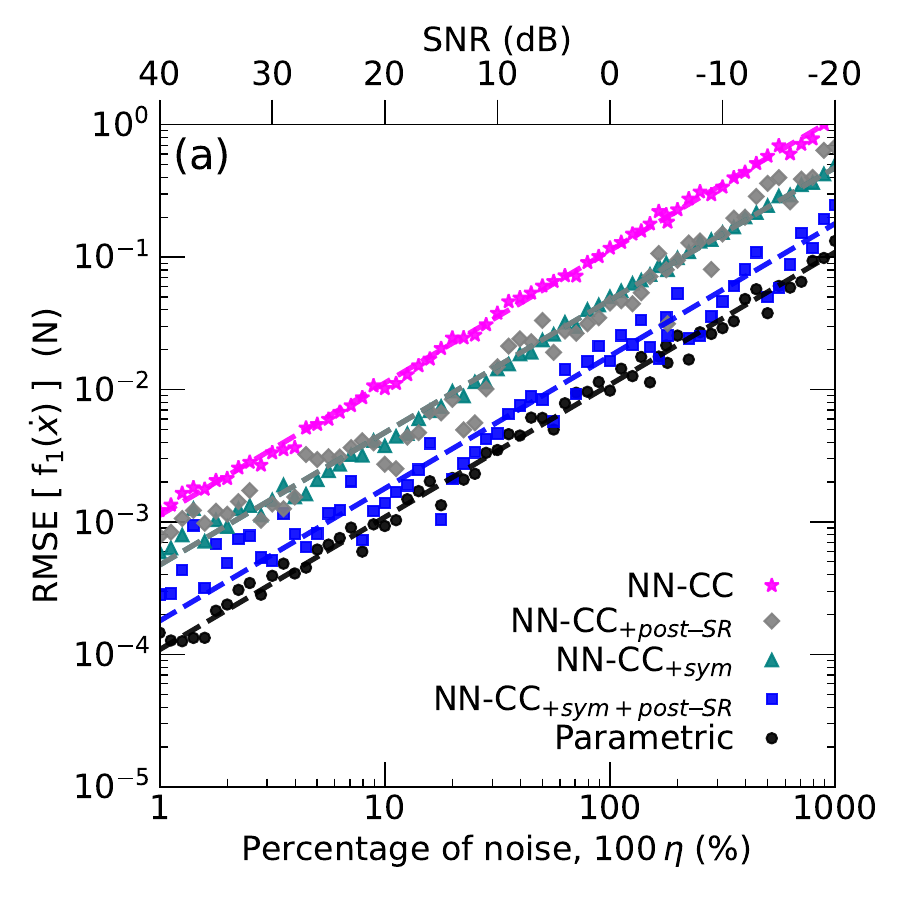}
    }
    \subfloat{
        \includegraphics[width=0.35\textwidth]{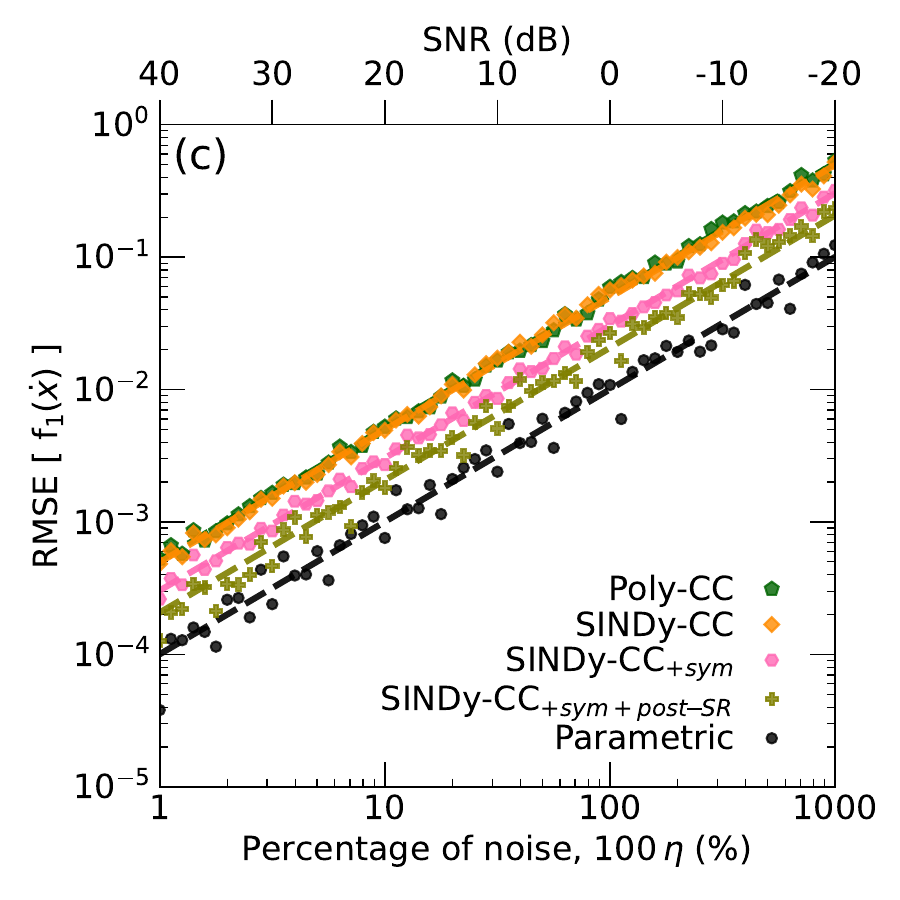}
    }
    \subfloat{
        \includegraphics[width=0.35\textwidth]{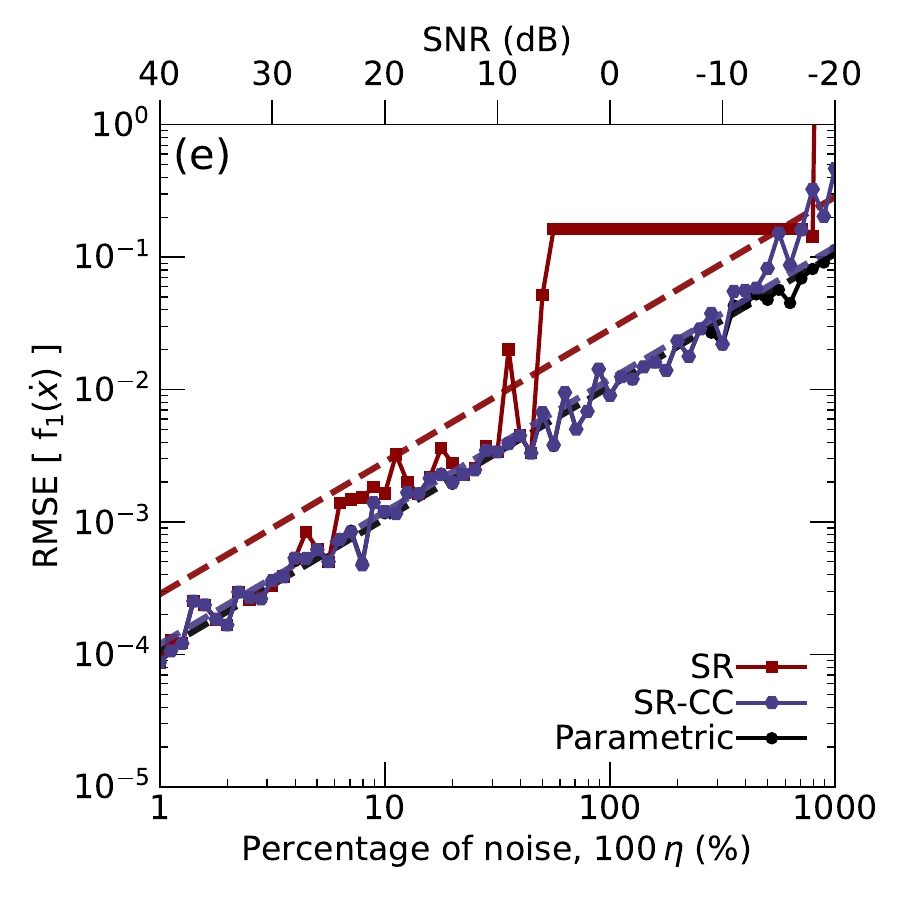}
    }\quad
    \subfloat{
        \includegraphics[width=0.35\textwidth]{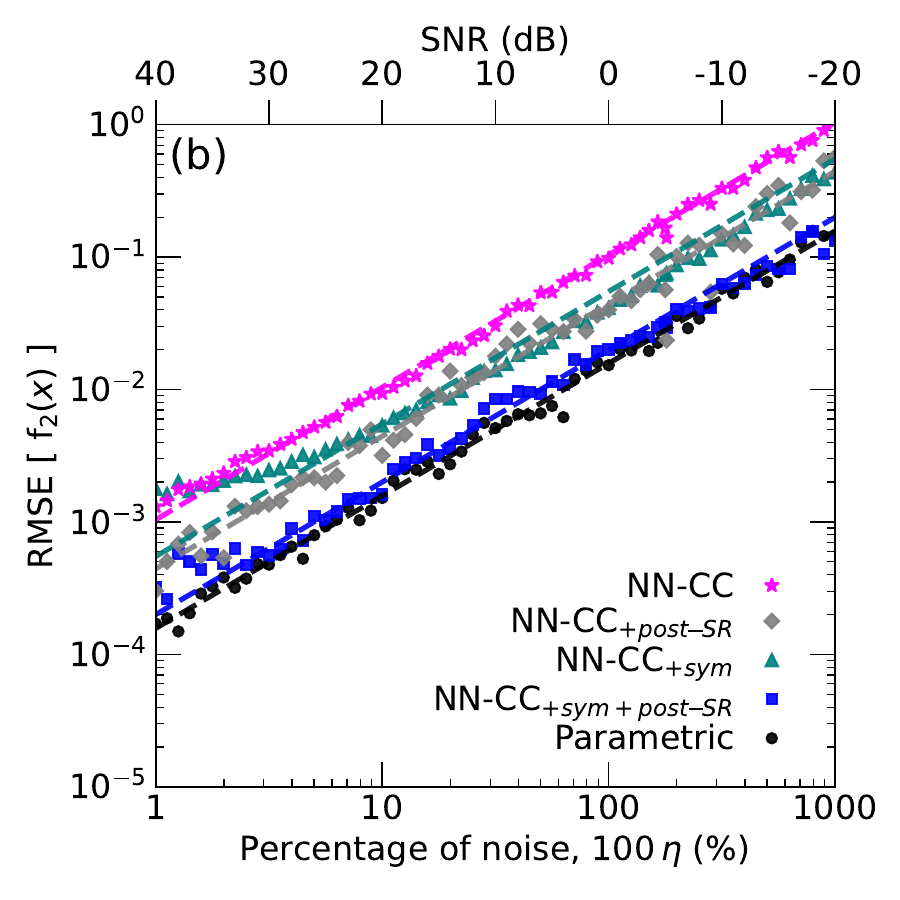}
    }
    \subfloat{
        \includegraphics[width=0.35\textwidth]{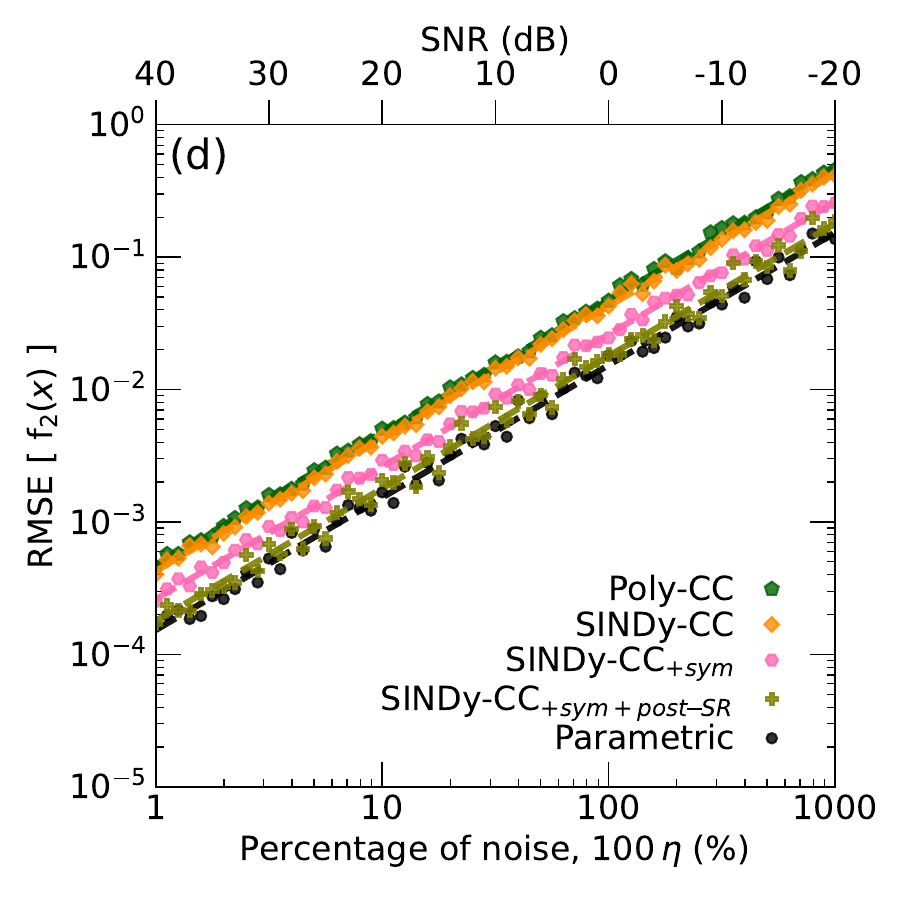}
    }
    \subfloat{
        \includegraphics[width=0.35\textwidth]{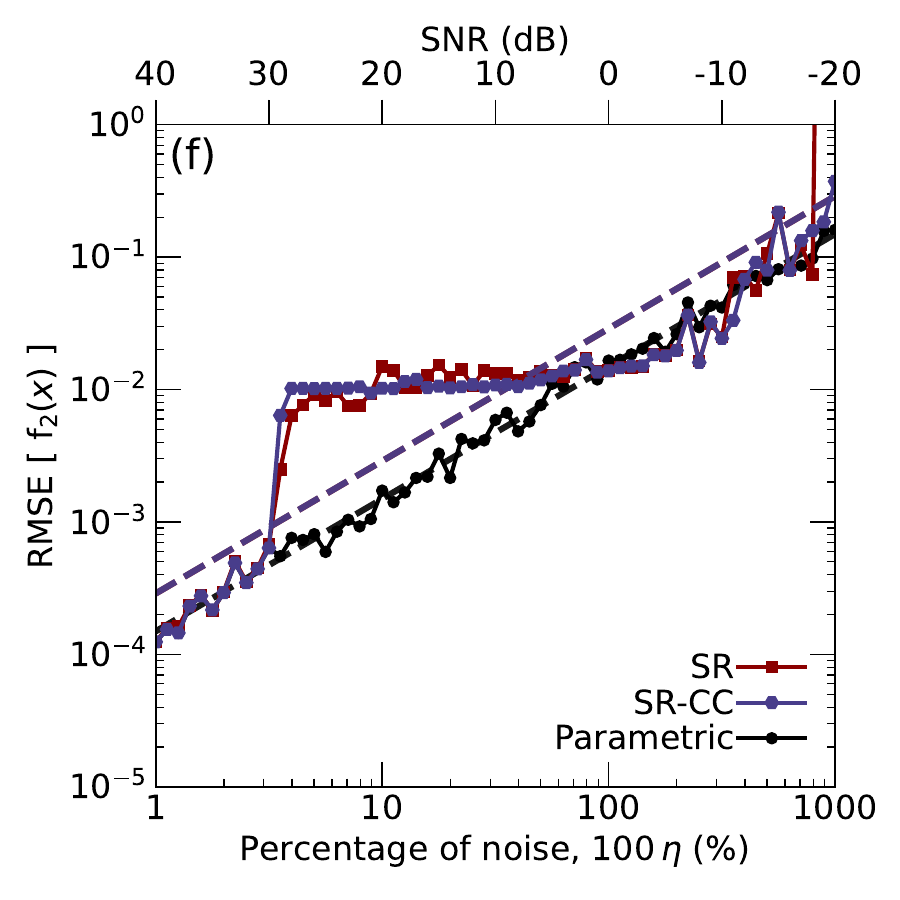}
    }
    \caption{RMSE analysis for the identified CCs ($f_1$ and $f_2$) for the Duffing system. Panels (a) and (b): NN-CC variants; (c) and (d): Polynomial basis variants; (e) and (f): SR variants. }
    \label{fig:duffing:sr}
\end{figure*}

Figure~\ref{fig:duffing:sr} presents the computed RMSE values across a broad range of noise levels (sampled at 1 dB intervals).  
Firstly, we observe that the Parametric method reaches the lowest RMSE values. This is not surprising since it incorporates as much information as possible about the system. By assuming complete a priori knowledge of the governing equation functional form (i.e., $\ddot{x}+\theta_1\,\dot{x}+\theta_2 \,x+\theta_3\,x^3=F_\text{ext}(t)$), it reduces the problem to a simple least-squares regression of three parameters ($M=3$). 
Since the variance of a linear estimator is proportional to the number of parameters $M$ (see \ref{app:noise_scaling}), the Parametric method yields the lowest possible RMSE achievable for the given dataset. Any method that discovers the structure (where the number of parameters $M$ is unknown or effectively larger) must necessarily have RMSE values equal to or greater than this baseline. Thus, the Parametric method can be considered as the theoretical lower bound for benchmarking the other methods.

Secondly, notice that the RMSE curves for the NN-CC and Polynomial variants (SINDy-CC and Poly-CC) exhibit a clear linear scaling on the log-log scale across the entire SNR range studied. 
This linear behavior is mathematically grounded in the asymptotic properties of least squares and maximum likelihood estimators. As detailed in \ref{app:noise_scaling}, for a linear parameter estimation problem under Gaussian noise, the standard deviation of the estimates in the noise-dominated regime scales linearly with the noise standard deviation.

Consistent with this theoretical expectation, we model the RMSE as a linear function of the noise fraction $\eta$:
\begin{equation}
    \text{RMSE}[f_i(z)]=A\;\eta\, .
\label{eq:fitting_rmse_empirical}
\end{equation}
On a log-log scale, this relationship is expressed as 
\begin{equation}
    \log\left\{\text{RMSE}[f_i(z)]\right\}=\log(A) + \;\log(\eta)
    \;.
\label{eq:fitting_rmse_duffing}
\end{equation}

The dashed lines in Fig.~\ref{fig:duffing:sr} correspond to fits using Eq.~\ref{eq:fitting_rmse_duffing}. 
The intercepts of these fits, represented by the parameter $A$, provide a direct metric for comparing model performance: lower $A$ values indicate superior accuracy and noise rejection. 
The results are summarized in Table~\ref{tab:fitting_duffing}, revealing a distinct performance hierarchy.
For the SR variants, which exhibit the jumps discussed later in this section, we provide an overall \(A\) value. 
This gain in precision involves a trade-off in terms of computational cost; a detailed comparison of the simulation times for each identification technique is provided in \ref{app:computational_cost}.

\begin{table}[!htpb]
    \centering
    \caption{Fitted parameters ($A$) for the Duffing oscillator. Lower values indicate smaller deviations from the theoretical CCs. Superindex $^{(*)}$ indicates the fit was performed only in the stable regime (SNR $\geq -18$ dB) where the correct model structure was identified.}
    \label{tab:fitting_duffing}
    \begin{tabular}{lcc}
        \toprule
        & \multicolumn{2}{c}{Fitted $A$ values} \\
        \cmidrule(lr){2-3}
        Model & $\mathrm{RMSE}[f_1]$ & $\mathrm{RMSE}[f_2]$ \\
           & ($10^{-2}$) & ($10^{-2}$)\\
        \midrule
        Parametric & $1.09$ & $1.58$ \\
        SR-CC & $1.19$ & $2.89$ \\
        SINDy-CC$_{\text{+sym+post-SR}}$ & $2.07$ & $ 1.79$ \\
        NN-CC$_{\text{+sym+post-SR}}$ & $1.79$ & $2.0$ \\
        SINDy-CC$_{\text{+sym}}$ & $3.03$ & $ 2.61$ \\
        SR$^{(*)}$ & $2.85$ & $2.88$ \\
        SINDy-CC & $5.1$ & $4.36$ \\      
        NN-CC$_{\text{+post-SR}}$ & $4.74$ & $4.44$ \\
        Poly-CC & $5.23$ & $4.75$ \\      
        NN-CC$_{\text{+sym}}$ & $4.73$ & $5.5$ \\
        NN-CC & $11.2$ & $10.4$ \\
        \bottomrule
    \end{tabular}
\end{table}

The NN-CC methods, shown in Fig.~\ref{fig:duffing:sr}, show that the baseline NN-CC model exhibits the highest $A$ values, which is not surprising given its wider search space. 
Introducing symmetry constraints (NN-CC$_{\text{+sym}}$) yields a measurable improvement ($\approx 50\%$ reduction in $A$). By enforcing parity, we effectively prune the space of admissible functions, acting as a symmetry-informed regularizer, thus reducing the search space by approximately one half. 

Applying post-SR alone (NN-CC$_{\text{+post-SR}}$) leads to a reduction of the $A$ values to nearly 60\% compared to the baseline NN-CC, yielding to a slightly better performance than using only symmetry constraints. 

The most significant performance is achieved by combining both strategies. The NN-CC$_{\text{+sym+post-SR}}$ model reaches a reduction in the $A$ values to $\approx$80\% compared to the baseline NN-CC. This result highlights a cumulative effect: the symmetry constraints add a strong bias that improves RMSE values, and the subsequent post-SR step smooths the CCs.

Figure~\ref{fig:duffing:sr} compares the different methods that use polynomial basis functions. Despite employing a polynomial basis capable of fully capturing the system dynamics, SINDy-CC and Poly-CC exhibits higher values of $A$ compared to the Parametric baseline. 
This difference reflects the cost of model-form uncertainty: while the library contains the true terms, the algorithm frequently retains spurious basis functions due to noise, resulting in an effective basis size larger than the minimal set (i.e., finding $M>3$) which naturally increases the estimation errors. 
Although advanced formulations (ensemble \cite{Fasel2022}, probabilistic \cite{Zhu2024sparse,Hsin2024,Meng2025}, or implicit \cite{Kaheman2020,Anvari2025}) exist to mitigate this issue, the fundamental challenge remains: distinguishing physics from noise in an unconstrained search is inherently more difficult than fitting the true Parametric structure.

This fragility is even more pronounced in the SR variants from Figs.~\ref{fig:duffing:sr} (e) and (f). In the high-SNR regime ($>30$ dB), SR and SR-CC successfully identifies the underlying theoretical values obtaining errors in the parameters below 0.1\$ thus obtaining similar RMSE than the Parametric method. 
However, for SNR$ > 30$ dB, both SR and SR-CC tend to oversimplify the stiffness term on $f_2$, reducing $\beta x^3$  to a unitary $x^3$, thus producing a practically constant plateau on Fig.~\ref{fig:duffing:sr}(f). A discrepancy is observed between SR and SR-CC for SNR$\lesssim 7$ dB. At this level, SR fails to identify the dynamical structure, collapsing $f_1(\dot{x})$ to a constant, whereas SR-CC successfully recovers the correct functional form $f_1(\dot{x})=c_1\,\dot{x}$. This behavior can be attributed to the trade-off between parsimony and error minimization. In the presence of high noise, standard SR discards the velocity term because the marginal reduction in error does not justify the penalty for increased model complexity. SR-CC circumvents this by explicitly constraining the search space to include a function of $\dot{x}$; this inductive bias forces the algorithm to fit the dissipative dynamics, which is orthogonal to the stiffness, even when the signal is heavily obscured by noise.

For SNR$<-18$ dB, the SR method identifies spurious cross-terms between $x$ and $\dot{x}$, rendering the decomposition of $f(x,\dot{x})$ into $f_1(\dot{x)}$ and $f_2(x)$ unfeasible. By constraining the search with the SR-CC method, we avoid these spurious terms by construction.

\begin{figure*}[!htpb]
    \centering
    \includegraphics[width=1.0\linewidth]{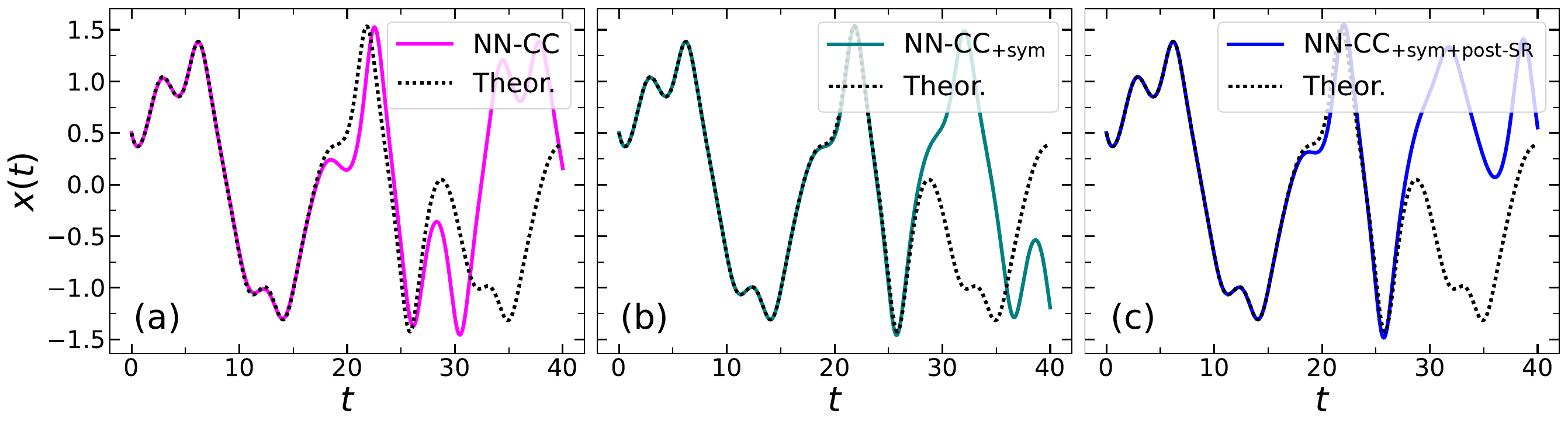}\\
    \includegraphics[width=1.0\linewidth]{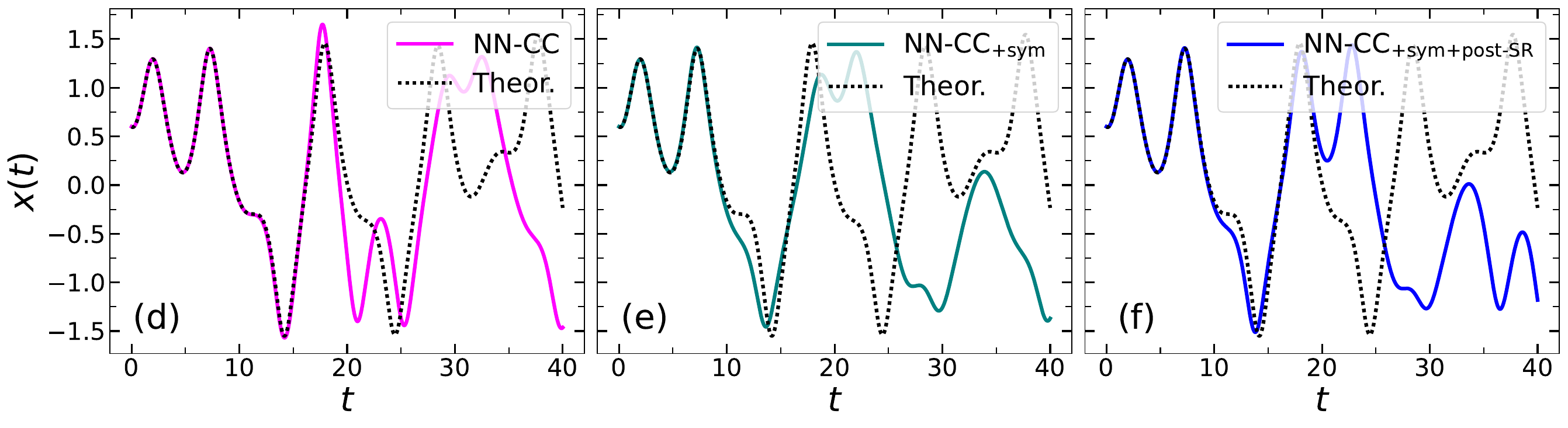}
    \caption{Forward integrations of different models and using different ICs: Panels (a-c) show integrations using the same ICs as those in the training dataset, while panels (d-f) show integrations with different ICs.}
    \label{fig:duffing_forward_simulations}
\end{figure*}

The analyzes presented thus far have focused on the direct error quantification of the identified CCs [according to Fig.~\ref{fig:schematic}(g)].  
Although this metric offers important structural insight, it is feasible only when the discovered model respects the structure of Eq.~\ref{eq:model:veloc}.  Methods such as SR and SINDy often yield models contaminated with spurious crossed-terms (e.g, $x\,\dot{x}$), which  prevent a direct comparison under the RMSE metric for the CCs.  
This structural violation was observed for SR in the low-SNR regime (SNR $<-18$ dB) and for SINDy at SNR $<10$ dB (the latter was omitted for clarity in Fig.~\ref{fig:duffing:sr}).

In order to compare all methods on an equal footing—regardless of their internal mathematical structure, we must adopt another metric based on forward integration [see Fig.~\ref{fig:schematic}(i)].

Figures~\ref{fig:duffing_forward_simulations}(a-c) show the forward integrations for the NN-CC variants compared to the ground truth, initialized with the same ICs used in the training dataset. The models appear to show strong agreement for a significant period of time. However, this visual agreement can be misleading. Since the Duffing oscillator operates in a chaotic regime, sensitivity to ICs is expected; yet, a model might `memorize' the specific training trajectory without learning the underlying grounded equations.
This risk is evidenced in Fig.~\ref{fig:duffing_forward_simulations}(b), where the NN-CC$_\text{+sym}$ model appears to track the truth accurately over a longer time interval compared to the other NN-CC variants. However, this performance is not sustained when the models are integrated with other ICs ($x_0=0.6$ and $v_0=-0.1$), as shown in Figs.~\ref{fig:duffing_forward_simulations}(d-f).  

These effects are intrinsically related to `dynamics interpolation' mentioned above, which measures the ability of the model to predict system dynamics in previously unexplored regions of the phase space but within the bounds of the training dataset. To evaluate this capability, we perform forward integrations with both the theoretical equation and the identified models using \textit{novel} ICs and driving forces, $F_\text{ext}(t)$, which are different from those used during training. Importantly, the parameters defining the CCs are kept fixed.  

To clearly distinguish interpolation dynamics from state-space extrapolation, we restrict these forward simulations to remain within the phase-space bounds $[x_{min},x_{max}]$ and $[\dot{x}_{min},\dot{x}_{max}]$ defined by the training database. Accordingly, we select ICs and driving forces such that the theoretical forward simulations remain within these bounds. This procedure ensures that the observed performance reflects only dynamics interpolation, without contamination from extrapolation effects.

\begin{figure}[!htpb]
    \centering
    \includegraphics[width=0.4\linewidth]{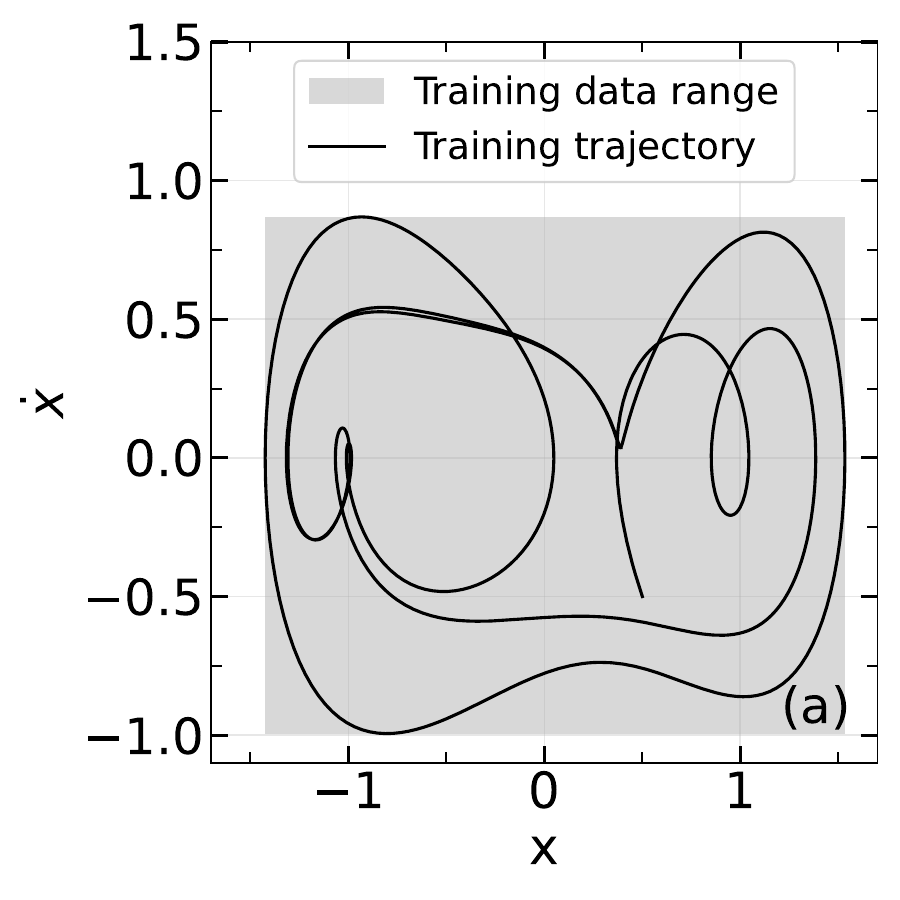}
    \includegraphics[width=0.4\linewidth]{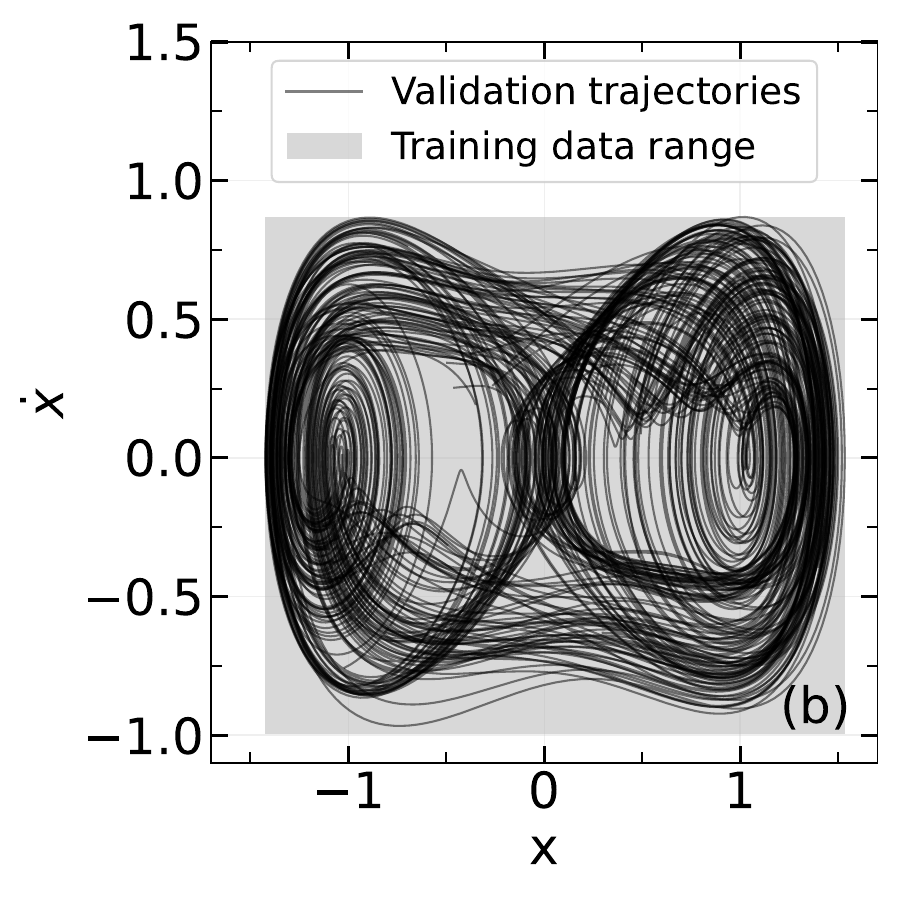}\caption{Illustration of the concept of dynamics interpolation on the phase space. (a) training dataset; (b) 10 validation trajectories generated by forward integrations of the theoretical equation using different ICs and driving forces. Gray zones indicate the phase-space range covered by the training data.}
    \label{fig:generalization_k2_phspace_duff}
\end{figure}

Figure~\ref{fig:generalization_k2_phspace_duff} illustrates the concept of dynamics interpolation. The training trajectory shown in panel (a) is generated using the parameters defined at the beginning of this section. 
To validate the interpolation capabilities of the models, we integrate 10 forward simulations using randomly selected ICs   and forcing amplitudes within the ranges $x_0,v_0 \in [-0.5,0.5]$, $A \in [0.45,0.5]$, $\Omega \in [1.1,1.3]$, as depicted in panel (b). Within this range of values, the system maintains its chaotic behavior. These theoretical trajectories are then compared with those generated by the identified models to evaluate their predictive performance in the interpolation regime.

To quantify this predictive performance, we will define a timescale relative to the inherent predictability horizon of the system.
Hence, to quantify the chaotic dynamics of the identified model, we computed the Largest Lyapunov Exponent (LLE) using the variational equation method \cite{Benettin1980_Part1,Benettin1980_Part2}. The identified system yields a positive exponent of $\lambda_\text{max}\approx 0.13$, confirming the presence of a strange attractor. This corresponds to a Lyapunov time (predictibility horizon) $T_L\approx 7.8$.

This time allows us to define a predicted accuracy horizon as the duration for which the integrated trajectory remains within a predefined tolerance of the theoretical solution. Specifically, 
we define the separation time ($t_\text{sep}$) as the smallest time $t$ such that 
\begin{equation}
|x_\mathrm{pred}(t) - x_{\mathrm{ref}}(t)| \geq \epsilon \;,
\end{equation}
where $\epsilon$ is a tolerance distance, set to $\epsilon=0.2$, which corresponds to approximately 15\% of the dynamic range of $x$.  
A simulation is classified as successful if the separation time satisfies $t_\text{sep}\geq T_L$.

This criterion ensures that the identified model reproduces the system dynamics up to the intrinsic predictability horizon, thereby distinguishing consistent system identification from mere short-term predictability. Although this benchmark is an empirical heuristic rather than a fundamental theoretical requirement, it provides a well-defined and practical metric for comparing different identification techniques under identical chaotic conditions.

To obtain a robust value for the $t_\text{sep}$ metric, we performed an average using a set of forward simulations as follows. 
For each SNR level, we generated $10$ datasets using different noise realizations, and performed a training run for each dataset.  Then, each of these models was subjected to $10$ distinct forward integration tests with randomized amplitudes and ICs within the intervals defined above. This yields a total of $100$ forward integrations per SNR value.

\begin{figure}[!htpb]
    \centering
    \includegraphics[width=0.33\linewidth]{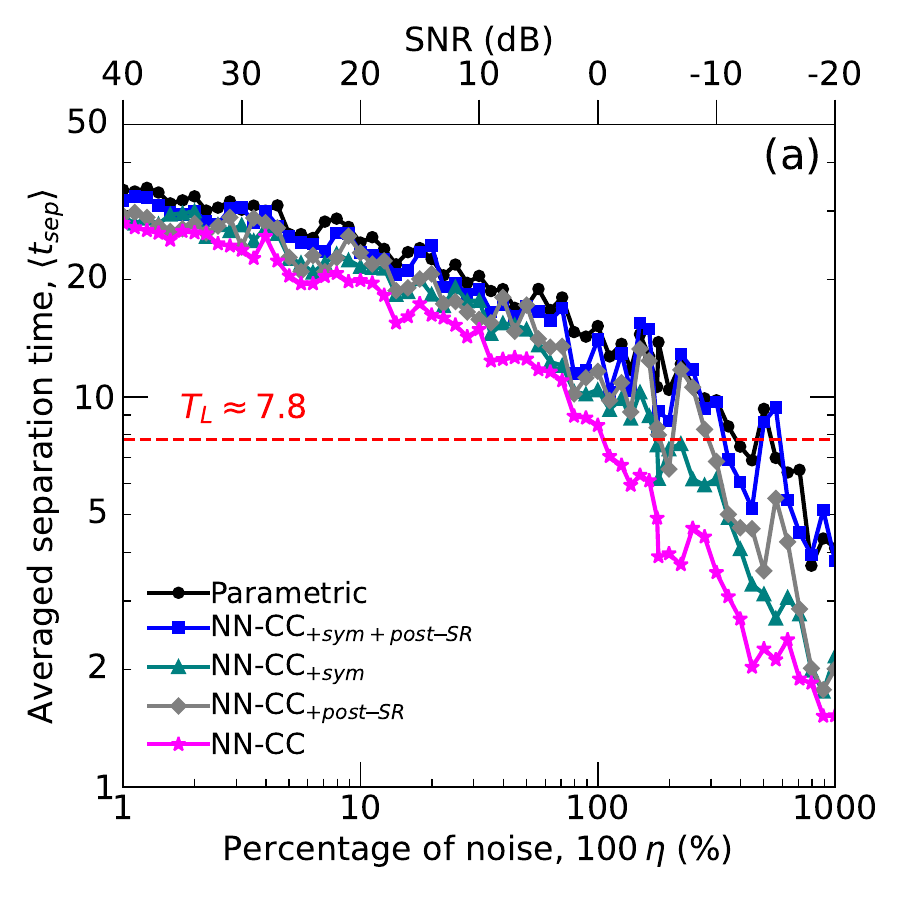}
    \includegraphics[width=0.33\linewidth]{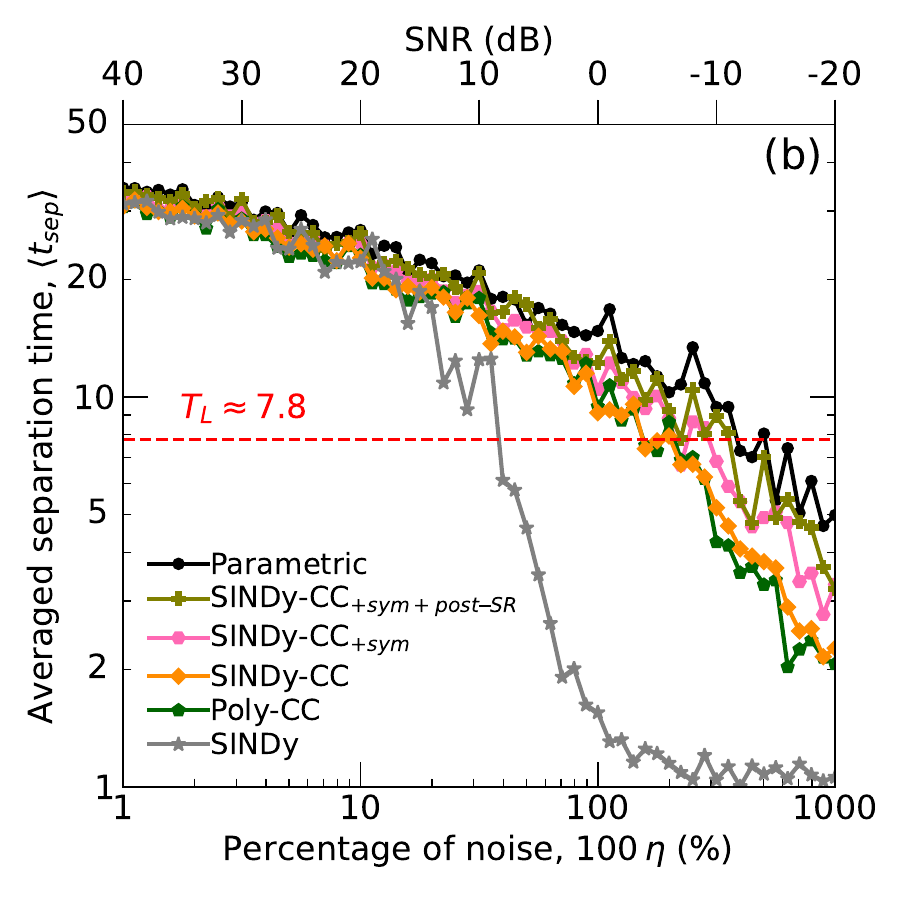}
    \includegraphics[width=0.33\linewidth]{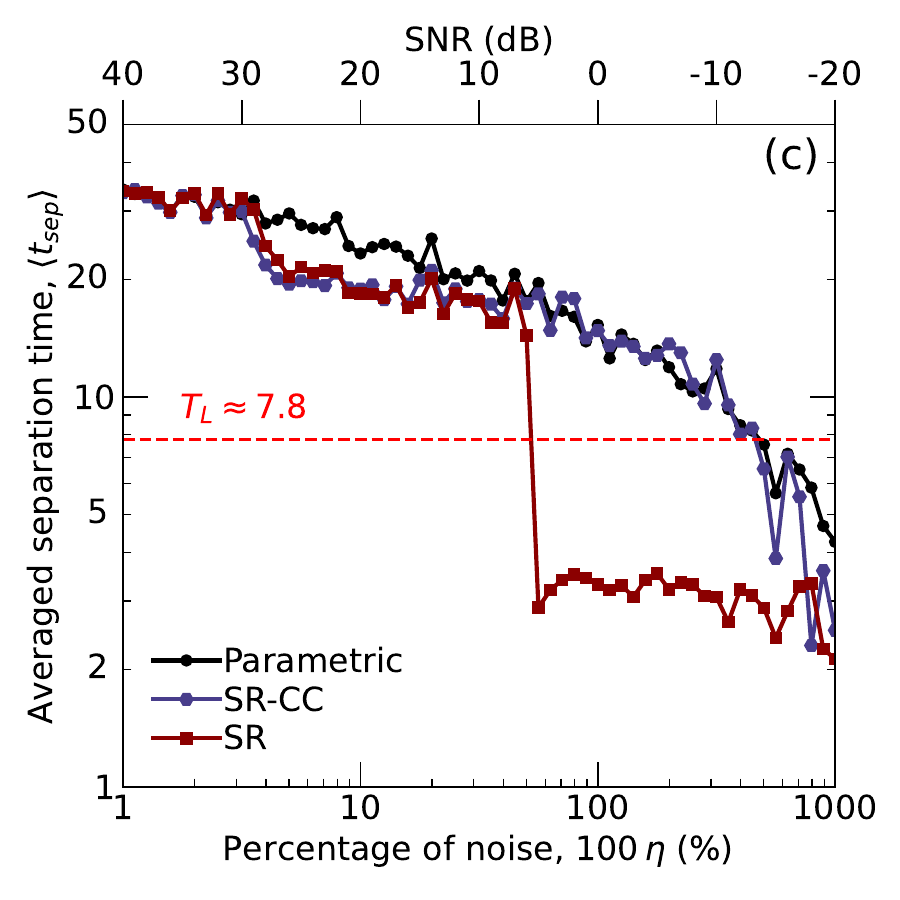}
    \caption{Averaged separation time as a function of noise percentage for the different identification models. This metric quantifies the predictive horizon (how long the model remains close to the theoretical trajectory before diverging due to the chaotic regime). The dashed red line indicates the Lyapunov time, $T_L\approx 7.8$. (a) NN-CC variants; (b) Sparse regression and polynomial basis variants; (c) SR variants.
    }
    \label{fig:duffing:time}
\end{figure}

The average separation times ($\langle t_{sep}\rangle$) as a function of the SNR values are presented in Fig.~\ref{fig:duffing:time}. The relatively smooth and monotonic degradation of these curves in the noise range studied demonstrates that $\langle t_\text{sep}\rangle$ is a robust metric that effectively captures model performance using forward integrations.

It is worth noting that SINDy does not strictly constrain the coefficient of external forcing to unity. Consequently, the discrepancies between SINDy and SINDy-CC are mainly attributable to deviations in this coefficient (e.g., identifying $0.95\, F_{ext}(t)$ instead of $1.0\,F_{ext}(t)$). Such deviations can drastically alter the dynamics during forward integration.

Using the intersection of the chaotic horizon $t_L$ with the separation time of each method, we can establish a clear performance hierarchy as a function of the noise threshold value. 
Ordering the methods from lowest to highest threshold, we obtain: (i) standard SINDy fails at SNR$\lesssim10$ dB; (ii) SR fails at SNR$\lesssim5$ dB; (iii) NN-CC fails at SNR$\lesssim 0$ dB; (iii) SINDy-CC, Poly-CC, NN-CC$_\text{+sym}$ and NN-CC$_\text{+post-SR}$  fail at SNR$\lesssim-5$ dB; (iv) NN-CC$_\text{+post-SR}$, SR-CC and Parametric fail at SNR$\lesssim-10$ dB.

Crucially, these dynamic results are fully consistent with the static RMSE analysis of the CCs. For SNR$\lesssim 5$ dB, the SR performance collapses abruptly due to structural instabilities consistent with the increasing RMSE values of Fig.~\ref{fig:duffing:sr}(e), exhibiting the characteristic `staircase' jumps often observed in the Pareto frontiers\cite{Schmidt2009}. Additionally, the performance of SINDy-CC, Poly-CC, NN-CC$_\text{+sym}$, and NN-CC$_\text{+post-SR}$ under the $\langle t_\text{sep}\rangle$ metric is similar, also in agreement with the CC analysis discussed above.
terms between $x$ and $\dot{x}$, which amplify integration errors and lead to earlier divergence during forward simulations.

\begin{figure}[!htpb]
    \centering
    \includegraphics[width=15cm]{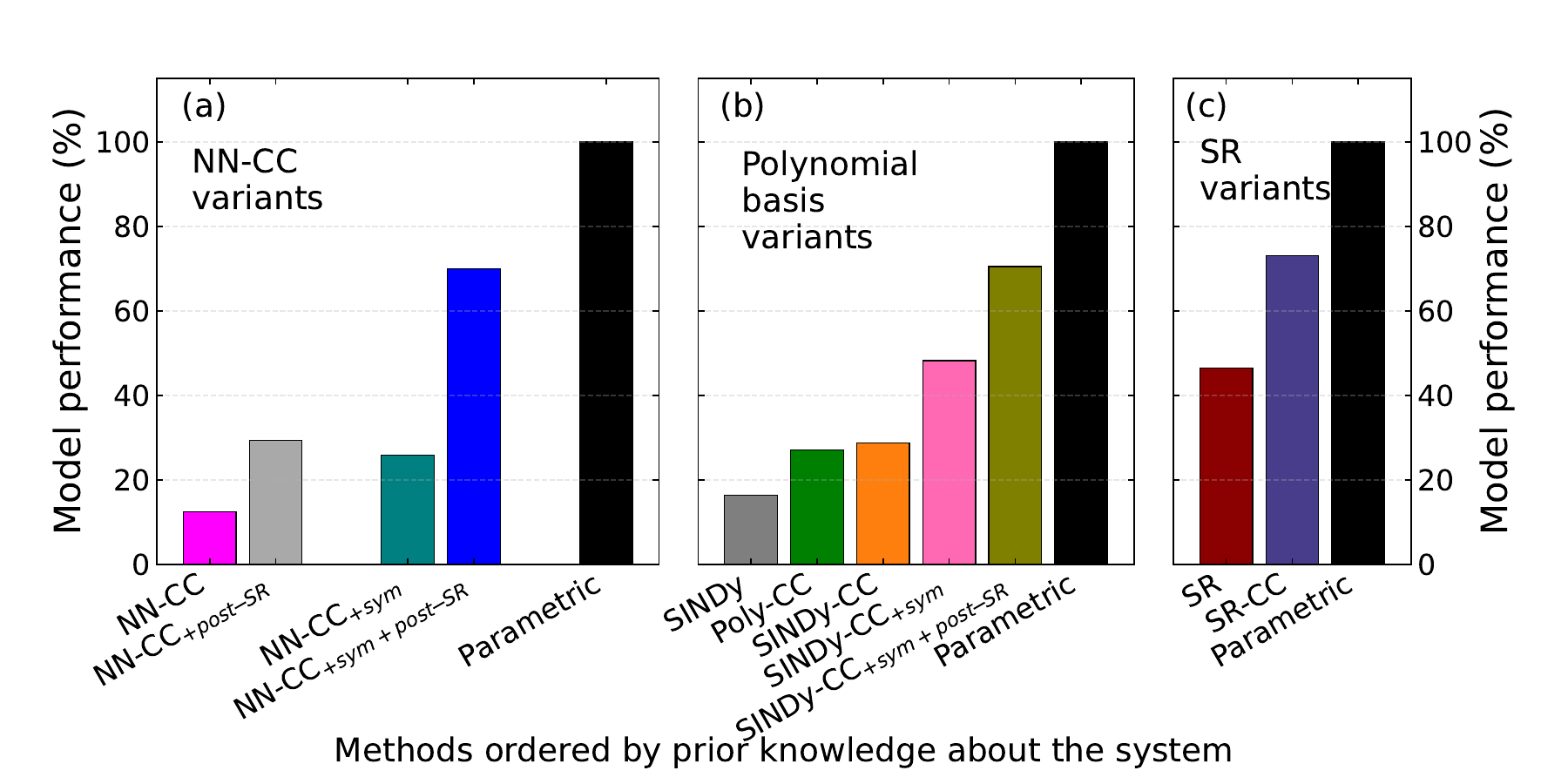}
\caption{Performance of the Duffing oscillator identification models, arranged by increasing levels of prior physical knowledge and compared to the baseline Parametric model: (a) NN-CC variants; (b) sparse regression variants; (c) symbolic regression variants.
}
    
    \label{fig:model_performance:duffing}
\end{figure}

The overall results for the Duffing oscillator are summarized in Fig.~\ref{fig:model_performance:duffing}.
In this figure, the methods are arranged along the horizontal axis according to the increasing amount of prior physical knowledge incorporated into the model structure, while the vertical axis displays a performance metric normalized to the theoretical baseline. The Parametric method serves as the 100\% reference standard, and the performance for all other methods is computed using the fitted $A$ values from
Table~\ref{tab:fitting_duffing} using the following equation:

\begin{equation}
    \text{Performance}(\%)=\frac{1}{2}\left(\frac{A_{\text{RMSE}[f_1]}^\text{Parametric}}{A_{\text{RMSE}[f_1]}^\text{method}}+\frac{A_{\text{RMSE}[f_2]}^\text{Parametric}}{A_{\text{RMSE}[f_2]}^\text{method}}\right)\,.100\;\% \; .
\end{equation}

Panel (a) illustrates the systematic improvement among the NN-CC variants by incorporating prior physical knowledge. The baseline NN-CC provides a flexible starting point, but its performance is visibly enhanced by incorporating the post-processing step (NN-CC$_\text{+post-SR}$) and symmetry constraints (NN-CC$_\text{+sym}$), with the combined approach (NN-CC$_\text{+sym+post-SR}$) demonstrating the cumulative benefit of both refinements for the CCs.

Panel (b) highlights the polynomial basis variants. For SINDy, which does not provide a direct equivalent in terms of $A$, we estimate a performance penalty of approximately 20\% relative to SINDy-CC, based on the separation-time analysis discussed above.  

Finally, panel (c) presents the SR variants. Notably, despite falling into the category with minimal prior knowledge, SR achieves very good performance. By framing the SR method within the CC-based formalism (SR-CC),  performance is further improved, reaching levels similar to the NN-CC$_\text{+sym+post-SR}$ method. However, this approach suffers from the observed jumps produced when the model structure is switched. This is a well-studied issue regarding the structural instability of the Pareto frontier, where minor variations in the selection criteria or data can lead to the selection of vastly different functional forms\cite{Smits2005,Cava2021,Kammerer2024,Muthyala2025}.

\subsection{Example 2: dry friction (stick-slip)}\label{sec:stickslip}

In this example, we consider a stick-slip system subject to Coulomb friction. Assuming a unit mass ($m=1$ kg), the grounded ODE is described by Eq.~\ref{eq:model:veloc}, where the CCs are defined as
\begin{equation}
\begin{cases}    
    f_1(\dot{x})&= c\;\dot{x}+ \mu \;N \; \text{sign}(\dot{x})\\
    f_2(x)&=k \, x \;.
\end{cases}
\label{eq:cc:stick_slip}
\end{equation}
In this benchmark, the physical parameters for the CCs are maintained fixed at $k=1.274$ N/m, $c=0.386$ Ns/m, and $\mu\, N=0.801$ N (similar values are used in Ref.~\cite{Gonzalez2025} for noiseless analysis).  
For the characterization of the CCs, we employed a single driving force with $A=2$ N and $\Omega=0.363$ rad/s, and ICs $x_0=-0.076$ m and $v_0=0.146$ m/s. For the forward simulations, we sample different ICs and driving forces, sampled from $\Omega \in [0.2,1],\, A\in [1,1
.5],\,  x_0,v_0\in [-0.5,0.5]$.

\begin{figure*}[!htpb]
\centering
\includegraphics[width=16 cm]{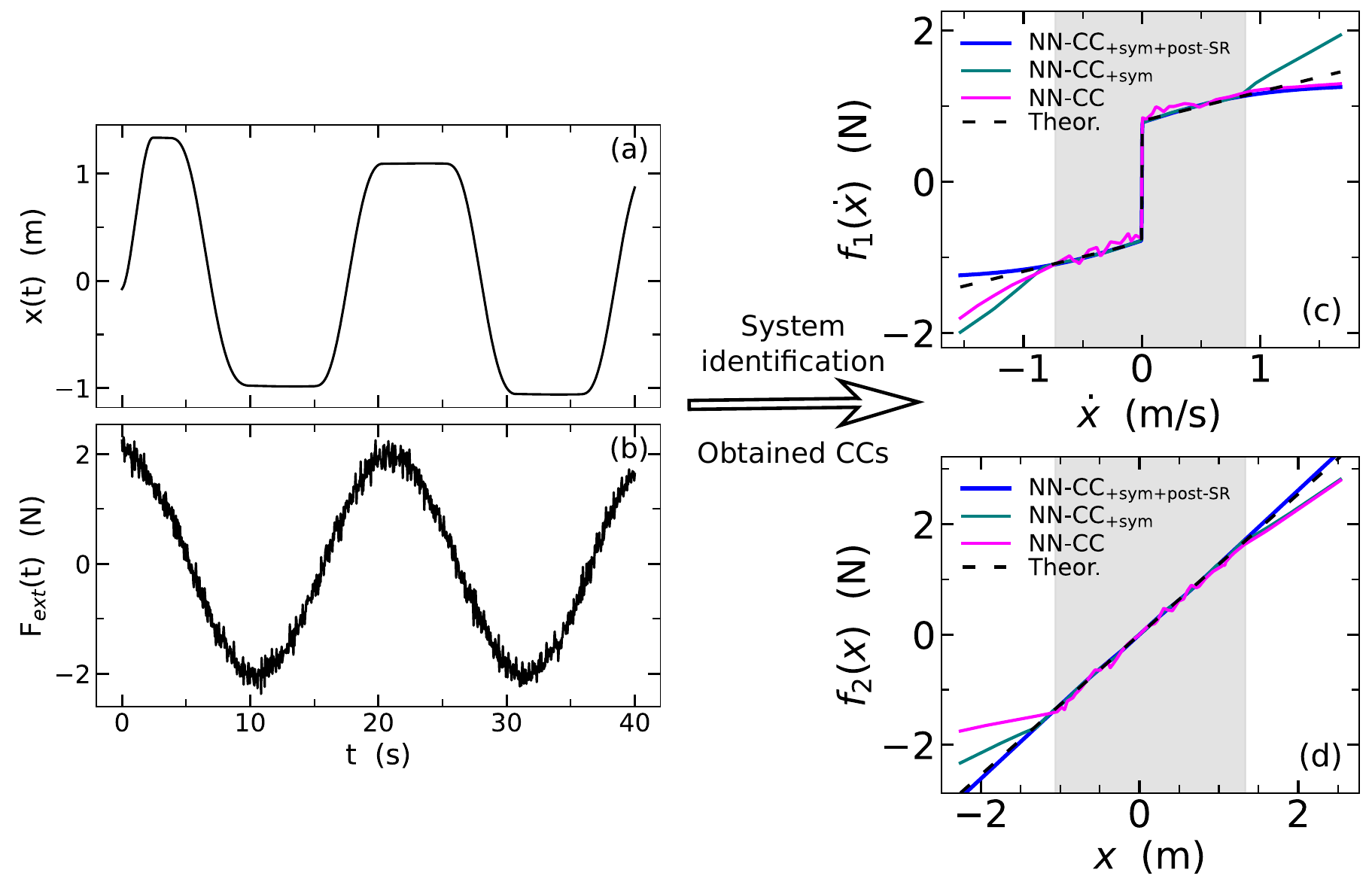}
\caption{System identification workflow for the stick-slip example. 
(a) and (b) show the input data: dynamical variable $x(t)$ and driving force $F_{ext}(t)$ with SNR~=~20~dB. (c) and (d) display the identified models $f_1(\dot{x})$ and $f_2(x)$, respectively, obtained using three NN-CC variants. The gray zones indicate the range of values of the training dataset.}
\label{fig:stick-slip:trajectory}
\end{figure*}

Figure~\ref{fig:stick-slip:trajectory} shows the identification results using a training dataset generated with SNR $=20$ dB [panels~(a) and (b)]. The CCs identified by the NN-CC, NN-CC$_{\text{+sym}}$, and NN-CC$_{\text{+sym+post-SR}}$ methods are shown in panels~(c) and (d). 
The gray regions highlight the data range from the training dataset. 
Following the approach of the Duffing example, we perform forward simulations within this range to evaluate the model accuracy in dynamics interpolation.

\begin{figure*}[!htpb]
    \centering
    \subfloat{\includegraphics[width=0.35\textwidth]{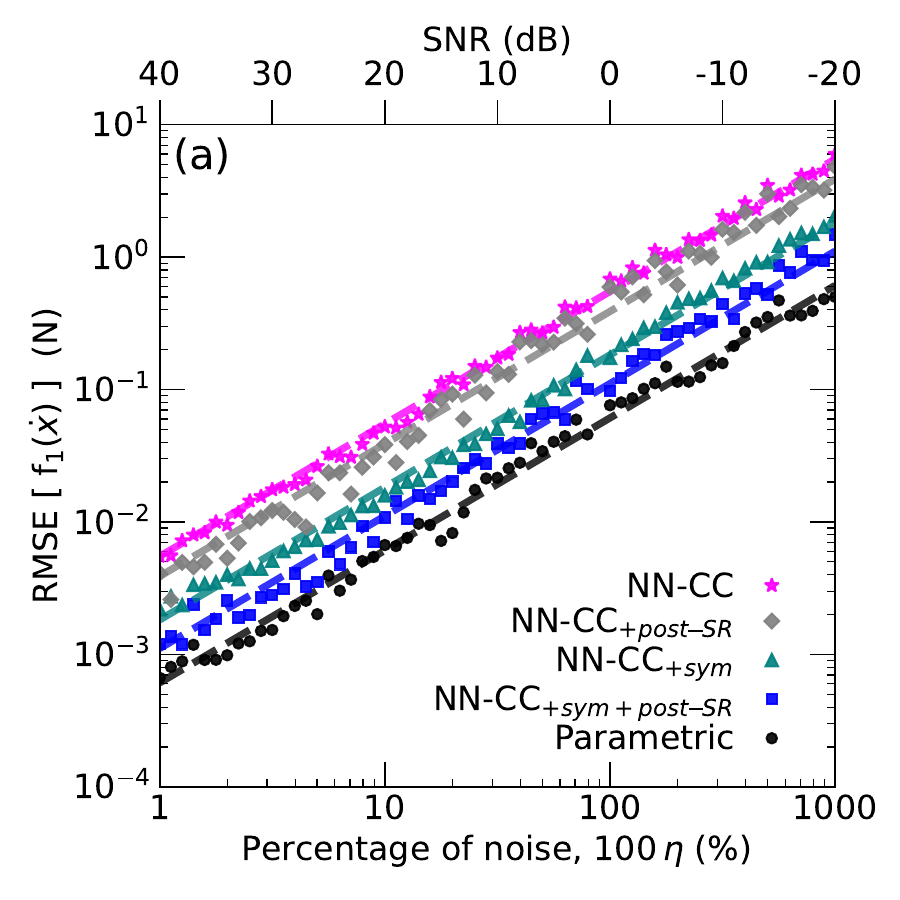}}
    \subfloat{\includegraphics[width=0.35\textwidth]{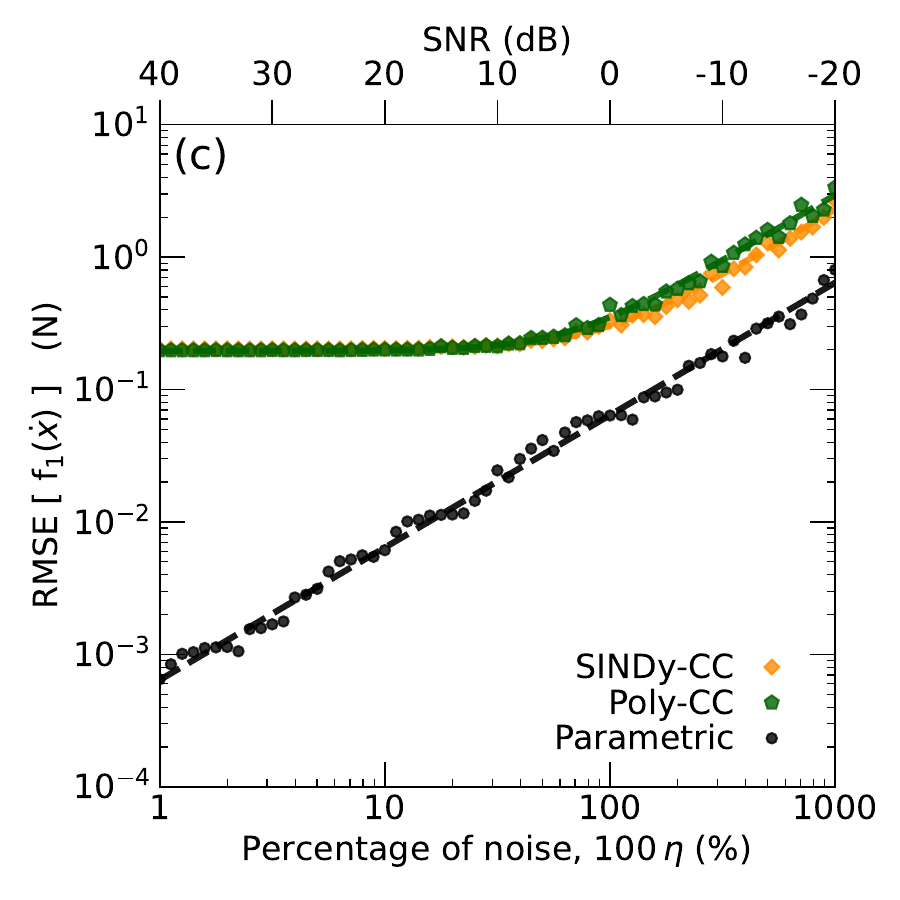}}
    \subfloat{\includegraphics[width=0.35\textwidth]{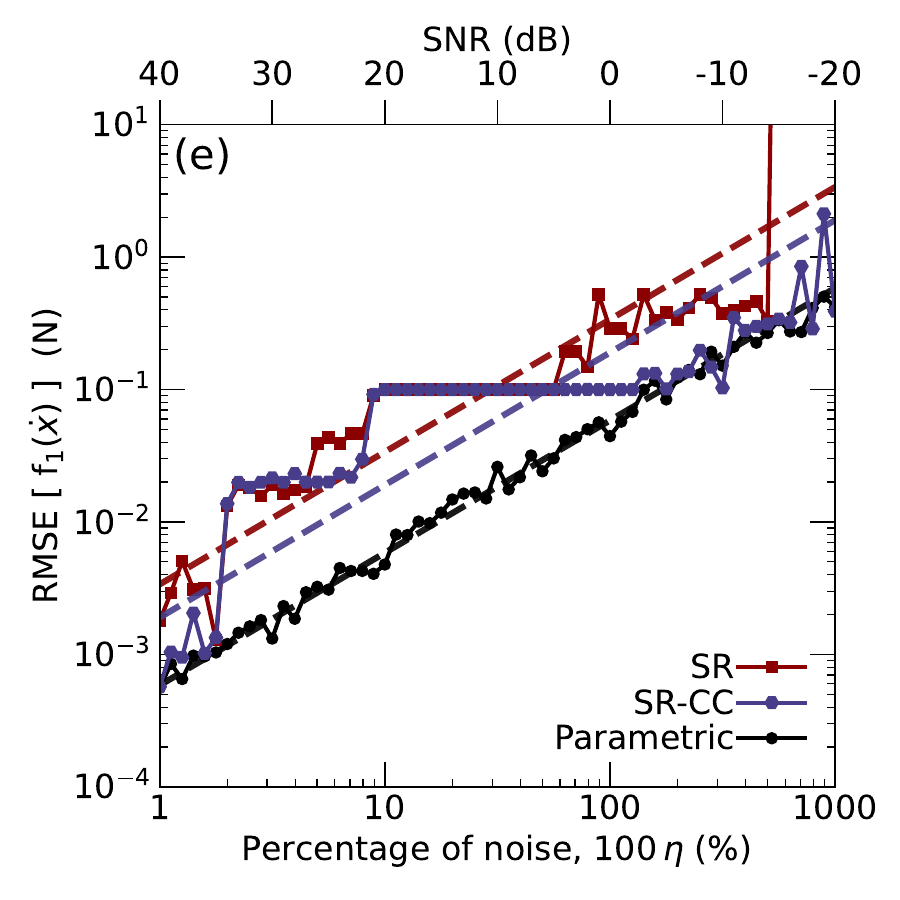}}\quad
    \subfloat{\includegraphics[width=0.35\textwidth]{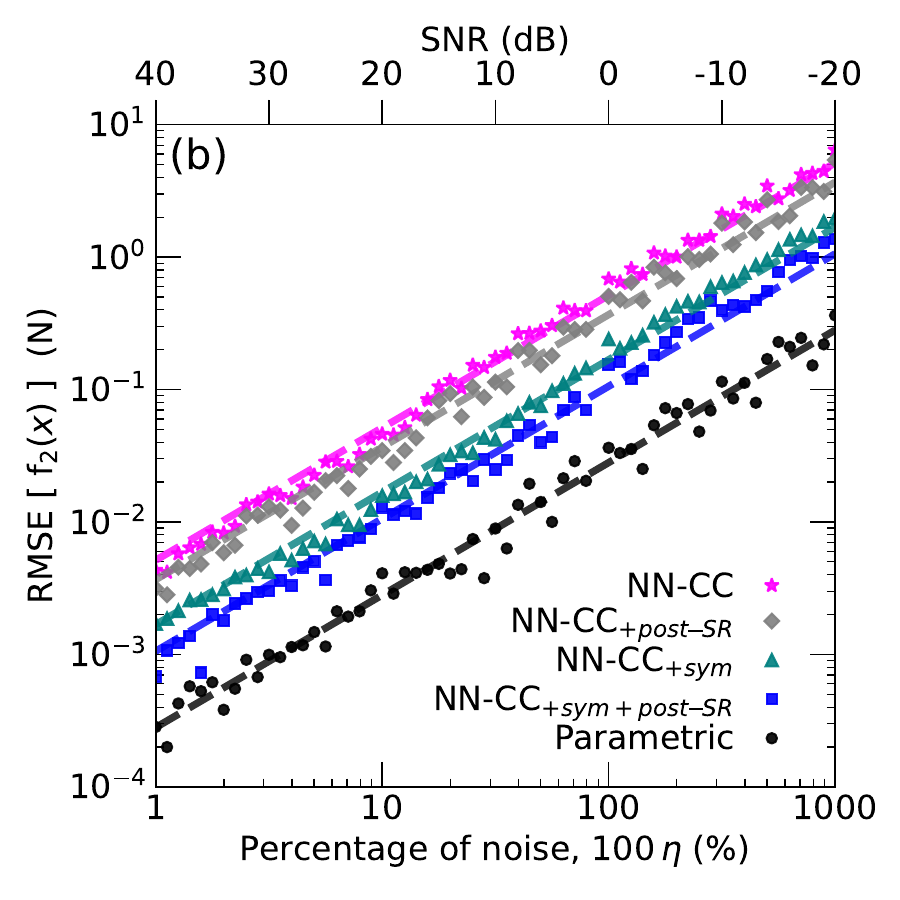}}  \subfloat{\includegraphics[width=0.35\textwidth]{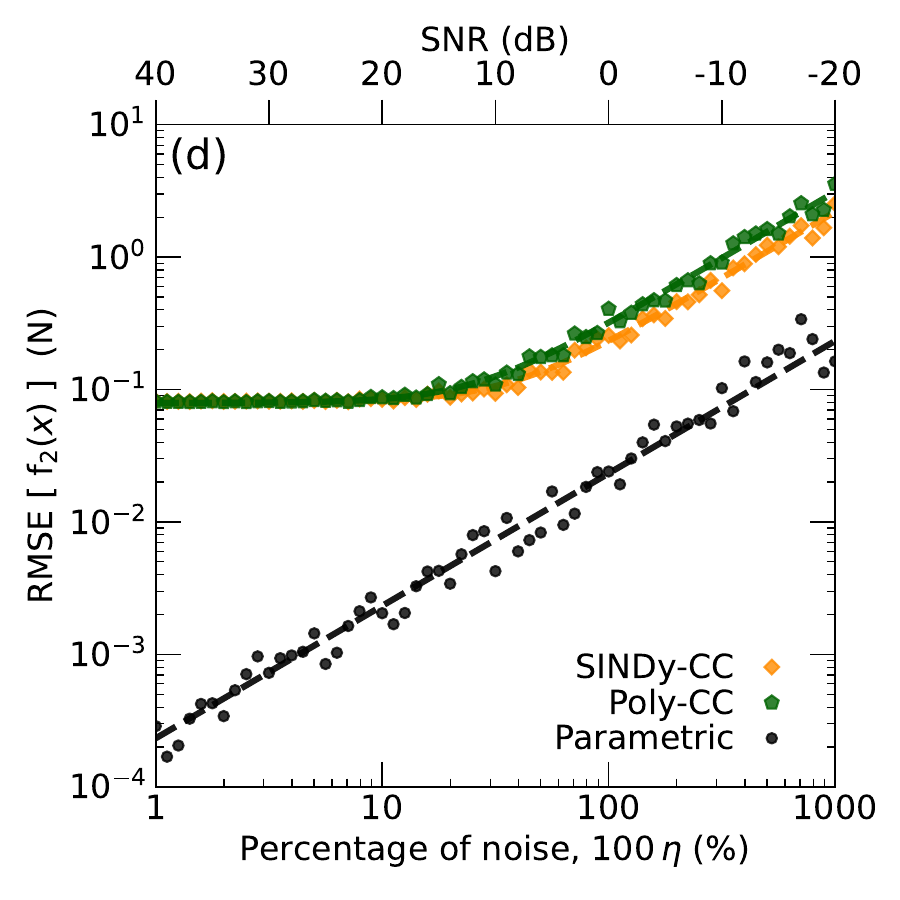}}
    \subfloat{\includegraphics[width=0.35\textwidth]{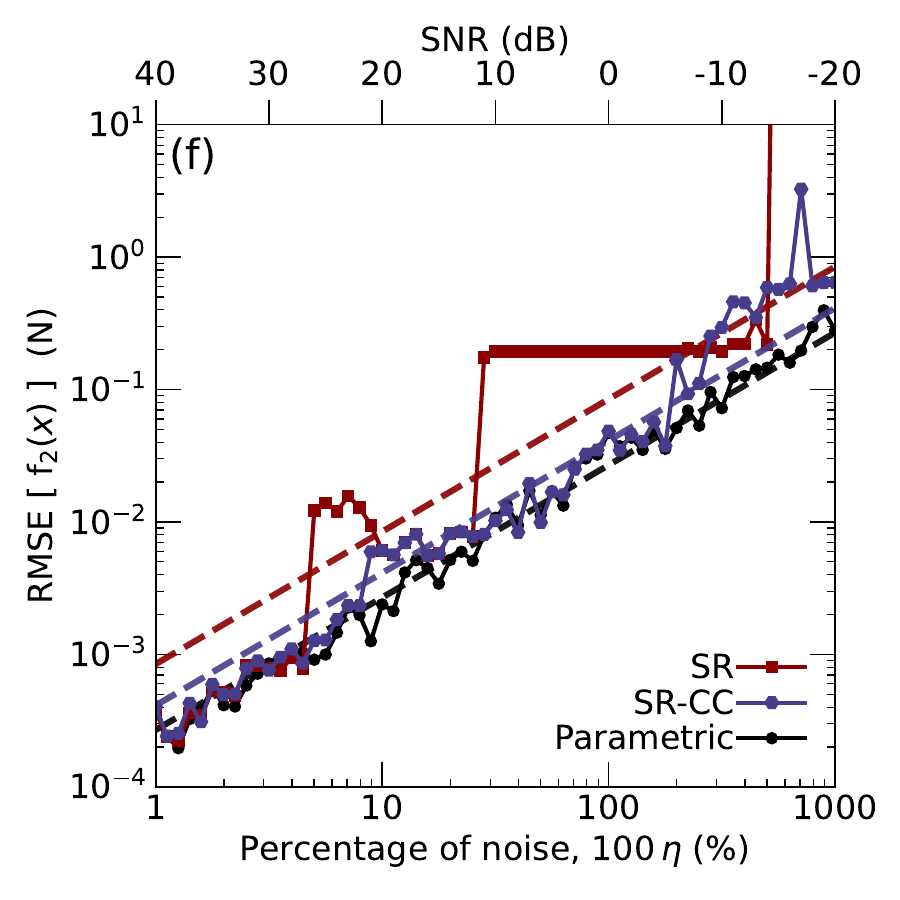}}

    \caption{RMSE analysis for the identified CCs ($f_1$ and $f_2$) of the stick-slip system. Panels (a) and (b): NN-CC variants; (c) and (d): Polynomial basis variants; (e) and (f): SR variants.}
    \label{fig:stickslip:cc}
\end{figure*}

Analogous to the Duffing example, we assessed the precision of the CCs by generating $10$ independent datasets (generated with distinct noise realizations) for each SNR value (with intervals of 1 dB). 
We performed a single training run per database and evaluated the ensemble errors in the identified CCs using the RMSE metric from Eq.~\ref{eq:rmse_fi}. These results are presented in Fig.~\ref{fig:stickslip:cc}.

As shown in Figs.~\ref{fig:stickslip:cc} (c) and (d), the RMSE for polynomial-based methods saturates for SNR $\gtrsim 0$ dB. 
This establishes an accuracy floor caused by the inability of smooth polynomials to capture the sharp discontinuities, a limitation intrinsically linked to the Gibbs phenomenon (see Refs.~\cite{Gonzalez2024,Gonzalez2025}).   
Consequently, a direct comparison of raw RMSE values between polynomial basis functions (SINDy, SINDy-CC, and Poly-CC) and universal approximators (NN-CC variants) is obscured by this structural rigidity.

To disentangle the structural error from the noise response, we use a complete expression (obtained in \ref{app:noise_scaling})  that contains a constant structural bias and the noise-limited term:
\begin{equation}
    \text{RMSE}[f_i(z)]=\sqrt{\text{bias}^2+(A\,\eta)^2}\;.
    \label{eq:rmse_bias_model}
\end{equation}
Here, the constant term represents the limit determined by the flexibility of the model.
If $A\,\eta\gg \text{bias}$ then Eq.~\ref{eq:rmse_bias_model} reduces to the linear expression from Eq.~\ref{eq:fitting_rmse_empirical}. 

Therefore, the polynomial-based models reach the saturation regime (where $\text{bias} \gg A\,\eta$) prematurely, stabilizing at SNR$\gtrsim10$ dB. 
In contrast, the NN-CC variants do not suffer from this structural rigidity; since the RMSE values at the limit of zero noise are below $10^{-4}$, the constant bias term remains negligible compared to the noise-limited term. 
This allows the NN-CC variants to continue scaling linearly down to at least 1\% noise.

To avoid the unfairness of comparing noise-limited methods against bias-limited methods, we evaluate performance based only on the noise-limited region. 
These asymptotic linear trends allow us to estimate the theoretical performance these methods would achieve in the absence of structural bias.


The behavior of the SR method, shown in Figs.~\ref{tab:fitting_stick-slip} (e) and (f), exhibits a staircase pattern, as was also observed in the Duffing example.

\begin{table}[!htpb]
    \centering
    \caption{Fitted $A$ parameters for the stick-slip system. Lower values indicate better performance. Superindex $^{(*)}$ indicates the fit was performed only in the stable regime (SNR $> -12$ dB). The bias parameter for the SINDy-CC and Poly-CC fitings are $\approx0.2$ for $f_1$ and $\approx 0.08$ for $f_2$  }
    \label{tab:fitting_stick-slip}
    \begin{tabular}{lccc}
        \toprule
        & \multicolumn{3}{c}{Fitted $A$ values} \\
        \cmidrule(lr){2-4}
        Model & $\mathrm{RMSE}[f_1]$& $\mathrm{RMSE}[f_2]$& $\mathrm{RMSE}[x(t)]$\\
         & ($ 10^{-2}$ N)& ($ 10^{-2}$ N) & ($ 10^{-2}$ m) \\
        \midrule
        Parametric & $6.1$ & $2.8$ & $6.5$ \\
        NN-CC$_{\text{+sym+post-SR}}$ & $11.1$ & $10.6$ & $13.3$ \\
        SR-CC & $19.0$ & $4.1$ & $24.8$ \\
        NN-CC$_{\text{+sym}}$ & $18.2$ & $16.8$ & $16.4$ \\
        SR$^{(*)}$ & $33.8$ & $8.5$ & $40.2$ \\
        NN-CC$_{\text{+post-SR}}$ & $38.9$ & $36.8$ & $22.5$ \\   
        SINDy-CC & $22.7$ & $21.9$ & $-$\\
        Poly-CC & $29.4$ & $31.0$ & $-$ \\
        NN-CC & $54.9$ & $51.4$ & $33.6$ \\
        \bottomrule
    \end{tabular}
\end{table}

The fitted $A$ values for the various methods (obtained using Eq.~\ref{eq:fitting_rmse_duffing}), are summarized in columns 2 and 3 of Table~\ref{tab:fitting_stick-slip}. 

Having characterized the CCs [according to the schematics in Fig.~\ref{fig:schematic}(g)], we now evaluate the performance with forward simulations [Fig.~\ref{fig:schematic}(i)]. 
Following the established protocol for the Duffing system, we generate $10$ models per SNR level and test each on $10$ novel driving forces and ICs sampled in the intervals mentioned at the beginning of this section.

We define an ensemble RMSE metric to analyze the integrated trajectories as:

\begin{equation}
\mathrm{RMSE}[x(t)] = \left\{
\frac{1}{N_{\mathrm{steps}}} 
\sum_{j=0}^{N_{\mathrm{steps}}-1} 
\frac{1}{N_{\text{runs}}} \sum_{k=1}^{N_{\text{runs}}}  
\frac{1}{N_{\mathrm{forw}}} 
\sum_{l=1}^{N_{\mathrm{forw}}} 
\left[ x_l^{k\mathrm{-th\, model}}(t_j) - x_l^{\mathrm{th}}(t_j) \right]^2 \right\}^{1/2}\; ,
\label{eq:rmse_x}
\end{equation}
where $x^{\mathrm{th}}$ denotes the theoretical forward integration, $x^{k\mathrm{-th\, model}}$ denotes the integrated trajectory obtained from the identified model using the $k\mathrm{-th}$ database, $N_{\text{runs}}=10$ is the number of training runs, $N_{\mathrm{forw}}=10$ is the number of forward integrations per model, and $N_{\mathrm{steps}}$ is the number of forward integration steps (chosen here as  $N_{\mathrm{steps}}=N_{\mathrm{data}}$).

\begin{figure*}[!htpb]
    \centering
    \includegraphics[width=0.32\linewidth]{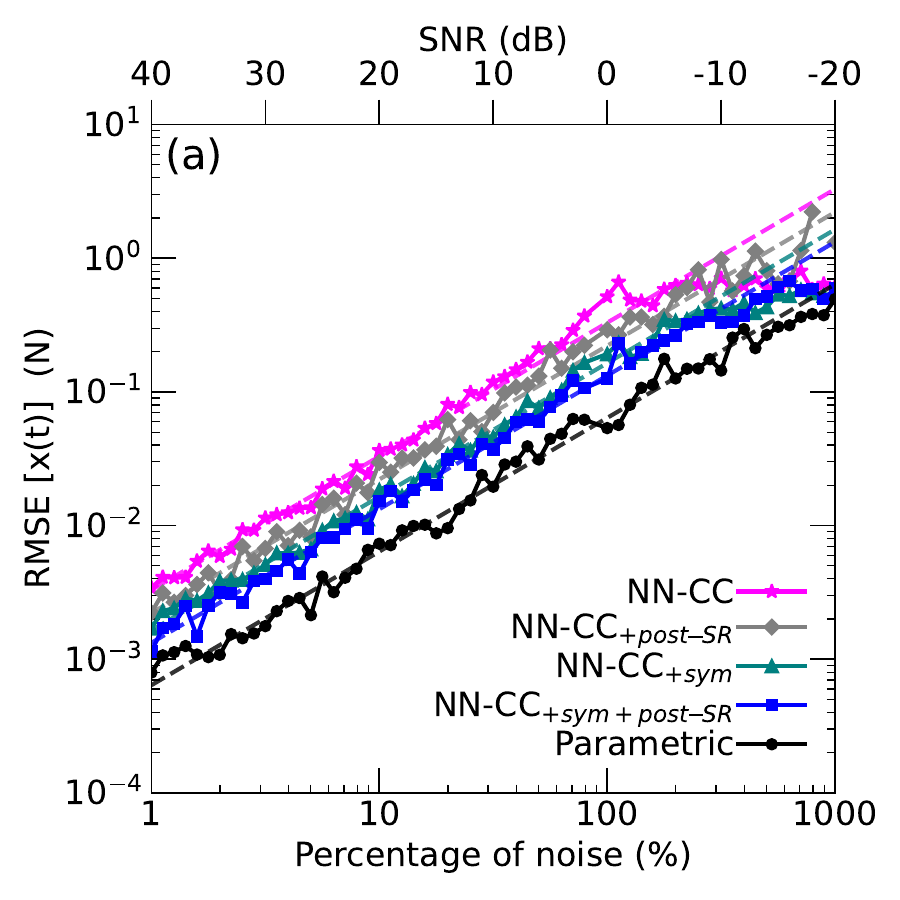}
    \includegraphics[width=0.32\linewidth]{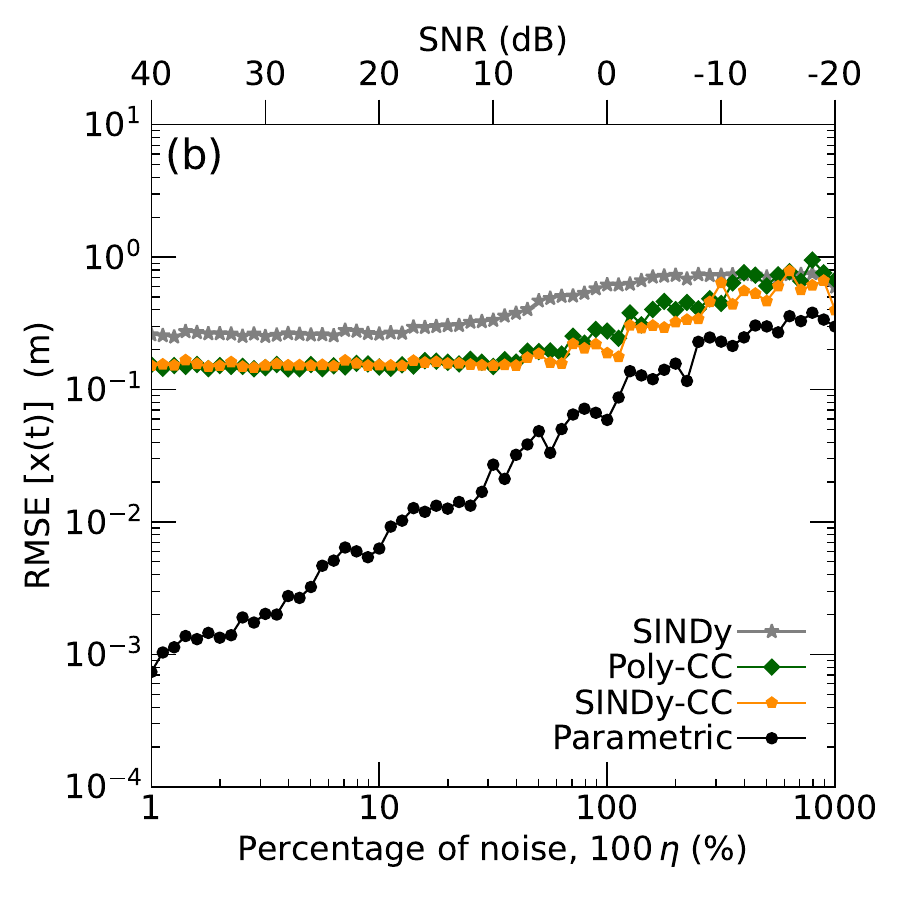}
    \includegraphics[width=0.32\linewidth]{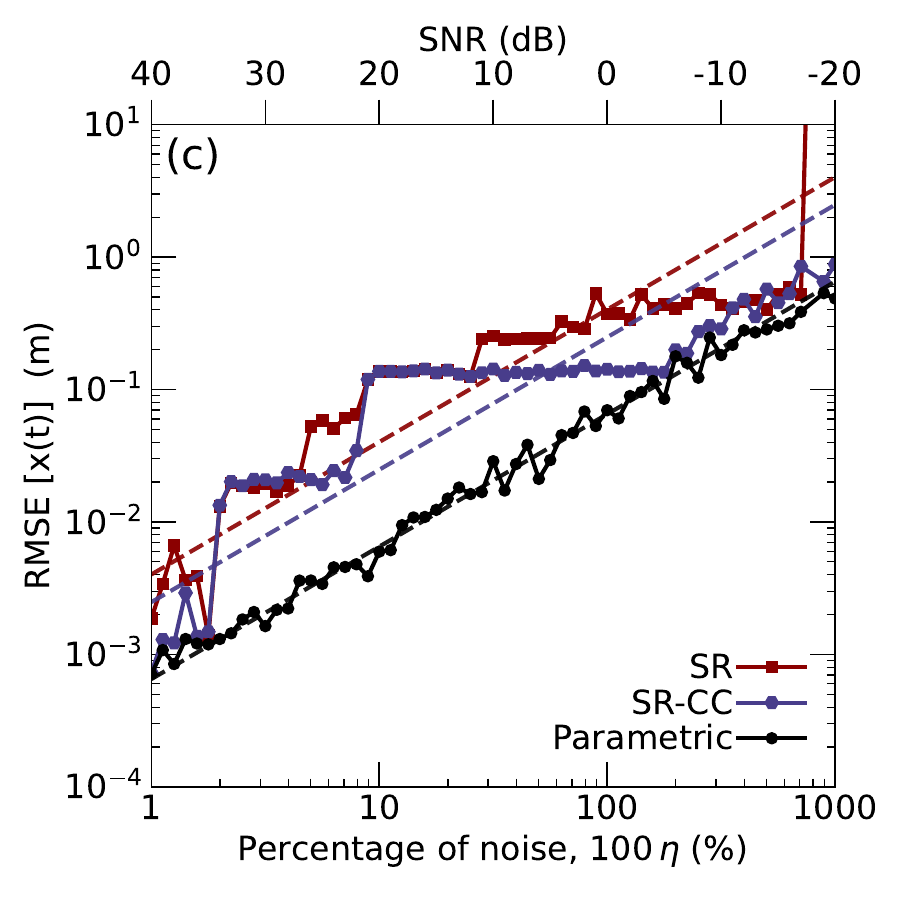}
    \caption{
RMSE values for the dynamical variable $x(t)$, evaluated from forward integrations of the identified models compared to the theoretical trajectory for the stick-slip system. (a) NN-CC variants; (b) Polynomial basis variants; (c) SR variants. 
    }
    \label{fig:stick_slip:noise}
\end{figure*}

The resulting RMSE values, depicted in Fig.~\ref{fig:stick_slip:noise}, exhibit trends similar to the RMSE analysis for the CCs. Specifically, the SR method displays the staircase pattern observed for the CCs, whereas the CC-based methods exhibit smoother linear trends. 
Column 4 of Table~\ref{tab:fitting_stick-slip} quantifies this behavior, showing ratios relative to the Parametric baseline that are consistent with the RMSE results for the CCs.

\begin{figure*}[!htpb]
    \centering
    \includegraphics[width=15cm]{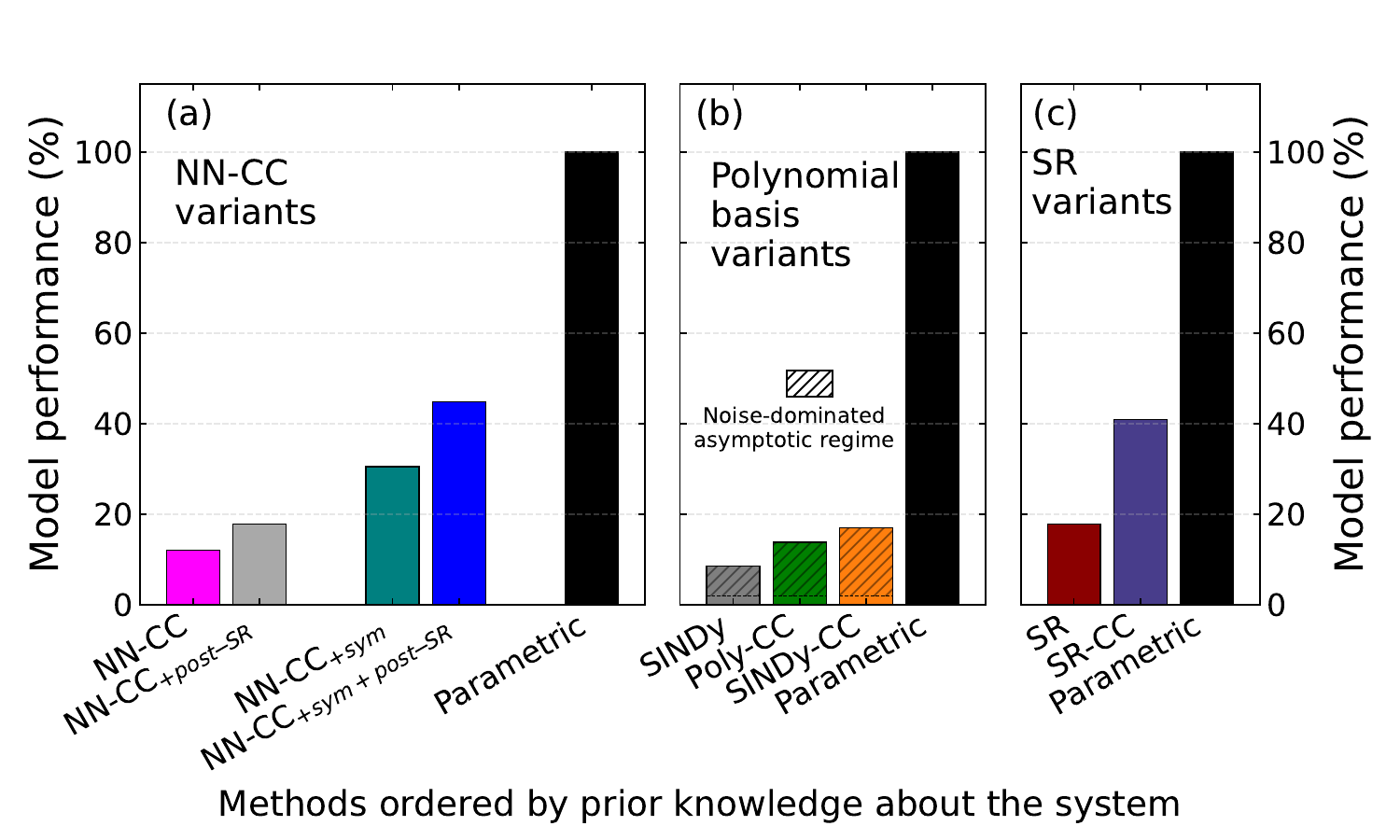}
\caption{Performance of the stick-slip identification models, arranged by increasing levels of prior physical knowledge and compared to the baseline Parametric model: (a) NN-CC variants; (b) sparse regression variants with polynomial basis functions; (c) polynomial methods and symbolic regression variants using polynomials basis functions. The hatched regions indicate the theoretical performance gap caused by the structural bias of using a continuous polynomial basis to approximate discontinuous friction.}
\label{fig:model_performance:stick-slip}
\end{figure*}

The overall performance for the stick-slip system is summarized in Fig.~\ref{fig:model_performance:stick-slip}, calculated using the $A$ values relative to the Parametric benchmark as:
\begin{equation}
    \text{Performance(\%)}=\frac{1}{3}\left(\frac{A_{\text{RMSE}[f_1]}^\text{Parametric}}{A_{\text{RMSE}[f_1]}^\text{method}}+\frac{A_{\text{RMSE}[f_2]}^\text{Parametric}}{A_{\text{RMSE}[f_2]}^\text{method}}+\frac{A_{\text{RMSE}[x]}^\text{Parametric}}{A_{\text{RMSE}[x]}^\text{method}}\right)\,.100\;\%
\end{equation}
The hatched regions in the bar charts for Poly-CC and SINDy-CC represent the theoretical asymptotic performance at the noise-limited regime. 

Figure~\ref{fig:model_performance:stick-slip} illustrate a clear trend: for the different approaches, incorporating prior physical information consistently improves model performance. 
The NN-CC$_{\text{+sym+post-SR}}$ method yields the best overall results among the NN-CC approaches, demonstrating that combining symmetry priors with symbolic refinement is a robust strategy for both Duffing and stick-slip systems. SR-CC achieves a performance comparable to that of the NN-CC$_{\text{+sym+post-SR}}$ method, although it suffers from the staircase artifacts absent in the NN-CC approaches.


\section{Discussion}\label{sec:discussion}

The findings of this study corroborate and extend a growing body of research demonstrating the efficacy of incorporating physical information (particularly symmetries) into data-driven system identification frameworks. The observation that such constraints enhance model performance, especially in the presence of noise, is consistent with foundational works in the field. For instance, approaches like Hamiltonian Neural Networks (HNNs) have successfully enforced energy conservation by design \cite{Greydanus2019}, while Lagrangian-based networks preserve variational symmetries inherent in physical laws \cite{Cranmer2020LNN}.

Our work builds upon these foundations by shifting the focus from enforcing scalar conservation laws to preserving the structural topology of the governing equations. While HNNs ensure that the system stays on a fixed energy manifold, the CC framework enforces the structure of the system to be compatible with a family of functions. In this context, the quantities that must be `conserved' during the learning process are the CCs themselves (univariate functions that uniquely represent the distinct physical mechanisms, such as stiffness and damping). By embedding this decomposition directly into the architecture, the CC approach ensures that the identified model is not just a curve-fit, but a decomposition of the system into physically meaningful components. Our results demonstrate that this structural prior, when combined with symmetry constraints, systematically improves performance as a function of the prior knowledge incorporated. 
Furthermore, this decoupling of the dynamics into univariate components enables a powerful post-processing strategy: the learned curves can be individually processed with SR (post-SR) to reduce the number of model parameters and to produce analytical laws, effectively bridging the gap between data-driven approximation and symbolic discovery.

Based on two representative benchmark systems, the chaotic Duffing and the discontinuous stick-slip model (Figs.~\ref{fig:model_performance:duffing} and \ref{fig:model_performance:stick-slip}), we reveal several key insights into the strengths and weaknesses of different identification strategies:

\begin{itemize}
\item[i)] \textit{The Parametric method as an upper bound on achievable performance}. Our comparative analysis begins by establishing a performance baseline. To this aim, we use the Parametric model, which assumes the maximum amount of prior information as known by explicitly specifying the full structural form of the governing equations. 
Consequently, it is immune to structural misidentification by definition and exceptionally robust to noise, the latter evidenced by the linear scaling of the RMSE with noise levels, as predicted by the Cramér-Rao bound for unbiased estimators (see \ref{app:noise_scaling}). 
However, the utility of the Parametric method is strictly limited to scenarios where the constitutive laws are fully known a priori. In exploratory contexts where the specific functional forms (e.g., the dissipation or stiffness constitutive relations) are unknown, this method serves not as a competitor, but as an upper bound for achievable accuracy.  

\item[ii)] \textit{The CC-based formalism} imposes a modular hypothesis: that the dynamics can be decomposed into one-dimensional CCs. 
To ensure a mathematically well-posed identification problem, it is mandatory to propose families of ODEs that satisfy specific uniqueness properties (see \ref{app:uniquiness_CC}). These properties guarantee that within a chosen model family, the mapping from the observed data to the governing functions is injective. This resolves the identifiability ambiguities typical of structure-agnostic modeling, where the lack of this uniqueness often leads to non-injective mappings (meaning multiple fitted models can reach similar loss error). In contrast, the uniqueness of CC-based approach ensures that a low loss error systematically converge to the ground system.

This theoretical guarantee has a direct practical consequence: Proposed model families can be treated as hypotheses that, once tested, allow for model selection or elimination (inner loop in Fig.~\ref{fig:schematic}, see also \ref{app:model_selection}). 
In practice, the low loss errors is the first indication that the model structure is correct but is not conclusive. Thus, two additional validations are discussed in this work (middle and mayor loops in Fig.~\ref{fig:schematic}). 
The middle loop, based on the invariance of the CCs, and the outer loop based on forward integrations and extra validations.
By cycling through the three validation loops (inner, middle, and outer), practitioners can systematically validate or reject these hypothesized families based on their domain expertise, effectively isolating the correct structural topology.

Consequently, this framework is ideal when the practitioner has prior knowledge about the mathematical structure (or wishes to test a hypothesized structure) but the specific expressions for constitutive relations are unknown, thus requiring exploratory research. Unlike general-purpose approaches that attempt to learn high-dimensional mappings directly from data, the CC approach takes advantage of this structural prior to simplify the optimization problem, making it a powerful tool for discovering interpretable laws in systems where the governing topology is suspected but the precise parametrization of the CCs remain to be found.

\item[iii)] \textit{NN-CC and its variants}. The CC-based approach can be applied based with multiple estimators for the CCs. Among them, the NN-CC method is promising due to the high flexibility of the NNs to fit one-dimensional functions (universal approximation theorem). 
Thus, this work focuses on extending the NN-CC approach by incorporating physical priors to improve identificability under noise and discontinuities. As major results and challenges of NN-CC, we can mention
\begin{itemize}
\vspace{-0.2 cm}\item  Symmetries as a physics-informed regularizer (NN-CC$_\text{+sym}$): By restricting the hypothesis space to physically admissible functions, symmetry constraints reduce the search space and prevent the NNs from overfitting to noise artifacts that violate fundamental invariances. This acts as a strong inductive bias, enhancing structural consistency without imposing the rigidity of fixed basis functions.
\vspace{-0.1 cm}\item Post-processing (NN-CC$_\text{+post-SR}$): The integration of SR as a post-processing step reduces the number of parameters, effectively acting as a transformation to a lower-dimensional space. This primarily serves to `filter' artifacts arising from the high parametrization of the neural networks, distilling the solution into a more parsimonious form. 

\vspace{-0.1 cm}\item The NN-CC$_\text{+sym+post-SR}$ method exploits both symmetries and post-SR, achieving the highest performance among NN-CC variants, showing a systematic improvement in model identification as additional prior knowledge is incorporated. Furthermore, while the SR-CC method attained comparable quantitative performance for both chaotic and discontinuous systems, the NN-CC variants did not exhibit the marked staircase effect observed for the SR-CC method.
\end{itemize}

\vspace{-0.2 cm} It is worth noting that the NN-CC implementation relies on iterative optimization and is therefore computationally more demanding than linear least-squares methods. However, as shown in \ref{app:computational_cost}, GPU acceleration can substantially mitigate this overhead, leading to a reduction in training times of $\approx$40\% for the datasets considered. Owing to the characteristics of GPU utilization in NN methods, we expect that such performance gains may become more pronounced as the number of data points increases; however, a systematic analysis of this scaling behavior is beyond the scope of the present work.



\item[iv)] \textit{Sparse identification (SINDy and its variants)}.  Our results demonstrate that the performance of sparse regression is systematically improved when framed within the CC-based formalism and when prior information is incorporated. Our implementation referred as SINDy often faces challenges in decoupling the coefficient of the external forcing from internal dynamics, particularly in high-noise regimes. The SINDy-CC formulation mitigates this by strictly enforcing the physical structure of the forcing ($k_0=1$). Furthermore, we observed a clear hierarchy of improvement with the addition of physical priors: SINDy-CC$_\text{+sym}$ eliminates spurious non-physical terms (e.g., even powers) by enforcing symmetry constraints, while the hybrid SINDy-CC$_\text{+sym+post-SR}$ achieves the highest accuracy among the sparse methods. This latter variant effectively combines the robustness of constrained optimization with the parsimony of symbolic regression, approaching the performance boundary established by the Parametric method for the Duffing system.

\item[v)] \textit{Symbolic regression (SR)}.  
By framing SR within the CC-based formalism (SR-CC), we achieve a systematic performance gain that stems directly from reducing the search space: instead of searching within a complex bivariate surface $f(x,\dot{x})$, the problem is decomposed into finding two uncoupled univariate functions, $f_1(\dot{x})$ and $f_2(x)$. This dimensional reduction enables SR-CC to attain quantitative accuracy comparable to the best NN-CC variant (NN-CC$_\text{+sym+post-SR}$) for both chaotic and discontinuous benchmarks. 
However, unlike the smooth error decay observed in NN-CC methods, SR-CC exhibits the characteristic `staircase' effect inherent to symbolic search. This phenomenon arises from the discrete nature of the search, where the algorithm locks onto specific functional forms to a range of noise levels before abruptly switching to a different model structure. Consequently, while the CC-based framework significantly enhances the robustness of SR, the remaining staircase in the Pareto frontier persists.

\end{itemize}

Building on these insights and the potential for deeper integration of geometric priors, we identify five promising avenues for future research:


\begin{itemize}
\item[i)] \textit{Extension to higher-dimensional systems.} A primary direction for future work is to extend the CC-based framework to higher-dimensional or multi-degree-of-freedom systems. 
A straightforward strategy is to define a list or hierarchy of model families 
that share uniqueness properties discussed in this work (three families are already incorporated within CC-based methodology, see \ref{app:uniquiness_CC}).

By systematically iterating through this `library of structures', practitioners could automatically select the optimal family or discard ill-posed hypotheses. However, a major challenge remains in defining a sufficiently comprehensive yet manageable list of candidate families a priori, suggesting the need for more automated structural search algorithms (see \ref{app:model_selection}).

\item[ii)] 
\textit{Automated symmetry discovery.} In this work, the symmetries were enforced within the NN-CC variants based on prior physical knowledge or visual inspection of the CCs obtained form the baseline NN-CC method. A future direction is to automate symmetry selection by integrating symmetry discovery methods \cite{Liu2022prl,Yang2023symmetry,Ouderaa2023,Otto2025} as a pre-processing step. By identifying relevant invariances (e.g., parity, scaling, or translational symmetries) directly from data as a pre-processing step, the framework could define the necessary constraints without requiring explicit user intervention.

\item[iii)] 
\textit{SR full integration into CC-based formalism.} There is significant potential in refining the SR-CC method. Future work should explore SR-CC implementations that allow symmetry constraints to be embedded directly into the symbolic search space, enabling a systematic comparison with NN-CC variants. Additionally, addressing the `staircase' instability observed in standard SR is critical; incorporating recent algorithms that utilize continuous optimization or probabilistic smoothing could mitigate the discrete jumps in model selection, offering a more robust transition between functional forms on the Pareto frontier. For example, future work could investigate recent advances on SR discovery such as the SyMANTIC approach\cite{Muthyala2025} that reformulates symbolic search as a sparse regression problem over a recursively expanded feature space, offering a potential pathway to smoother Pareto frontiers.

\item[iv)] \textit{Enhancing basis functions and solvers within SINDy-CC} 
In this work, SINDy-CC was implemented using polynomial basis functions, which naturally struggles for the discontinuous stick-slip dynamics due to the Gibbs phenomenon. Future work may explore SINDy-CC implementations using more flexible candidate libraries. While identifying the optimal basis set \textit{a priori} remains a non-trivial challenge, solving this would allow SINDy-CC to rival the flexibility of the NN-CC approach while retaining the computational efficiency of sparse regression. It is also important to assess in future directions different optimizers and optimization strategies.   
For instance, integrating Ensemble-SINDy\cite{Fasel2022} (bagging) or Weak-SINDy\cite{Messenger2021} (integral formulations) 
within the CC-based decomposition is a promising direction. 
Furthermore, probabilistic approaches such as MCMC-SINDy\cite{Zhou2024} would be particularly valuable, as they would enable the quantification of uncertainty in the identified CCs.

\item[v)] \textit{Integration with Bayesian approaches.} 
The modular nature of the CC formalism allows for replacing the core approximator with recent probabilistic regression techniques (see Refs.~\cite{Stock2024,Niven2024,Fung2025,Guimera2025,Wolff2025}). Future implementations could employ Bayesian estimators to provide inherent uncertainty quantification (e.g., confidence intervals for the CCs). 
Potential candidates range from non-parametric approaches, such as Gaussian Process Regression (GPR)~\cite{Rasmussen2006}, to sparse parametric strategies like Sparse Bayesian Learning (SBL)~\cite{Tipping2001} or Bayesian Neural Networks (see, e.g., Refs.~\cite{Niven2024,Fung2025,Guimera2025}).

\item[vi)] \textit{Practical and real-world applications.} The successful identification of chaotic and discontinuous dynamics under a wide range of noise levels establishes a strong foundation for applying the CC-based formalism to experimental settings. A promising next step is to employ this framework in scenarios with complex noise environments, such as colored noise or outliers common in real applications such as in fault diagnosis and system monitoring\cite{Van2016,Bounemeur2023,Abdelhamid2024,Chen2025} . 
However, although we dedicated an \ref{app:noise_x} to address noise on $x(t)$, other studies could be investigated such as taking into account robust differentiation schemes or filtering protocols directly into the pipeline (probably as a pre-processing step)\cite{Kaheman2022,Egan2024,Strebel2026}. the framework can be adapted to raw experimental recordings. 
Future research will be necessary to fully assess the efficacy of CC-based algorithms for applications.

\end{itemize}

Ultimately, the overarching goal is to bridge purely data-driven algorithms and first-principles modeling. The CC-based formalism establishes a rigorous framework that guarantees structural uniqueness of physically motivated model families, while also facilitating the integration of symmetries and post-processing tools. 
In particular, its implementation via NN-CC takes advantage of the expressive power of NNs to capture the complexity of the constitutive relations. 
This work represents a step toward a more robust paradigm of scientific discovery, one capable of autonomously extracting interpretable, physically consistent, and trustworthy models from complex observational data.


\section{Conclusions}\label{sec:conclusions}

In this work, we investigated a strategy to bridge the gap between purely data-driven learning and physics-based modeling based on the concept of characteristic curves (CCs). This approach decomposes complex dynamical systems into modular univariate functions, each corresponding to a distinct physical component defined by a constitutive relation.
This formalism transforms an ill-posed black-box learning problem into a tractable and well-posed discovery task by enforcing a structural \textit{skeleton}, which constrains the model to a physically motivated family of admissible systems and shifts the objective from unrestricted fitting to the validation of hypothesized mathematical structures.

A fundamental insight emerging from the CC-based methods is the critical role of structural identifiability. In \ref{app:uniquiness_CC}, we formally proved that for the families considered, the proposed structural templates satisfy specific uniqueness conditions. Formally, we demonstrate that when these topological requirements are met, the mapping from observed data to the underlying constitutive functions becomes injective. In practice, this establishes a rigorous basis for model selection: a low identification error, combined with the stability of the obtained CCs and successful forward validation, implies convergence to the true physical laws. Conversely, significant residual error during the optimizations or instability in the identified CCs suggests that the proposed template does not adequately describe the system dynamics. Ultimately, this uniqueness allows the formalism to serve as decisive tool for the validation or rejection of physical hypotheses.

Among the various implementations of CC-based methods, the NN-CC variants benefit from the universal approximation capabilities of NNs to capture unknown CCs without assuming rigid functional forms or basis functions. Unlike SR variants, this approach avoids reliance on rigid functional forms and discrete search, resulting in smoother approximations that degrade more gradually as noise levels increase.

Furthermore, we showed how the CC-based formalism  allows practitioners to conveniently incorporate prior knowledge or physical hypotheses, thereby refining the discovery process beyond basic structural constraints. Specifically, we integrated: 
\begin{itemize}
\vspace{-0.2 cm}
    \item[i)] Symmetry constraints, which enable the direct injection of known geometric properties (e.g., parity) into the learning process. By restricting the search space to physically admissible functions, this inductive bias significantly enhances robustness against noise and ensures that the identified dynamics respect fundamental invariances.    
\vspace{-0.2 cm}
    \item[ii)] Post-processing with symbolic regression (post-SR), which incorporates the principle of parsimony into the final model representation. By projecting the learned CCs onto compact analytical expressions, this stage drastically reduces the dimensionality of the parameter space, effectively filtering out approximation artifacts and also recovering analytical forms for each CC.
\end{itemize} 
\vspace{-0.2 cm}

Ultimately, this work presents a well-defined framework for extracting interpretable and physically consistent models from observational data. By taking advantage of the universal approximation capabilities of NNs within a structure that ensures uniqueness, the CC-based approach emerges as a promising candidate to address the identifiability issues common in purely data-driven methods. This methodology enables the integration of prior or hypothesized knowledge into the learning process and post-processing stages, offering a practical pathway to uncover the governing laws of complex dynamical systems even when theoretical understanding is incomplete.


\section*{Acknowledgments}
We acknowledge the computer time provided by CCT-Rosario and UNC Computational Centers, member of the High Performance Computing National System (SNCAD, ME-Argentina). Special thanks are given to Luis P. Lara for insightful discussions.

\section*{Funding}
This work was partially supported by CONICET (Consejo Nacional de Investigaciones Científicas y Técnicas, Argentina) under Project PIP 1679.


\section*{Competing interests}
The author declares no competing interests.

\section*{Data availability}

The data supporting the findings of this study and the source codes for the CC-based methods implemented herein are openly available in the GitHub repository: https://github.com/FedejGon/NN-CC.



\appendix

\section{Uniqueness of the CC-based formalism}
\label{app:uniquiness_CC}

The CC-based methodology, depicted in Fig.~1 of the main text, is applicable to a wide range of dynamical systems. 
However, to guarantee physical interpretability, it is essential to verify that the decomposition of the dynamics into constituent single valued function is structurally well-posed. 
In this section, we examine four specific families of systems and formally demonstrate the structural identifiability (uniqueness) of the inferred functions.

Although we formulate the derivation within the context of the NN-CC approach, the resulting uniqueness conditions are general and apply equally to other universal approximators or complete basis sets used to represent the functions $f_i$.

\subsection{Second-order velocity-dependent friction model}

Consider the family of second-order ODEs governed by the equation
\begin{equation}
    F_{ext}(t)=\ddot{x}+ f_1(\dot{x})+f_2(x)\;,
    \label{eq:2nd_veloc_th}
\end{equation}
where $x=x(t)$ is a time dependent state variable and $F_{ext}(t)$ is a time dependent external or driving force. The functions $f_1(\dot{x})$ and $f_2(x)$  are unknown nonlinear functions that define the internal elements of the system, such as dissipative (damping) and elastic (stiffness) components, respectively\cite{Gonzalez2025}.

In the NN-CC formalism, the unknown $f_i$ functions are approximated by NNs, denoted as NN$_i(\theta_i)$, where $\theta_i$ represents the trainable parameters (weights and biases).  
The proposed model takes the form:
\begin{equation}
    \hat{F}_{ext}(t)=\ddot{x}+ \text{NN}_1(\dot{x};\theta_1)+\text{NN}_2(x,\theta_2)\;,
    \label{eq:2nd_veloc_model}
\end{equation}
where $\hat{F}_{ext}(t)$ is the predicted external force.

The problem of uniqueness is motivated by this question: \textit{Does the minimization of the error between predicted and observed forces guarantee that each NN$_i$ function converges to the corresponding true $f_i$ function?} 

Let us assume that two independent training processes yield two parameter sets, $\{\theta_1^A, \theta_2^A\}$ and $ \{\theta_1^B, \theta_2^B\}$. Both sets are assumed to successfully reproduce the observed dynamics such that the training loss approaches zero. Consequently, the models satisfy the  equations:

\begin{align}
    \hat{F}_{\text{ext}}^A(t) &= \ddot{x} + \text{NN}_1^A(\dot{x}; \theta_1^A) + \text{NN}_2^A(x; \theta_2^A) \label{eq:modelA} \\
    \hat{F}_{\text{ext}}^B(t) &= \ddot{x} + \text{NN}_1^B(\dot{x}; \theta_1^B) + \text{NN}_2^B(x; \theta_2^B) \;, \label{eq:modelB}
\end{align}

where $\hat{F}_{\text{ext}}^A(t)\approx F_{ext}(t)$ and $\hat{F}_{\text{ext}}^B(t)\approx F_{ext}(t)$, and therefore $\hat{F}_{\text{ext}}^A(t)\approx \hat{F}_{\text{ext}}^B(t)$. By using this result and subtracting Eq.~\ref{eq:modelB} from Eq.~\ref{eq:modelA}) and rearranging terms, we obtain: 

\begin{equation}
    \left[ \text{NN}_1^A(\dot{x}; \theta_1^A) - \text{NN}_1^B(\dot{x}; \theta_1^B) \right] + \left[ \text{NN}_2^A(x; \theta_2^A) - \text{NN}_2^B(x; \theta_2^B) \right] = 0 \;.
    \label{eq:subtraction}
\end{equation}

Let us define the difference functions 
\begin{equation}
     \begin{aligned}
 \delta_1(\dot{x}) &= \text{NN}_1^A(\dot{x}; \theta_1^A) - \text{NN}_1^B(\dot{x}; \theta_1^B) \\      
 \delta_2(x) &= \text{NN}_2^A(x; \theta_2^A) - \text{NN}_2^B(x; \theta_2^B)
     \end{aligned}
\end{equation}
Equation~\ref{eq:subtraction} then simplifies to:
\begin{equation}
    \delta_1(\dot{x}) + \delta_2(x) = 0 \quad \forall t \;.
    \label{eq:delta_sum}
\end{equation}

For a dynamical system, the trajectory explores a manifold in the phase space $(x, \dot{x})$. Crucially, $x$ and $\dot{x}$ act as independent coordinates in this space (i.e., $\dot{x}$ is not a single-valued function of $x$). 
Equation~\ref{eq:delta_sum} implies $\delta_1(\dot{x}) = -\delta_2(x)$, thus, since the left-hand side depends strictly on velocity and the right-hand side strictly on position, both terms must be equal to a constant over all combinations of $x$ and $\dot{x}$ visited by the system:
\begin{equation}
    \delta_1(\dot{x}) = C \quad \text{and} \quad \delta_2(x) = -C \;,
    \label{eq:delta_constant}
\end{equation}
where $C$ is an arbitrary scalar constant. A strictly mathematical justification of Eq.~\ref{eq:delta_constant} can be addressed by differentiating Eq.~\ref{eq:delta_sum} by $\dot{x}$, obtaining:
\begin{equation}
    \frac{\partial}{\partial \dot{x}}[\delta_1(\dot{x})]= -   \left.\frac{\partial}{\partial \dot{x}} [\delta_2(x)]\right|_{x=\text{constant}} \,.
\end{equation}
As the right-hand size depends only on $x$, so its derivative with respect to $\dot{x}$ is zero, then we obtain:
\begin{equation}
    \frac{d}{d \dot{x}}\delta_1(\dot{x})=0 \;,
\end{equation}
yielding to Eq.~\ref{eq:delta_constant}.

This result leads to two important consequences:
\begin{itemize}
   \vspace{-0.2 cm}\item[(i)] Inter-model consistency: Any two sufficiently converged  training instances will be identical up to a constant shift: $\text{NN}_1^A(\dot{x}\,; \theta_1^A) = \text{NN}_1^B(\dot{x}\,; \theta_1^B) + C$ and $\text{NN}_2^A(x\,; \theta_i^A) = \text{NN}_2^B(x\,; \theta_i^B) - C$. Thus, for instance, if $\text{NN}_1^A(\dot{x}\,; \theta_i^A)$ has converged to a parabolic shape, then $\text{NN}_1^B(\dot{x}\,; \theta_i^B)$ must converge to the same parabola (up to a constant). This implies that each NN describes the same physical curve (except by a constant vertical shift). 
   It is crucial to emphasize that this functional equivalence between the identified CCs does not imply equivalence between the parameter sets $\theta^A$ and  $\theta^B$.  Due to the over-parameterized nature of NNs, the mapping from parameter space to function space is many-to-one; thus, distinct weight and bias values can yield identical physical approximations.

    \vspace{-0.2 cm} \item[(ii)] Convergence to true functions: The true physical functions $f_1(\dot{x})$ and $f_2(x)$ satisfies the governing ODE (Eq.~\ref{eq:2nd_veloc_th}). 
    By equating Eqs.~\ref{eq:2nd_veloc_th} and \ref{eq:2nd_veloc_model}, and following an analogous derivation, we obtain $\text{NN}_1(\dot{x}\,; \theta_1) = f_1(\dot{x}) + D$ and $\text{NN}_2(x\,; \theta_2) = f_2(x) - D$, where $D$ is a constant that is not necessarily equal to the constant $C$ defined above.  
    This shows that, for any training instance that converges with a sufficiently small loss error, the resulting NN$_1$ and NN$_2$ functions must converge to the true underlying CCs $f_1$ and $f_2$ (up to the constant $D$). 
\end{itemize}

In practice, the shift constants $C$ and $D$ are not a major issue, as they can be eliminated by imposing an appropriate physical constraint. For example, if we require $f_{1}(0) = 0$, this condition can be enforced by adding a corresponding penalty term to the loss function during training. Alternatively, the constant shift can be removed through a post-processing correction of the learned functions. In this case, enforcing $f_{1}(0) = 0$ requires redefining $\text{NN}_1(\dot{x}\,;\theta)$ as $\text{NN}_1(\dot{x}\,;\theta) - \text{NN}_1(\dot{x}=0\,;\theta)$ and $\text{NN}_2(x\,;\theta)$ as $\text{NN}_2(x\,;\theta) + \text{NN}_1(\dot{x}=0\,;\theta)$.
Thus, the NN-CC methodology applied to this family of equations uniquely identifies the underlying physical laws governing the system dynamics, irrespective of the network initialization or parameterization, provided that the training loss is sufficiently small (i.e., $\hat{F}_{\text{ext}}(t)\approx F_{\text{ext}}(t)$).

\subsection{Second-order position-dependent friction model}

Consider the family of systems where the dissipative force is nonlinearly dependent on position, referred to as position-dependent friction model\cite{Gonzalez2025}: 

\begin{equation} 
F_{ext}(t) = \ddot{x} + f_1(x)\dot{x} + f_2(x) \;, \label{eq:pos_damp_gov} 
\end{equation}
where $f_1(x)$ represents a position-dependent damping coefficient and $f_2(x)$ represents the stiffness. In the NN-CC formalism, we propose the estimator: 
\begin{equation} 
\hat{F}_{ext}(t) = \ddot{x} + \text{NN}_1(x;\theta_1)\dot{x} + \text{NN}_2(x;\theta_2) \;. 
\end{equation}

To analyze the uniqueness, we again assume two distinct parameter sets, A and B, that successfully reproduce the external force. Subtracting the model equations yields: 
\begin{equation} 
\left[ \text{NN}_1^A(x; \theta_1^A) - \text{NN}_1^B(x; \theta_1^B) \right]\;\dot{x} + \left[ \text{NN}_2^A(x; \theta_2^A) - \text{NN}_2^B(x; \theta_2^B) \right] = 0 \;. 
\label{eq:app:uniqueness:position_dependence}
\end{equation} 

Defining the difference functions 
$\delta_i(x)=\text{NN}_i^A(x\,;\theta_i^A)- \text{NN}_i^B(x\,;\theta_i^B)$, with $i=\{1,2\}$, Eq.~\ref{eq:app:uniqueness:position_dependence} simplifies to:
\begin{equation} 
\delta_1(x)\;\dot{x} + \delta_2(x) = 0 \;. 
\label{eq:delta_pos_damp} 
\end{equation} 
Since $x$ and $\dot{x}$ are treated as independent variables in the phase space formulation, we differentiate Eq.~\ref{eq:delta_pos_damp} with respect to $\dot{x}$ (holding $x$ constant):
\begin{equation} 
\frac{\partial}{\partial \dot{x}} [\delta_1(x)\dot{x}] + \frac{\partial}{\partial \dot{x}} [\delta_2(x)] = 0 \implies \delta_1(x) = 0 \;. 
\end{equation} 
Since $\delta_1(x)=0$, it immediately follows from Eq.~\ref{eq:delta_pos_damp} that $\delta_2(x)=0$.

This result demonstrates strict structural uniqueness: $\text{NN}_1^A(x; \theta_1^A)=\text{NN}_1^B(x; \theta_1^B)$ and $\text{NN}_2^A(x; \theta_2^A)=\text{NN}_2^B(x; \theta_2^B)$. Unlike the second-order velocity-dependent case, there are no arbitrary constants. Provided that the training loss error obtained after minimization is sufficiently small, this uniqueness guarantees that NN$_1$ and NN$_2$ converge exactly to the true physical functions $f_1$ and $f_2$, respectively.

\subsection{A family of first-order ODEs}

Consider the family of first-order systems governed by the equation: 

\begin{equation} F_{ext}(t) = f_1(x) + f_2(x)\;\dot{x} \,, \label{eq:1st_order_gov} 
\end{equation} 
where $x$ is a time dependent state variable, and $F_{ext}(t)$ is a external or driving force. The terms $f_1(x)$ and $f_2(x)$  are nonlinear functions that define the internal dynamics of the system, such as dissipative and elastic components, as discussed in Refs.~\cite{Gonzalez2023,Gonzalez2024}.

In the NN-CC formalism, we approximate the unknown functions as: \begin{equation} 
\hat{F}_{ext}(t) = \text{NN}_1(x;\theta_1) + \text{NN}_2(x;\theta_2)\;\dot{x} \;. 
\end{equation}

To address the uniqueness of the identification, we assume two distinct trained models, A and B, both satisfying the governing equation with negligible loss error. Subtracting the model equations ($\hat{F}_{ext}^A\approx \hat{F}_{ext}^B \approx F_{ext}$) leads to: \begin{equation} 
\left[ \text{NN}_1^A(x; \theta_1^A) - \text{NN}_1^B(x; \theta_1^B) \right] + \left[ \text{NN}_2^A(x; \theta_2^A) - \text{NN}_2^B(x; \theta_2^B) \right]\;\dot{x} = 0 \;. 
\end{equation}

Defining the difference functions $\delta_i(x)=\text{NN}_i^A(x\,;\theta_i^A)-\text{NN}_i^B(x\,;\theta_i^B)$, with $i=\{1,2\}$, we obtain: 
\begin{equation} 
\delta_1(x) + \delta_2(x)\;\dot{x} = 0 \;. 
\label{eq:delta_1st_order} 
\end{equation}

We can differentiate Eq.~\ref{eq:delta_1st_order} with respect to $\dot{x}$ while treating $x$ as constant, thus obtaining: 
\begin{equation} 
\frac{\partial}{\partial \dot{x}} [\delta_1(x)] + \frac{\partial}{\partial \dot{x}} [\delta_2(x)\dot{x}] = 0  \implies \delta_2(x)=0\;. 
\end{equation} 
Substituting this result back into Eq.~\ref{eq:delta_1st_order} immediately yields $\delta_1(x)=0$.

This result demonstrates strict structural uniqueness: $\text{NN}_1^A(x; \theta_1^A)=\text{NN}_1^B(x; \theta_1^B)$ and $\text{NN}_2^A(x; \theta_2^A)=\text{NN}_2^B(x; \theta_2^B)$. Unlike the second-order velocity-dependent case, there are no arbitrary constants. Provided that the training loss error obtained after minimization is sufficiently small, this uniqueness guarantees that NN$_1$ and NN$_2$ converge exactly to the true physical functions $f_1$ and $f_2$, respectively.

\subsection{Multi-degree-of-freedom systems}
\label{app:uniqueness_MDOF}

In this section, we consider the application of the CC-based formalism to multi-degree-of-freedom (MDOF) systems. We focus on a coupled two-degree-of-freedom (2DOF) system, which serves as a canonical model for coupled nonlinear oscillators in structural dynamics and lattice systems~\cite{Nayfeh_2004}. The governing equations are defined as: 
\begin{align} 
F_{ext,1}(t) &= \ddot{x}_1 + f_1(\dot{x_1}) + f_2(x_1 - x_2) \label{eq:mdof_gov1} \\ 
F_{ext,2}(t) &= \ddot{x}_2 + f_3(\dot{x_2}) + f_4(x_2 - x_1) \label{eq:mdof_gov2} 
\end{align}
where $x_1$ and $x_2$ are the time-dependent state variables of each oscillator.
The functions $f_1(\dot{x_1})$ and $f_3(\dot{x_2})$ represent local velocity-dependent dissipation (damping), while $f_2(x_1-x_2)$ and $f_4(x_2-x_1)$ represent position-dependent coupling forces.

In the NN-CC formalism, we approximate these unknown functions using separate neural networks. The estimator model takes the form: 
\begin{align} 
\hat{F}_{ext,1}(t) &= \ddot{x}_1 + \text{NN}_1(\dot{x}_1\,; \theta_1) + \text{NN}_2(x_1 - x_2\,; \theta_2) \label{eq:mdof_gov1_NN}\\ 
\hat{F}_{ext,2}(t) &= \ddot{x}_2 + \text{NN}_3(\dot{x}_2\,; \theta_3) + \text{NN}_4(x_2 - x_1\,; \theta_4)\; . 
\label{eq:mdof_gov2_NN}
\end{align}

To analyze the uniqueness, we assume that two independent training instances yield parameter sets $\{\theta^A\}$ and $\{\theta^B\}$ that successfully reproduce the observed external forces for both degrees of freedom. Consequently, $\hat{F}_{ext,1}^A(t)\approx \hat{F}_{ext,1}^B(t)$ for $k=\{1,2\}$.  
Subtracting the model equations for the first degree of freedom (Eq.~\ref{eq:mdof_gov1_NN}) yields: 
\begin{equation} 
\left[ \text{NN}_1^A(\dot{x}_1\,; \theta_1^A) - \text{NN}_1^B(\dot{x}_1\,; \theta_1^B) \right] + \left[ \text{NN}_2^A(r\,; \theta_2^A) - \text{NN}_2^B(r\,; \theta_2^B) \right] = 0 \;, 
\label{eq:mdof_subtraction} 
\end{equation}

where $r=x_1-x_2$ represents the relative displacement. Defining the difference functions $\delta_i(\cdot)=\text{NN}_i^A(\cdot\,;\theta^A)-\text{NN}_i^B(\cdot\,;\theta^B)$, 
 Eq.~\ref{eq:mdof_subtraction} simplifies to: \begin{equation} 
 \delta_1(\dot{x}_1) + \delta_2(r) = 0 \;. \label{eq:delta_mdof} 
 \end{equation}
In the phase space of the system, the absolute velocity $\dot{x_1}$ and the relative position $r$ act as independent coordinates. We can differentiate Eq.~\ref{eq:delta_mdof} with respect to $\dot{x_1}$ while treating r as constant:

\begin{equation} 
\frac{d}{d \dot{x}_1}\delta_1(\dot{x}_1) = - \left. \frac{\partial}{\partial \dot{x}_1} [\delta_2(r)] \right|_{r=\text{constant}} = 0 \;. 
\end{equation}

Since the derivative is zero, $\delta_1(\dot{x_1})$ must be a constant $C_1$.  It immediately follows from Eq.~\ref{eq:delta_mdof} that $\delta_2(r)=-C_1$. Applying the same derivation to the second degree of freedom yields 
$\delta_3(\dot{x_2})=C_2$ and $\delta_4(-r)=-C_2$.

This result leads to the following consequences: \begin{itemize} 
\vspace{-0.2 cm}\item[(i)] Inter-model consistency: The identified functions are unique up to additive constants: 
\begin{equation}
\begin{aligned} 
\text{NN}_1^A(\dot{x}_1) &= \text{NN}_1^B(\dot{x}_1) + C_1 \\ \text{NN}_2^A(r) &= \text{NN}_2^B(r) - C_1 \\ 
\text{NN}_3^A(\dot{x}_2) &= \text{NN}_3^B(\dot{x}_2) + C_2 \\
\text{NN}_4^A(-r) &= \text{NN}_4^B(-r) - C_2 \,.
\end{aligned}
\end{equation}
This implies that for any sufficiently converged training instance, the NNs will describe the same physical curves, differing only by vertical shifts $C_1$ and $C_2$.
\vspace{-0.2 cm} \item[(ii)] Physical constraints: As with the single-degree-of-freedom case, these arbitrary constants can be eliminated by imposing physical boundary conditions. For instance, enforcing zero damping at rest ($f_{damping}(0)=0$) and zero coupling at equilibrium ($f_{coupling}(0)=0$) allows us to set $C_1=0$ and $C_2=0$.
\end{itemize}

Thus, provided the system trajectory does not collapse onto a synchronization manifold (where $x_1(t)=x_2(t)$ for all $t$), the NN-CC methodology uniquely identifies the underlying interaction and damping laws governing the MDOF system.

\section{Model selection}
\label{app:model_selection}

A core challenge in system identification is distinguishing between a model that merely fits the data (overfitting) and one that captures the true underlying physical mechanism. Unlike library-based sparse regression methods that select terms from a list of candidates, the CC-based formalism operates by hypothesizing a structural family of models (e.g., Eq.~\ref{eq:2nd_veloc_th}) and validating its consistency. In this section, we formalize the hierarchical selection process depicted in the schematic of Fig.~\ref{fig:schematic}, detailing the rejection criteria used to discard incorrect model structures.

The selection process is governed by three nested feedback loops (inner, middle, and outer), each serving as a filter for increasingly subtle structural errors.

\begin{itemize}
\item[i)] Inner loop: Optimization and representational flexibility. The first criterion is the ability of the proposed family to represent the training data. This is assessed via the training loss $L_\text{tot}$ (step (f) in Fig.~\ref{fig:schematic}). 
\begin{itemize} \item Criterion: if the converged loss $L_\text{tot}$ remains significantly high (relative to some noise floor), it implies that the chosen family of models lacks the representational capacity to describe the system. For example, employing a first-order model to represent second-order systems results in systematic errors that cannot be eliminated through further optimization. 
\item Action: if training error is high, the current model structure is rejected. The practitioner then shifts to a more sophisticated or alternative model type, as discussed below.
\end{itemize}

\item[ii)] Middle loop: Invariance under the constitutive relations. A low training loss is necessary but not sufficient; a highly flexible NN might `memorize' a specific trajectory even with an incorrect physical structure. The middle loop serves as a structural consistency check based on the uniqueness properties derived in \ref{app:uniquiness_CC}. 
\begin{itemize} 
\item Criterion: if the proposed structural family matches the ground truth, the mapping from data to the constitutive functions ($f_i$) is injective. Consequently, if the model is trained on multiple independent datasets (generated with different initial conditions or driving forces), the identified CCs must converge toward the same invariant functions (within the limits of noise-induced uncertainty).  
This invariance of the CCs could be rigorously assessed in future works using statistical resampling techniques such as \textit{k-fold cross-validation} or \textit{bootstrapping}. By training models on resampled datasets, the practitioner can generate confidence intervals for the CCs. Alternatively, adopting a \textit{Bayesian perspective} (e.g., via Bayesian NNs or ensembles) allows for estimating the full posterior distribution of the CCs, directly quantifying the epistemic uncertainty associated with the structural hypothesis.

\item Indicator of failure: If the identified CCs exhibit significant variations or shifts across different datasets (e.g., the damping curve $f_1(\dot{x})$ changes shape when the driving amplitude $A$ changes), this indicates `structural compensation'. The NNs are deforming to compensate for missing terms (e.g., a missing cross-term $x \,\dot{x}$), and thus, the obtained models are dependent on the specific trajectory rather than being state invariants. 

\item Action: divergence of CCs across datasets triggers model rejection, even if the training error is low. 
\end{itemize}

\item[iii)] Outer loop: generalization and dynamic consistency. The final validation assesses the predictive power of the model on unseen regions of the phase space (dynamics interpolation). 
\begin{itemize} 
\item Criterion: the identified model is integrated forward in time using initial conditions and forcing functions distinct from those in the training set.
A metric should be defined based on the obtained trajectories, for example we used the separation time ($t_\text{sep}$) for chaotic Duffing system and RMSE for the stick-slip system.
\item Action: the hypothesized family is rejected if the model demonstrates poor generalization on novel test cases according to the predefined metric. This criterion ensures the recovery of the global vector field, distinguishing true system identification from simple interpolation of the training data.
\end{itemize}
\end{itemize}

To systematically identify the governing equations without prior knowledge, the validation protocol described above can be deployed within a hierarchical search strategy. In this approach, the practitioner defines a library of candidate families $\mathcal{F} = \{ \mathcal{S}_1, \mathcal{S}_2, \dots, \mathcal{S}_K \}$, ordered by complexity (parsimony). The search iterates through this library, testing the simplest structures first. Based on the uniqueness proofs provided in \ref{app:uniquiness_CC}, a standard hierarchy for single-degree-of-freedom oscillators is:
\begin{itemize}
    \item[i)] Family I (First-order decomposition): $f_1(x) + f_2(x)\dot{x} = F_\text{ext}$. Suitable for overdamped regimes or chemical kinetics\cite{Gonzalez2023,Gonzalez2024}.
    \item[ii)] Family II (additively separable): $\ddot{x} + f_1(\dot{x}) + f_2(x) = F_\text{ext}$. This covers standard mechanical systems\cite{Gonzalez2025}.
    \item[iii)] Family III (position-dependent friction): $\ddot{x} + f_1(x)\dot{x} + f_2(x) = F_\text{ext}$. This covers systems with variable damping, such as the Van der Pol oscillator\cite{Gonzalez2025}.
    
\end{itemize}
The procedure accepts the first family $\mathcal{S}_i$ that satisfies all three validation loops. This hierarchical approach presents distinct advantages and disadvantages compared to fully automated library-based methods:

\textit{Advantages:}
\begin{itemize}
    \item[i)] Well-posedness and uniqueness: by restricting the search to families with proven injective mappings, the method guarantees that the identified functions are unique physical invariants rather than degenerate solutions.
    \item[ii)] Avoidance of combinatorial explosion: unlike sparse regression, which searches a combinatorial space of $2^N$ candidate terms, this approach searches a small set of $K$ structural families, which in parallel implementations could drastically reduce computational cost.
    \item[iii)] Physical interpretability: the imposed structures enforce physical consistency (e.g., separation of dissipative and conservative forces) by design.
\end{itemize}

\textit{Disadvantages:}
\begin{itemize}
    \item[i)] Library dependency: the success of the method is contingent on the true governing law belonging to one of the hypothesized families in $\mathcal{F}$. If the system dynamics lie outside the defined library, the method will fail to converge to a valid solution. However, this limitation also serves as a robust mechanism for hypothesis testing: due to the uniqueness property, the method can definitively discard incorrect structures. If a proposed family fails the validation loops, the practitioner can confidently rule out that physical description, systematically pruning the search space of candidate theories.
    \item[ii)] Semi-automated nature: unlike black-box discovery tools, this strategy requires user intervention to define the library and the ordering of complexity, introducing a degree of inductive bias. 
    However, in practical engineering and scientific applications, the practitioner rarely approaches a problem without prior assumptions, frequently utilizing domain expertise to narrow the search to a specific class of models (e.g., mechanical oscillators vs. chemical kinetics). The NN-CC framework is ideal for these grey-box scenarios, as it allows explicitly framing and validating these structural hypotheses, aiming to bridge the gap between pure theory and data-driven discovery.
\end{itemize}

Finally, the CC formalism offers a distinct advantage through the direct visualization of the identified CCs. Unlike opaque black-box models, the obtained CCs can be inspected visually to identify qualitative features, such as parity or monotonicity. This transparency allows the practitioner to incorporate prior domain expertise or to hypothesize symmetries based on preliminary training runs (e.g., observing an odd symmetry in the friction constitutive relation). By formally embedding these symmetries into the loss function, the method effectively constrains the search space. This acts as a rigorous inductive bias within the model selection process, rejecting physically inconsistent solutions and ensuring that the final model respects the fundamental geometric properties of the system.

\section{Sensitivity analysis}\label{app:sensitivity_analysis}

In this section, we perform a sensitivity analysis to quantify the dependence of the identified models on the NN-CC hyperparameters. 
Our primary objective is to define a \textit{unified} set of hyperparameters capable of accurately describing both continuous (Duffing) and discontinuous (stick-slip) systems without requiring system-specific tuning. These results serve as a baseline for subsequent parameter refinement.

It is important to mention that the training loss function ($L_\text{tot}$) is not always a conclusive metric for parameter selection.  
As discussed in the main text, a lower training loss may indicate overfitting to specific noise realizations rather than better physical identification. Therefore, we also rely on the RMSE values of the recovered CCs
against the ground truth as a more decisive metric for hyperparameter tuning.  
All models in this section were trained using datasets generated with $\text{SNR}=20$ dB, using the system parameters defined in the main text.

Table~\ref{tab:nn_performance_optimizers} presents the performance comparison for different optimizers. The stochastic gradient descent (SGD) optimizer fails to converge to a good solution, exhibiting high errors across all momentum values. In contrast, adaptive moment estimation methods (Adam and AdamW) demonstrate superior performance. Specifically, Adam and AdamW (the latter with weight decay $<10^{-4}$) yield the lowest RMSE values. To maintain simplicity without sacrificing accuracy, we select the Adam optimizer in this work.

\begin{table}[!htpb]
\centering
\caption{Performance comparison for different optimizers using the NN-CC method. Fixed parameters: 100 neurons per layer, learning rate $10^{-4}$, ReLU activation.}
\label{tab:nn_performance_optimizers}
\begin{tabular}{l ccc ccc}
\toprule
\multirow{2}{*}{{Optimizer}}  & \multicolumn{3}{c}{Duffing} & \multicolumn{3}{c}{Stick-slip}  \\
\cmidrule(lr){2-4} \cmidrule(lr){5-7} 
 &  L$_\text{tot}$ & RMSE[$f_1$] & RMSE[$f_2$] & L$_\text{tot}$ & RMSE[$f_1$] & RMSE[$f_2$]  \\
 &  ($\times10^{-3}$) & ($\times10^{-3}$) & ($\times10^{-3}$) & ($\times10^{-3}$) & ($\times10^{-3}$) & ($\times10^{-3}$)  \\
\midrule
\textit{Adam:}    & 1.1      & 7  & 10  & 17 & 46 & 25 \\ 
\textit{AdamW:}    &       &   &   &  &  &  \\
\quad Weight decay$=10^{-6}$    & 1.12      & 8  & 10  & 18 & 44 & 44 \\
\quad Weight decay$=10^{-4}$    & 1.11      & 8  & 9  & 17 & 38 & 41 \\
\quad Weight decay$=10^{-3}$    & 1.11      & 8  & 10  & 18 & 50 & 61 \\
\quad Weight decay$=10^{-2}$    & 1.11      & 9 & 8  & 16 & 58 & 63 \\
\quad Weight decay$=10^{-1}$    & 1.12      & 9 & 11  & 17 & 43 & 53 \\
\quad Weight decay$=10^{0}$    & 1.14      & 13 & 16  & 19 & 72 & 63 \\
\textit{SGD:}    &       &   &   &  &  &  \\
\quad Momentum$=0.5$    & 120      & 98  & 422  & 172 & 446 & 366 \\
\quad Momentum$=0.9$    & 4      & 294  & 298  & 130 & 207 & 68 \\
\quad Momentum$=0.99$    & 1.2      & 150  & 190  & 43 & 86 & 71 \\
\bottomrule
\end{tabular}%
\end{table}

Table~\ref{tab:nn_performance_activ_func} compares various activation functions. While Tanh and RReLU show competitive performance, we observe that Softplus performs significantly worse for the stick-slip system (RMSE[$f_1$] of 60 vs 46 for ReLU). This is probably because Softplus enforces a smooth approximation that struggles to capture the sharp discontinuity of the Coulomb friction. Consequently, we select ReLU as it offers a robust trade-off, handling the discontinuity of stick-slip dynamics well while remaining accurate for the continuous Duffing system. Other activations not listed here were excluded based on poor performance in noise-free tests~\cite{Gonzalez2025}.

\begin{table}[!htpb]
\centering
\caption{Performance comparison for different activation functions (NN-CC method). Fixed parameters: 100 neurons per layer, learning rate $10^{-4}$.}
\label{tab:nn_performance_activ_func}
\begin{tabular}{c ccc ccc}
\toprule
\multirow{2}{*}{{Activation}}  & \multicolumn{3}{c}{Duffing} & \multicolumn{3}{c}{Stick-slip}  \\
\cmidrule(lr){2-4} \cmidrule(lr){5-7} 
function &  L$_\text{tot}$ & RMSE[$f_1$] & RMSE[$f_2$] & L$_\text{tot}$ & RMSE[$f_1$] & RMSE[$f_2$]  \\
 &  ($\times10^{-3}$) & ($\times10^{-3}$) & ($\times10^{-3}$) & ($\times10^{-3}$) & ($\times10^{-3}$) & ($\times10^{-3}$)  \\
\midrule
ReLU    & 1.1      & 7  & 10  & 17 & 46 & 25 \\
LeakyReLU    & 1.1      & 8  & 10  & 18 & 57 & 60 \\
Tanh    & 1.2      & 4  & 6  & 19 & 40 & 37 \\
RReLU    & 1.14      & 8  & 8  & 18 & 42 & 34 \\
Softplus    & 1.6      & 12  & 12  & 30 & 60 & 70 \\
\bottomrule
\end{tabular}%
\end{table}

Table~\ref{tab:nn_performance_learning_rate} analyzes the sensitivity to the learning rate. The methods exhibit a stable plateau of good performance for learning rates between $10^{-5}$ and $10^{-3}$. Rates larger than $10^{-2}$ lead to instability. We select a value of $\mathbf{10^{-4}}$ as it lies safely within the stable regime for both systems.

\begin{table}[!htpb]
\centering
\caption{Performance comparison for different learning rates (NN-CC method).}
\label{tab:nn_performance_learning_rate}
\begin{tabular}{c ccc ccc}
\toprule
\multirow{2}{*}{{Learning}}  & \multicolumn{3}{c}{Duffing} & \multicolumn{3}{c}{Stick-slip}  \\
\cmidrule(lr){2-4} \cmidrule(lr){5-7} 
rate &  L$_\text{tot}$ & RMSE[$f_1$] & RMSE[$f_2$] & L$_\text{tot}$ & RMSE[$f_1$] & RMSE[$f_2$]  \\
 &  ($\times10^{-3}$) & ($\times10^{-3}$) & ($\times10^{-3}$) & ($\times10^{-3}$) & ($\times10^{-3}$) & ($\times10^{-3}$)  \\
\midrule
$10^{-6}$    & 1.15      & 7  & 10  & 210 & 390 & 62 \\
$10^{-5}$    & 1.15      & 7  & 10  & 20 & 30 & 25 \\
$10^{-4}$    & 1.1      & 7  & 10  & 17 & 46 & 25 \\
$10^{-3}$    & 1.05      & 13  & 11  & 18 & 65 & 54 \\
$10^{-2}$    & 1.1      & 10  & 14  & 17 & 87 & 76 \\
$10^{-1}$    & 250      & 171  & 555  & 220 & 278 & 240 \\
\bottomrule
\end{tabular}%
\end{table}

We now analyze the impact of the symmetry regularization weight $\lambda_\text{sym}$. The total loss function is defined as $L_\text{tot}=L_\text{data}+ \lambda_\text{sym}L_\text{sym}$. Table~\ref{tab:nn_performance_lambda_sym} shows the model performance for different $\lambda_\text{sym}$ values.

For low values ($\lambda_\text{sym} \lesssim 10^{-3}$), the symmetry term is negligible, and the model behaves like the unconstrained NN-CC. However, for $\lambda_\text{sym} \gtrsim 10^{-2}$, we observe a significant reduction in the RMSE of the identified CCs, particularly for the stick-slip system where the RMSE[$f_1$] drops from $\approx 60$ to $\approx 13$. A robust range where both systems perform optimally is between $10^{0}$ and $10^{2}$; we select ${\lambda_\text{sym}=10}$ for our final models.

It is important to note that the optimal parameters do not correspond to the lowest total loss $L_\text{tot}$. As shown in Table~\ref{tab:nn_performance_lambda_sym}, $L_\text{tot}$ is minimal for $\lambda_\text{sym} \approx 0$. 
Although $L_\text{data}$ increases with $\lambda_\text{sym}$), the RMSE metric reduce. This highlights the necessity of using validation metrics beyond simple training loss when optimizing these physics-informed models.

\begin{table}[!htpb]
\centering
\caption{Performance comparison for different symmetry hyperparameter values for the NN-CC$_\text{+sym}$ method.}
\label{tab:nn_performance_lambda_sym}
\begin{tabular}{c cccc cccc}
\toprule
\multirow{2}{*}{{Symmetry}}  & \multicolumn{4}{c}{Duffing} & \multicolumn{4}{c}{Stick-slip}  \\
\cmidrule(lr){2-5} \cmidrule(lr){6-9} 
hyperparameter &  $L_\text{data}$ & $\lambda_\text{sym}\,L_\text{sym}$ & RMSE[$f_1$] & RMSE[$f_2$] & $L_\text{data}$ & $\lambda_\text{sym}\,L_\text{sym}$ & RMSE[$f_1$] & RMSE[$f_2$]  \\
($\lambda_\text{sym}$) &  ($\times10^{-3}$) & ($\times10^{-3}$) & ($\times10^{-3}$) & ($\times10^{-3}$) & ($\times10^{-3}$) & ($\times10^{-3}$) & ($\times10^{-3}$) & ($\times10^{-3}$)  \\
\midrule
$0$    & 1.1      & 2$\times10^{-5}$  & 7  & 10 & 17 &  4$\times10^{-5}$ & 46 & 25 \\
$10^{-6}$    & 1.09      & 2$\times10^{-5}$  & 13  & 10 & 17 &  4$\times10^{-5}$ & 66 & 30 \\
$10^{-5}$    & 1.09      & 4$\times10^{-5}$  & 9  & 9 & 18 &  2$\times10^{-3}$ & 61 & 29 \\
$10^{-4}$    & 1.09      & 3$\times10^{-5}$  & 9  & 10 & 18 &  3$\times10^{-3}$ & 59 & 32 \\
$10^{-3}$    & 1.09      & 4$\times10^{-5}$  & 9  & 9 & 18 &  2$\times10^{-3}$ & 56 & 21 \\
$10^{-2}$    & 1.1      & 3$\times10^{-3}$  & 8  & 10 & 17 &  3$\times10^{-2}$ & 22 & 25 \\
$10^{-1}$    & 1.11      & 2$\times10^{-2}$  & 8  & 10 & 18 &  1$\times10^{-1}$ & 28 & 19 \\
$10^{0}$    & 1.16      & 1$\times10^{-2}$  & 8  & 10 & 18 &  1$\times10^{-1}$ & 16 & 18 \\
$10^{1}$    & 12      & 3$\times10^{-3}$  & 3  & 6 & 19 &  2$\times10^{-1}$ & 16 & 9 \\
$10^{2}$    & 1.29      & 2$\times10^{-2}$  & 3  & 9 & 19 &  4$\times10^{-1}$ & 13 & 13 \\
$10^{3}$    & 3      & 4$\times10^{-2}$  & 4  & 45 & 110 &  2 & 106 & 66 \\
\bottomrule
\end{tabular}%
\end{table}

Finally, we evaluate the effect of network width (neurons per layer, $n$). Tables~\ref{tab:nn_performance_neurons_duffing} and \ref{tab:nn_performance_neurons_stick_slip} summarize the results. The number of learnable parameters for NN-CC and NN-CC$_\text{+sym}$ methods is given by $2\, (2\,n^2+5\,n+1)$ (accounting for two separate networks).

For very small networks ($n \le 5$), the model lacks the expressiveness to capture the dynamics, resulting in high RMSE values for the CCs. 
The theoretical number of parameters is 3 for the Duffing system ($\delta$, $\alpha$, $\beta$) and effectively 4 for the stick-slip system. For the latter, in addition to the physical parameters $c$, $\mu\,N$, and $k$, the symbolic regression search identifies a scaling parameter $c_1$ within the approximation $ \text{sign}(\dot{x})\sim\tanh(c_1\,\dot{x})$. This $c_1$ parameter ($c_1\gg 1$) governs the steepness of the transition at zero velocity, allowing the smooth basis functions to approximate the discontinuity.
Notably, for $n \lesssim 10$, the post-SR step (NN-CC$_\text{+sym+post-SR}$) occasionally identifies the wrong number of terms (e.g., 5 parameters instead of the theoretical 3 for Duffing or lower than 4 for stick-slip) due to errors propagated from the under-fitted NNs.

Conversely, large networks ($n\geq 500$) show higher RMSE values for the stick-slip system.  Based on these results, we identify a broad optimal range between $n=25$ and $n=250$. We select $n=100$ as a balanced choice that ensures sufficient expressivity for both continuous and discontinuous systems while maintaining a manageable parameter count.

\begin{table}[!htpb]
\centering
\caption{Performance comparison as a function of the neurons per layer ($n$) for the Duffing system.}
\label{tab:nn_performance_neurons_duffing}
\resizebox{\textwidth}{!}{%
\begin{tabular}{cc ccc ccc ccc}
\toprule
\multirow{2}{*}{{Neurons}} & \multirow{2}{*}{{NN}} & \multicolumn{3}{c}{NN-CC} & \multicolumn{3}{c}{NN-CC$_\text{+sym}$} & \multicolumn{3}{c}{NN-CC$_\text{+sym+post-SR}$} \\
\cmidrule(lr){3-5} \cmidrule(lr){6-8} \cmidrule(lr){9-11} per layer & params & L$_\text{tot}$ & RMSE[$f_1$] & RMSE[$f_2$] & L$_\text{tot}$ & RMSE[$f_1$] & RMSE[$f_2$] & Params & RMSE[$f_1$] & RMSE[$f_2$] \\
($n$) & $2\,(2\,n^2+5\,n+1)$ & ($\times10^{-3}$) & ($\times10^{-3}$) & ($\times10^{-3}$) & ($\times10^{-3}$) & ($\times10^{-3}$) & ($\times10^{-3}$) &  & ($\times10^{-3}$) & ($\times10^{-3}$) \\
\midrule
2    & 38      & 240  & 18  & 500 & 240  & 8   & 500 & 5 & 8   & 500 \\
5    & 152     & 2.1  & 8   & 32  & 6.9  & 4   & 80  & 2 & 4   & 9   \\
10   & 502     & 1.2  & 6   & 13  & 1.7  & 2.9 & 20  & 4 & 2   & 1.6 \\
25   & 2752    & 1.15 & 7   & 10  & 1.25 & 2.7 & 10  & 3 & 1.4 & 2.5 \\
50   & 10502   & 1.14 & 9   & 8.5 & 1.2  & 3   & 6   & 3 & 1.3 & 2.5 \\
100  & 41002   & 1.1  & 7  & 10   & 1.2  & 3   & 6   & 3 & 0.9 & 2   \\
250  & 252502  & 1.09  & 10  & 10   & 1.2  & 4   & 6  & 3 & 1   & 2   \\
500  & 1005002 & 1.08  & 10  & 10  & 1.2  & 5   & 4   & 3 & 1   & 2   \\
1000 & 4010002 & 1.07 & 9.7 & 9.3 & 1.1  & 5   & 5   & 3 &  1   & 2   \\
\bottomrule
\end{tabular}%
}
\end{table}

\begin{table}[!htpb]
\centering
\caption{Performance comparison as a function of the neurons per layer ($n$) for the stick-slip system.}
\label{tab:nn_performance_neurons_stick_slip}
\resizebox{\textwidth}{!}{%
\begin{tabular}{cc ccc ccc ccc}
\toprule
\multirow{2}{*}{{Neurons}} & \multirow{2}{*}{{NN}} & \multicolumn{3}{c}{NN-CC} & \multicolumn{3}{c}{NN-CC$_\text{+sym}$} & \multicolumn{3}{c}{NN-CC$_\text{+sym+post-SR}$} \\
\cmidrule(lr){3-5} \cmidrule(lr){6-8} \cmidrule(lr){9-11} per layer & params & L$_\text{tot}$ & RMSE[$f_1$] & RMSE[$f_2$] & L$_\text{tot}$ & RMSE[$f_1$] & RMSE[$f_2$] & Params & RMSE[$f_1$] & RMSE[$f_2$] \\
($n$) & $2\,(2\,n^2+5\,n+1)$ & ($\times10^{-3}$) & ($\times10^{-3}$) & ($\times10^{-3}$) & ($\times10^{-3}$) & ($\times10^{-3}$) & ($\times10^{-3}$) &  & ($\times10^{-3}$) & ($\times10^{-3}$) \\
\midrule
2    & 38      & 650  & 950  & 182 & 2200  & 960   & 899 & 1 & 961   & 898 \\
5    & 152     & 23  & 69   & 70  & 1500  & 147   & 898  & 2 & 353   & 890   \\
10   & 502     & 18  & 53   & 39  & 28  & 14 & 20  & 4 & 6   & 2 \\
25   & 2752    & 18 & 57   & 39  & 19 & 17 & 7  & 4 & 8 & 2 \\
50   & 10502   & 18 &  50   & 39 & 19  & 11     & 9   & 4 & 8 & 4 \\
100  & 41002   & 17  & 45  & 25   & 19  & 16   & 9   & 4 & 8& 3   \\
250  & 252502  & 18  & 48  & 37   & 19  & 18   & 10  & 4 & 9& 4   \\
500  & 1005002 & 17  & 55  & 50  & 18  & 19   & 15   & 5 & 9   & 3   \\
1000 & 4010002 & 17 & 81 & 65    & 18  & 18  & 29   & 5 & 13   & 2   \\
\bottomrule
\end{tabular}%
}
\end{table}

\section{Noise on the state variables}
\label{app:noise_x}

In the main text, we analyzed the noise in the external forcing $F_{ext}(t)$. However, in experimental settings, measurement noise inevitably corrupts the state variable $x(t)$ and its derivatives. Standard finite difference methods amplify this high-frequency noise, rendering the estimation of derivatives ($\dot{x}, \ddot{x}$) ill-posed.

Addressing this issue is an active area of research. 
Recent literature highlights that identification results are highly sensitive to the methodology used for derivative computation\cite{Khilchuk2025}. 
Sophisticated strategies to mitigate this include approaches that benefit from automatic differentiation \cite{Paszke2017, Baydin2018, Both2021}, weak-form formulations that use linear transformations and variance reduction techniques\cite{Messenger2021}, and methods that use automatic differentiation and time-stepping constraints\cite{Kaheman2022,Rudy2019}. Furthermore, robust differentiation schemes based on multi-objective optimization \cite{VanBreugel2020}, 
regularized schemes by combining discretization-based regularization with optimized smoothing\cite{Strebel2026},  integrating denoising techniques to smooth the signal\cite{Egan2024}, total variation regularization (TVR)\cite{Chartrand2011}, and probabilistic smoothing via Gaussian processes \cite{Wolff2025} have shown significant promise. 



We acknowledge that optimal real-world performance requires the integration of these advanced protocols; however, systematically evaluating their influence on the proposed formalism constitutes a separate study beyond the scope of this work. 
For the present analysis, we restrict ourselves to a standard pre-processing step [within block (b) of schematic Fig.~\ref{fig:schematic}] using a Savitzky-Golay (SG) filter. This allows us to investigate the identification capabilities under noisy conditions using a well-established differentiation technique.

We model the noisy dynamical variable ($x_{meas}(t)$) measurement as:
\begin{equation}
    x_{meas}(t) = x_{true}(t) + \epsilon(t), \quad \epsilon(t) \sim \mathcal{N}(0, \sigma_{x}^2),
\end{equation}
where $\sigma_{x}$ is determined by the desired SNR analogously to the description in the main text. The objective is to recover estimates for position, velocity, and acceleration $[\hat{x}, \hat{\dot{x}}, \hat{\ddot{x}}]$ from $x_{meas}$.

The SG filter fits a local polynomial of order $p$ to a window of $w$ adjacent points via linear least squares. This method preserves features of the underlying signal (such as relative maxima and minima) better than standard moving averages while effectively suppressing noise. 
Based on a sensitivity analysis of the differentiation error, we fixed the polynomial order at $p=3$ to allow for smooth estimation of the second derivative (acceleration), and the window length at $w=51$.
The filter is applied to $x_{meas}(t)$ to directly estimate the smoothed position $\hat{x}$, as well as the velocity ($\hat{\dot{x}}$) and acceleration ($\hat{\ddot{x}}$) utilizing the obtained derivative coefficients of the filter. 
These filtered variables $[\hat{x}, \hat{\dot{x}}, \hat{\ddot{x}}]$ replace the clean theoretical variables and define the database used to train the models. 

In the following two subsections, we show the training results for the Duffing and stick-slip systems.

\subsection{Duffing system}
\begin{figure}[!htpb]
    \centering
     \includegraphics[width=0.5\textwidth]{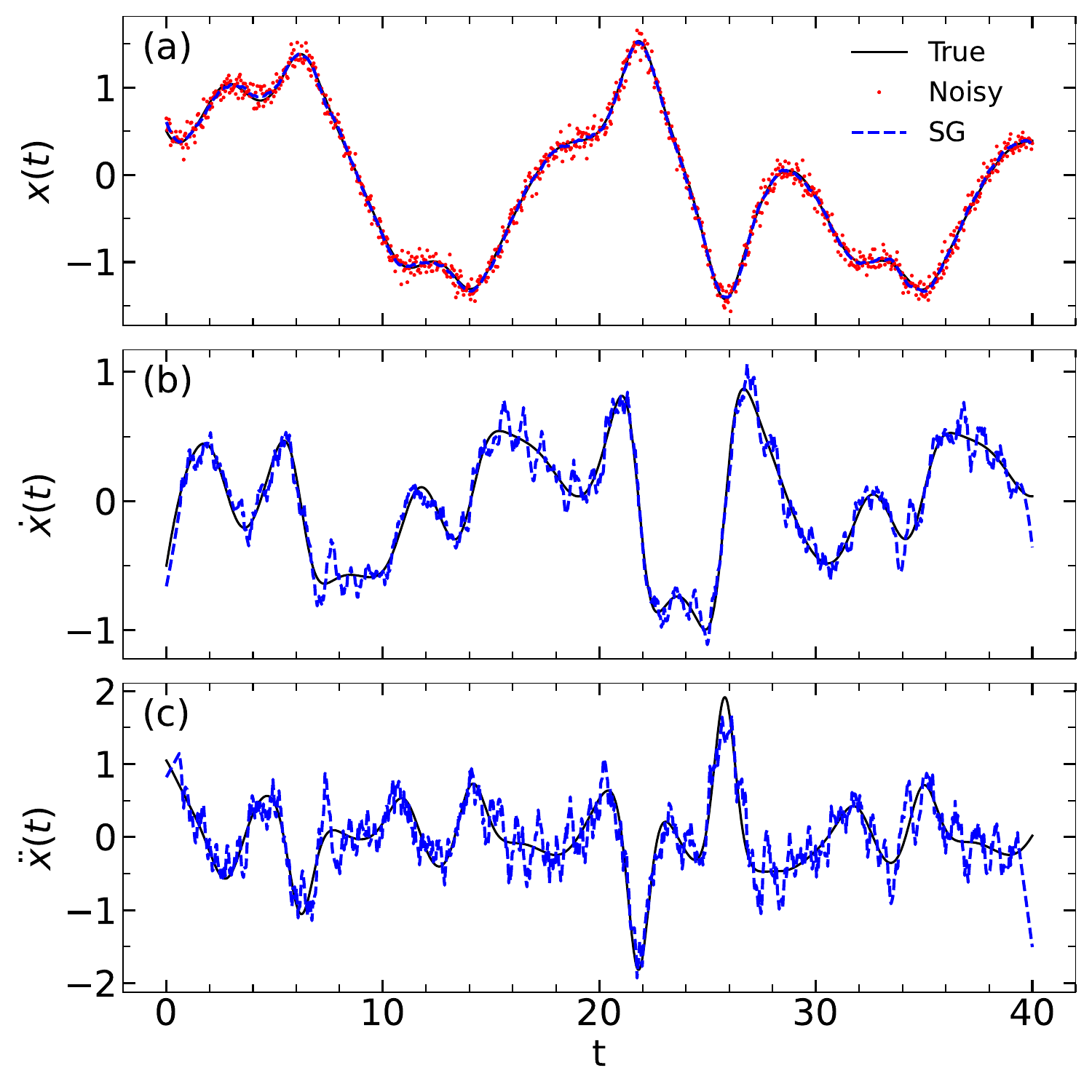}
    \caption{Assessment of the differentiation scheme using the SG filter for the Duffing system at SNR$=20$ dB. (a) Position $x(t)$, (b) velocity $\dot{x}(t)$, and (c) acceleration $\ddot{x}(t)$. The red dots indicate noisy measurements ($x_{meas}$), the blue dashed lines represent the SG estimates ($\hat{x}, \hat{\dot{x}}, \hat{\ddot{x}}$), and black lines represent the ground truth.}
    \label{fig:sg_differentiation:duffing}
\end{figure}

Figure~\ref{fig:sg_differentiation:duffing} illustrates the pre-processing results for the Duffing system at SNR$=20$ dB. While the filter effectively recovers a smooth position profile, velocity and acceleration estimates exhibit inevitable residual noise due to the amplification inherent in higher-order differentiation, though the underlying trend is preserved. 


\begin{figure*}[!htpb]
    \centering
    \subfloat{
        \includegraphics[width=0.35\textwidth]{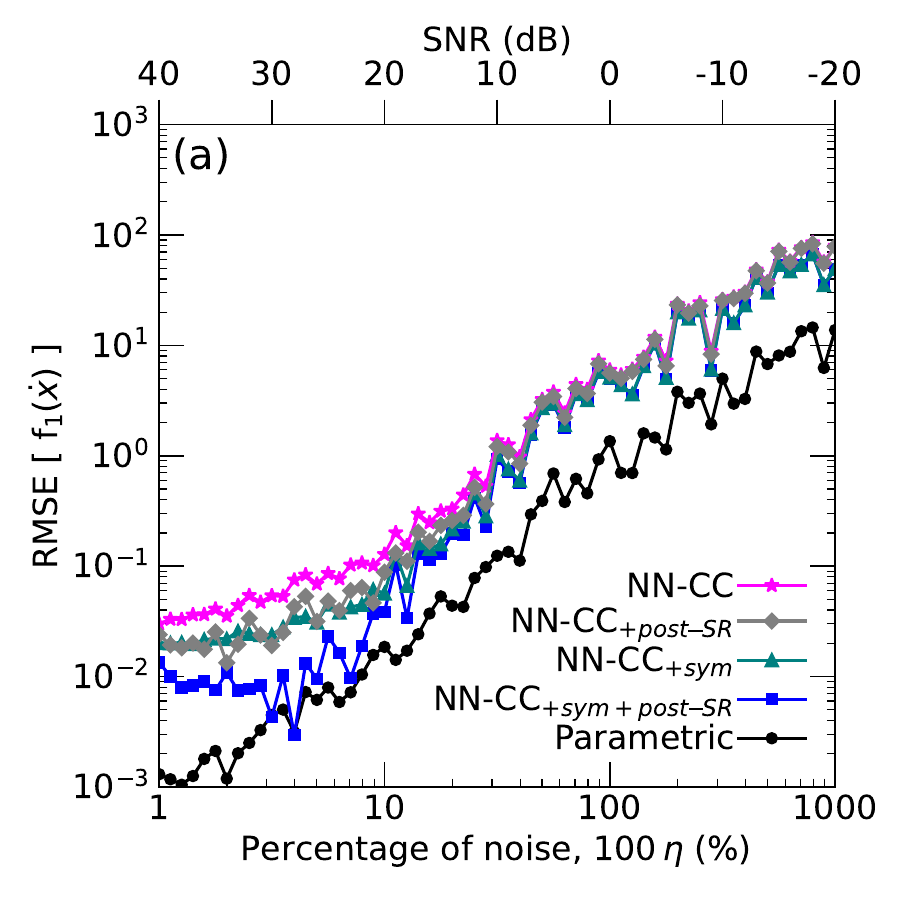}
    }
    \subfloat{
        \includegraphics[width=0.35\textwidth]{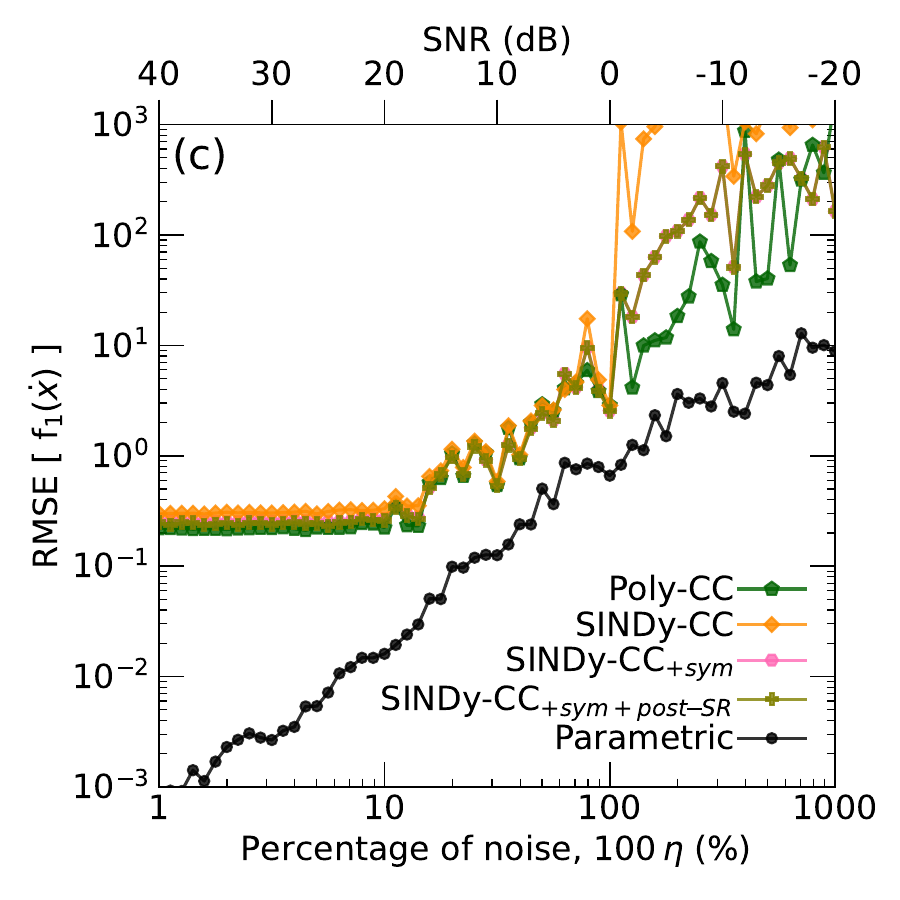}
    }
    \subfloat{
        \includegraphics[width=0.35\textwidth]{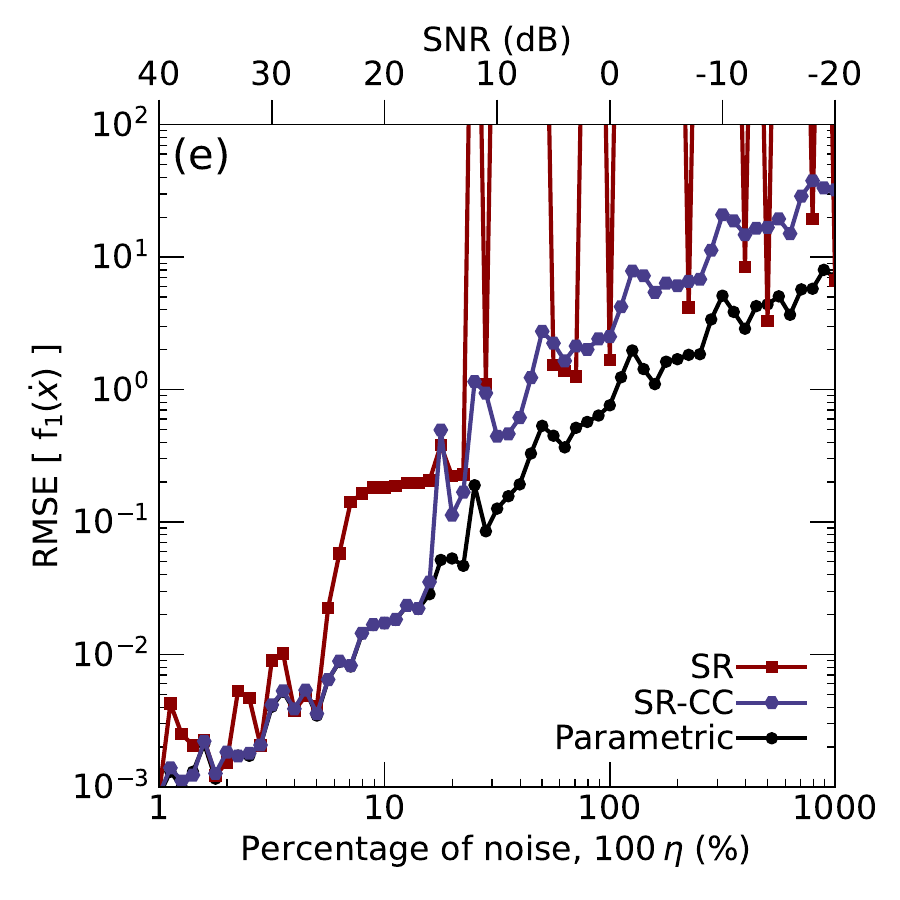}
    }\quad
    \subfloat{
        \includegraphics[width=0.35\textwidth]{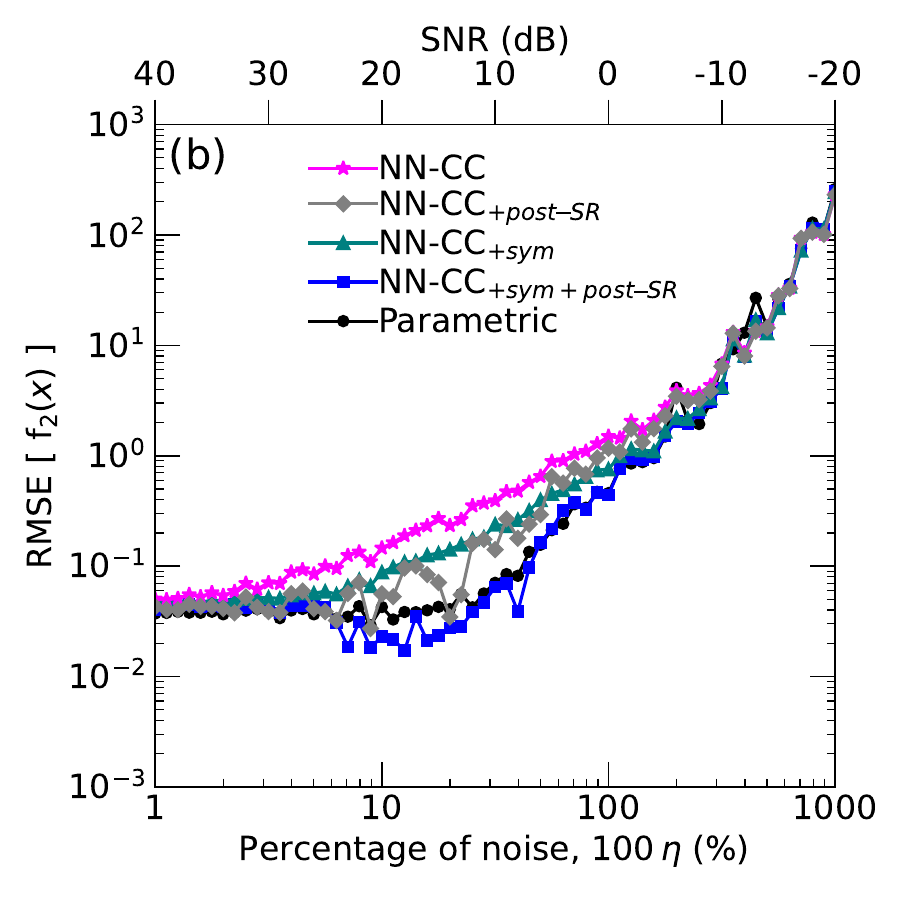}
    }
    \subfloat{
        \includegraphics[width=0.35\textwidth]{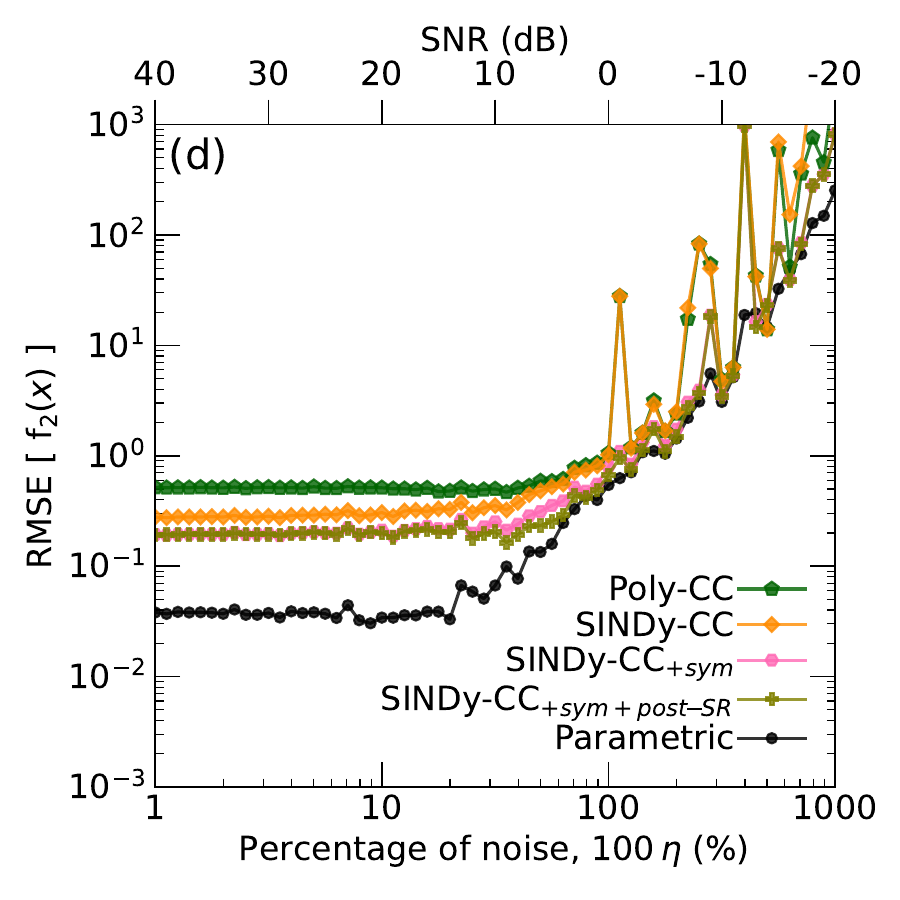}
    }
    \subfloat{
        \includegraphics[width=0.35\textwidth]{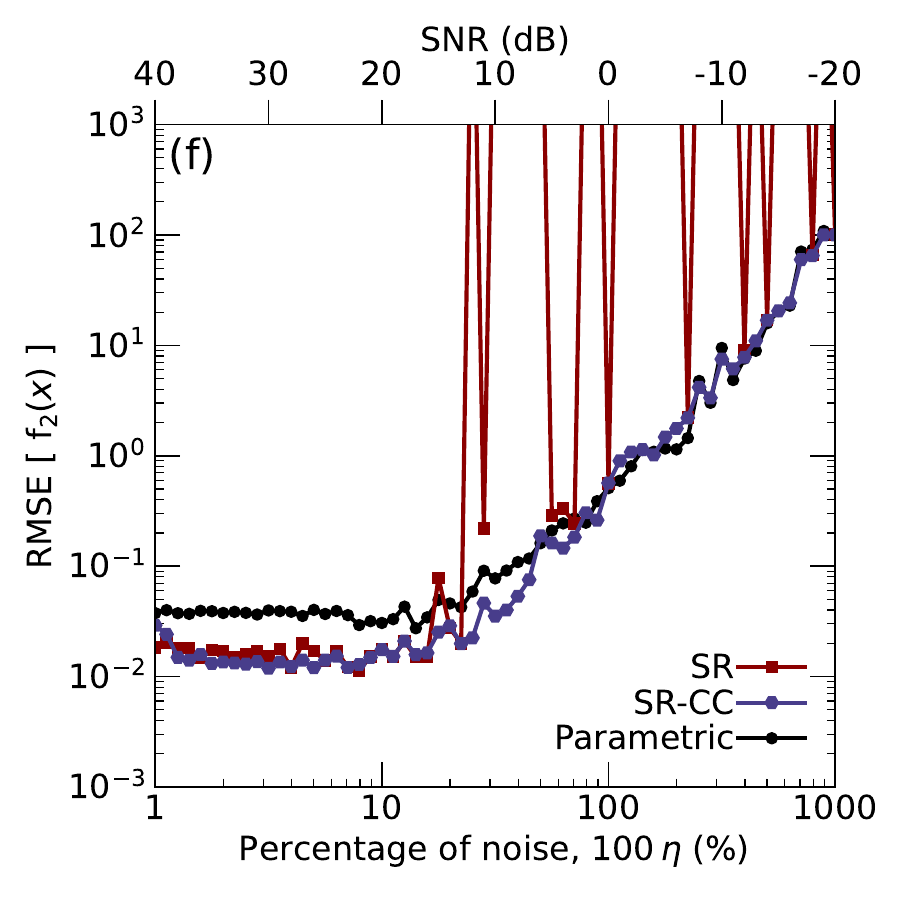}
    }
    \caption{RMSE analysis for the identified CCs ($f_1$ and $f_2$) for the Duffing system. Panels (a) and (b): NN-CC variants; (c) and (d): Polynomial basis variants; (e) and (f): SR variants. }
    \label{fig:appF_duffingCCs}
\end{figure*}

Figure~\ref{fig:appF_duffingCCs} shows the RMSE of the identified CCs as a function of SNR, averaged over of 10 independent noise realizations (i.e., 10 databases), following the same methodology as the main text. At high SNR (low noise), the RMSE values for $f_2(x)$ saturate.  This plateau represents the bias introduced by the SG differentiation errors, which effectively impose a lower bound on identifiability, as evidenced by the saturation of the Parametric baseline.

The NN-CC variants, shown in Figs.~\ref{fig:appF_duffingCCs}(a) and (b), demonstrate a clear hierarchy of improvement: adding symmetry constraints reduces the error, and the post-SR refinement yields further accuracy. This trend is consistent with the results reported in the main manuscript regarding process noise on the forcing term. 

In contrast, the Poly-CC and SINDy-CC methods [Figs.~\ref{fig:appF_duffingCCs}(c) and (d)] exhibit higher RMSE values. This is attributed to their tendency to overfit the noise amplified by the differentiation step, resulting in spuriously large coefficients for high-order polynomial terms. 

Interestingly, Figs.~\ref{fig:appF_duffingCCs}(e) and (f) suggest that, for SNR$\lesssim 20$ dB, the SR and SR-CC methods outperform the Parametric baseline. 
However, a closer inspection reveals this is an artifact of model parsimony. 
In this regime, SR variants favor the oversimplified expression $f_2(x)\approx -x^3$, discarding the linear term with $x$. Conversely, the Parametric method attempts to fit the full form $f_2(x)=\hat{\alpha} \,x^3+\hat\beta \, x$. 
For instance, at SNR$=30$ dB, the Parametric fit yields $\hat{\alpha}\approx-0.956$ and $\hat{\beta}\approx0.938$. Since the ground truth is $\alpha=-1.05$, the simplified SR coefficient (implicitly -1) is numerically closer to the true value than the Parametric estimate. Thus, the lower RMSE for SR in this specific range is a fortuitous result of simplification rather than superior physics extraction. 

Furthermore, at lower SNR values (SNR$\lesssim 13$ dB), the SR method fails, identifying spurious cross-terms between $x$ and $\dot{x}$, which are visualized by divergent RMSE values. While the SR-CC framework eliminates these divergences, the characteristic `staircase' pattern in the error metric (visible at SNR$\approx 15$ dB) persists, reflecting the discrete nature of the symbolic search space.

\subsection{Stick-slip system}
\begin{figure}[!htpb]
    \centering
     \includegraphics[width=0.5\textwidth]{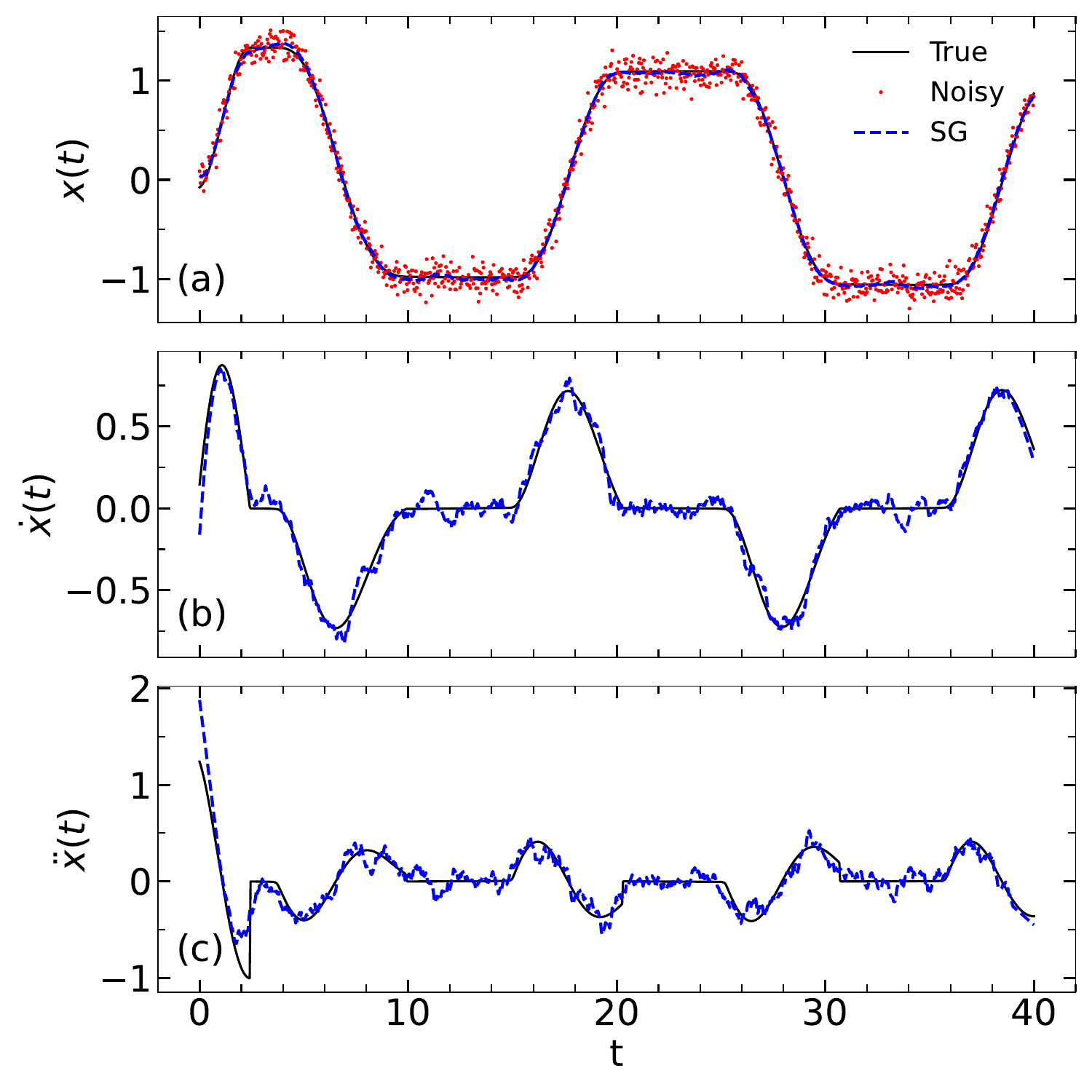}
    \caption{Assessment of the differentiation scheme using the SG filter for the stick-slip system at SNR $=20$ dB. (a) Position $x(t)$, (b) velocity $\dot{x}(t)$, and (c) acceleration $\ddot{x}(t)$. The red dots indicate noisy measurements ($x_{meas}$), the blue dashed lines represent the SG estimates ($\hat{x}, \hat{\dot{x}}, \hat{\ddot{x}}$), and black lines represent the ground truth.}
    \label{fig:sg_differentiation:stick-slip}
\end{figure}

Figure~\ref{fig:sg_differentiation:stick-slip} illustrates the pre-processing results for the stick-slip system at SNR$=20$ dB. Similarly to the Duffing system, while the filter effectively recovers a smooth position profile, velocity and acceleration estimates exhibit inevitable residual noise due to the amplification inherent in higher-order differentiation, though the underlying trend is preserved. 


\begin{figure*}[!htpb]
    \centering
    \subfloat{
        \includegraphics[width=0.35\textwidth]{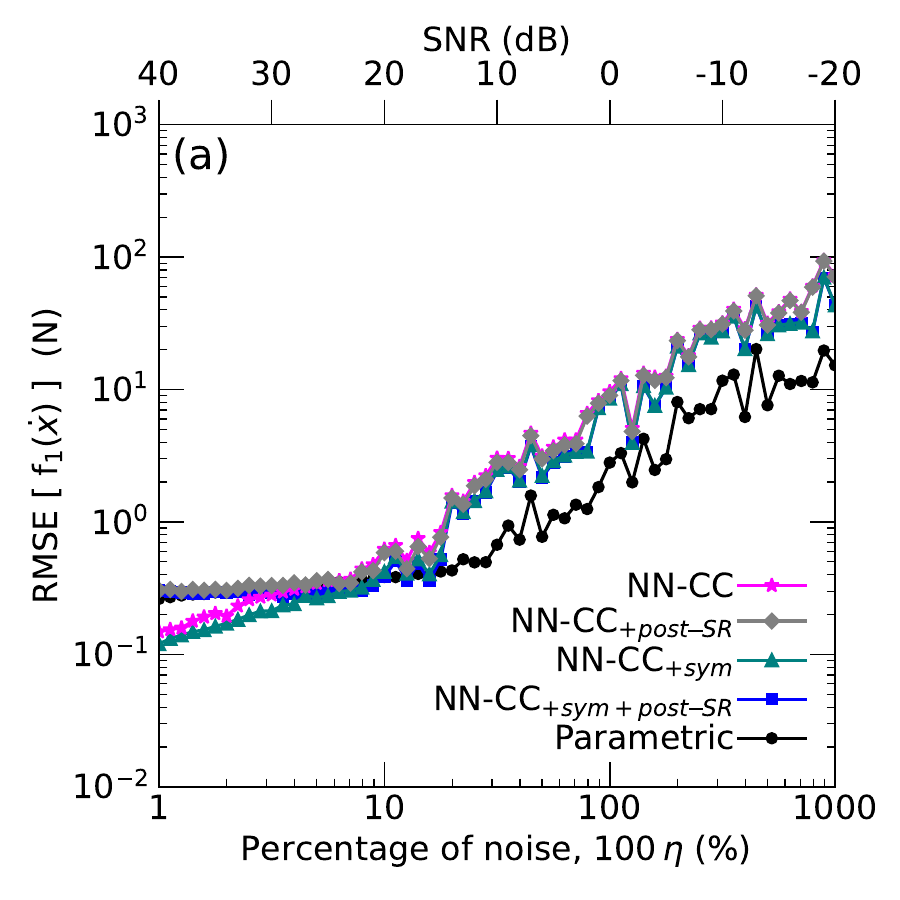}
    }
    \subfloat{
        \includegraphics[width=0.35\textwidth]{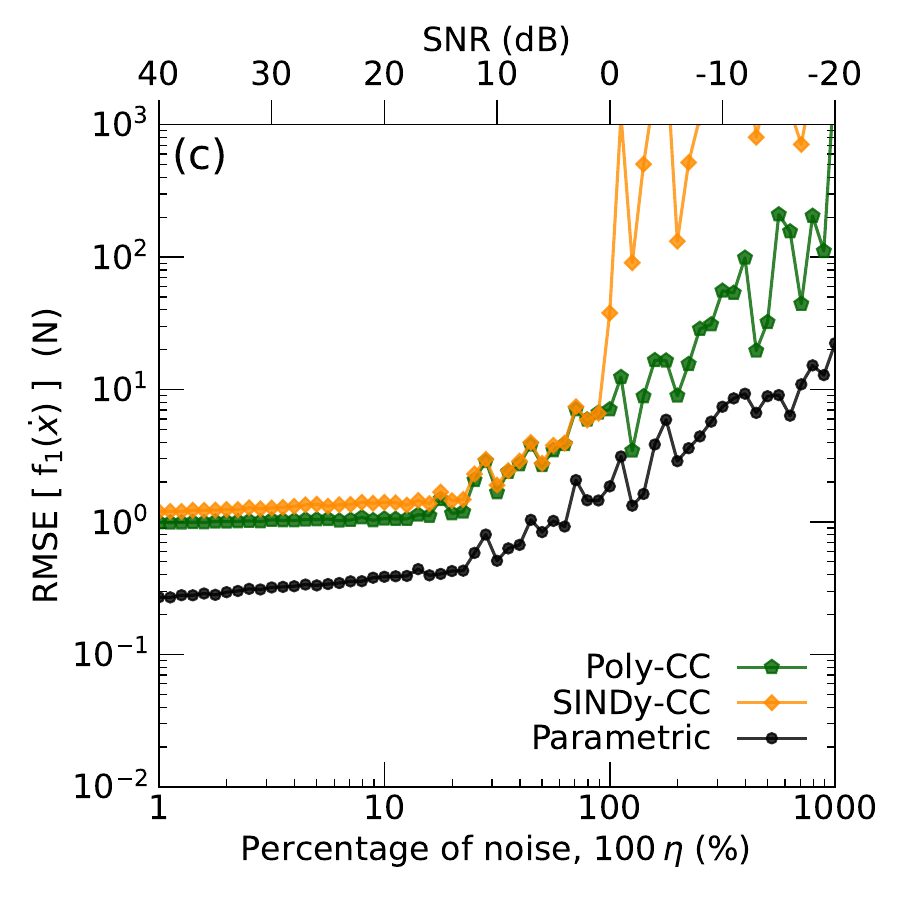}
    }
    \subfloat{
        \includegraphics[width=0.35\textwidth]{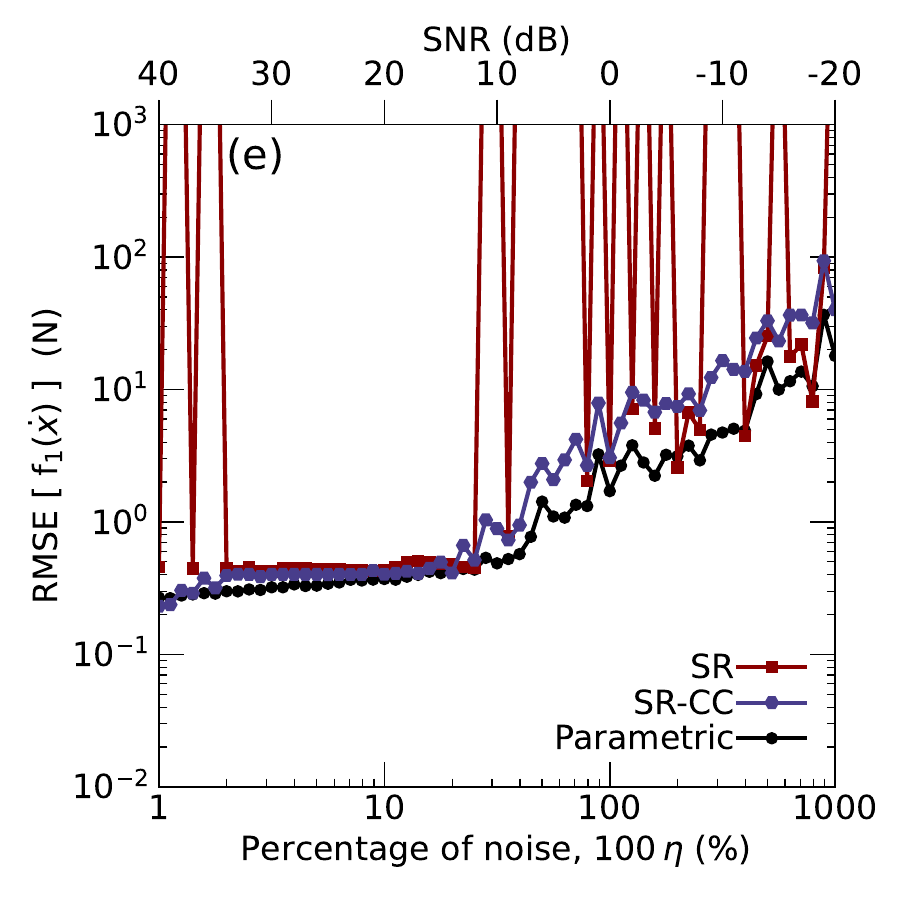}
    }\quad
    \subfloat{
        \includegraphics[width=0.35\textwidth]{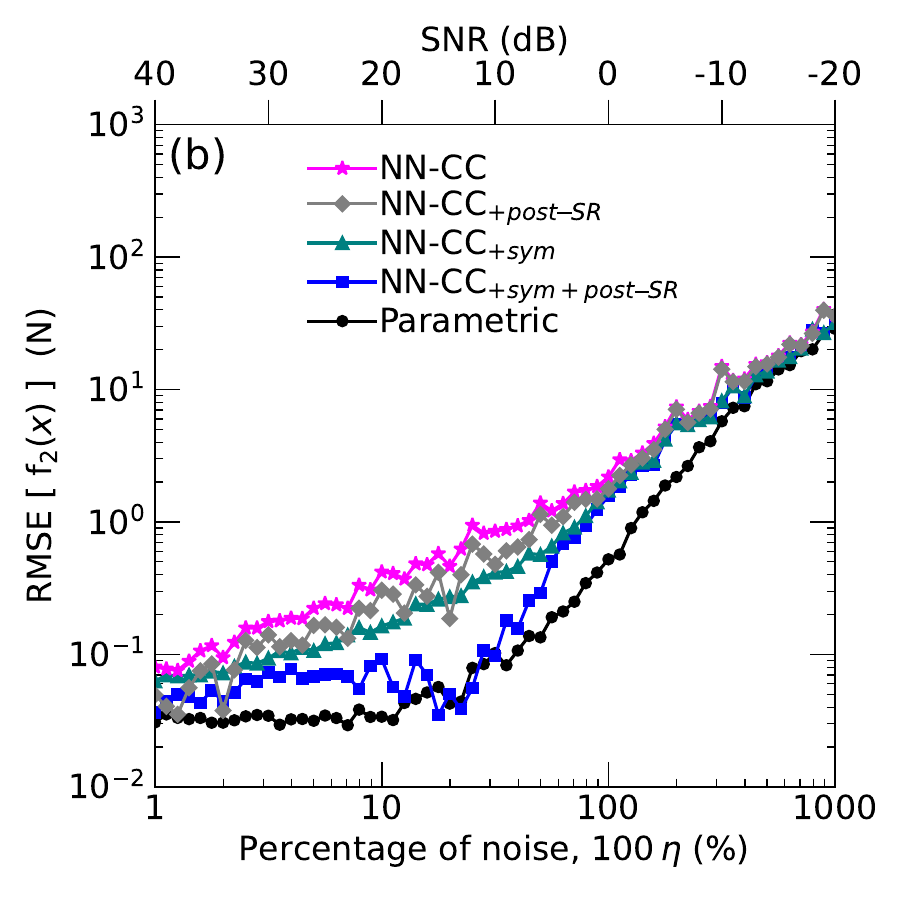}
    }
    \subfloat{
        \includegraphics[width=0.35\textwidth]{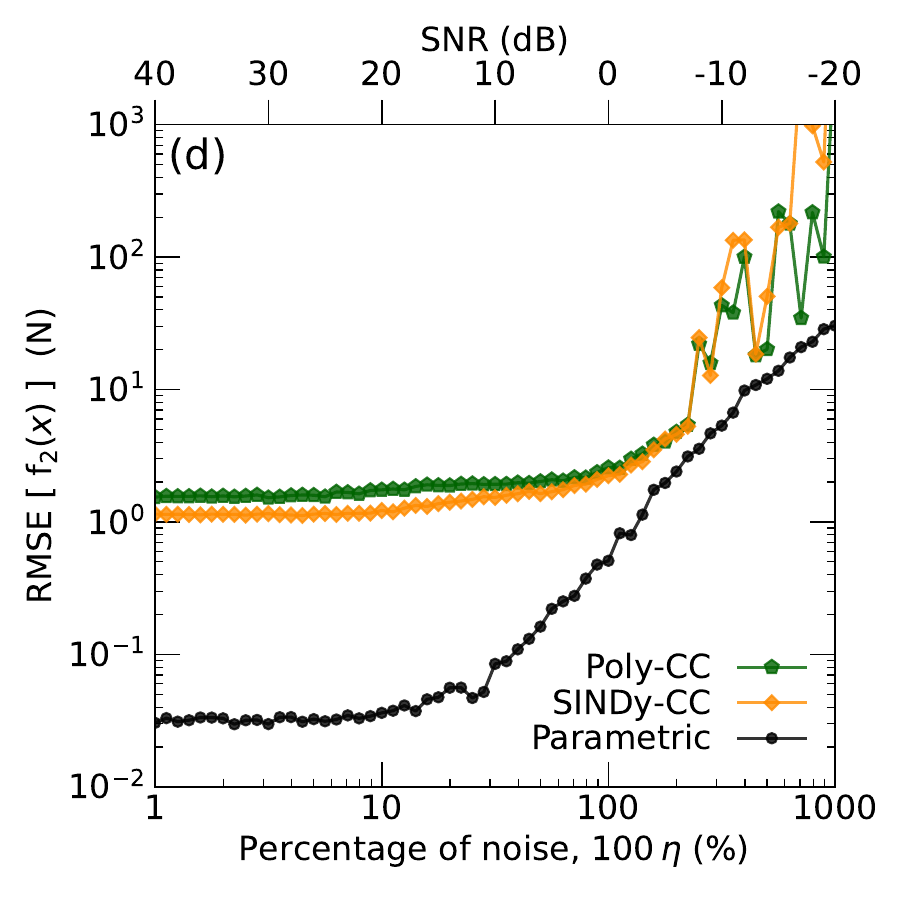}
    }
    \subfloat{
        \includegraphics[width=0.35\textwidth]{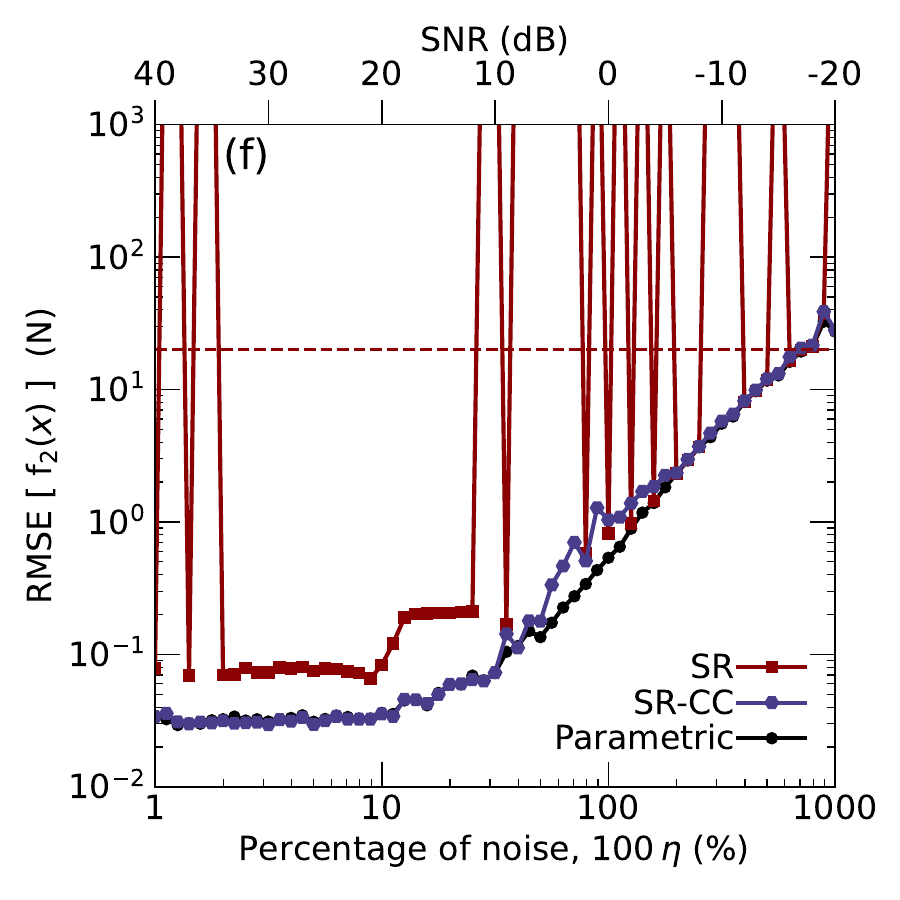}
    }
    \caption{RMSE analysis for the identified CCs ($f_1$ and $f_2$) for the stick-slip system. Panels (a) and (b): NN-CC variants; (c) and (d): Polynomial basis variants; (e) and (f): SR variants. }
    \label{fig:appF_stick-slipCCs}
\end{figure*}

Figure~\ref{fig:appF_stick-slipCCs} shows the RMSE of the identified CCs as a function of SNR, averaged over of 10 independent noise realizations (i.e., 10 databases), consistent with the methodology in the main text. In this case, the RMSE values for both $f_1(\dot{x})$ and $f_2(x)$ functions saturate at high SNR (low noise).  This plateau reflects the bias introduced by the SG differentiation errors, which effectively impose a lower bound on identifiability, as evidenced by the saturation of the Parametric baseline.

For the NN-CC variants, shown in Figs.~\ref{fig:appF_duffingCCs}(a) and (b), it is suspicious that the $f_1(\dot{x})$ function for the NN-CC$_\text{+sym+post-SR}$ at SNR $\gtrsim 20$ dB have higher RMSE values than the other methods, contrarily to the expected hierarchy of improvement. However, the $f_1(\dot{x})$ function identified after the post-SR process for the the NN-CC$_\text{+sym+post-SR}$ method identifies a $\tanh$ function and thus the RMSE values become similar to the Parametric case.


As anticipated, the Poly-CC and SINDy-CC methods [Figs.~\ref{fig:appF_duffingCCs}(c) and (d)] exhibit higher RMSE values compared to the NN-CC variants. This performance gap arises from the limited capacity of polynomials to accurately approximate the discontinuity near the origin.

Interestingly, Figs.~\ref{fig:appF_duffingCCs}(e) and (f) reveal that while standard SR fails even at very low noise levels, incorporating SR within the CC-based formalism yields a substantial improvement in performance.

\section{RMSE scaling under additive noise}
\label{app:noise_scaling}

This appendix provides a theoretical justification for the empirical observation that the RMSE of the reconstructed functions scales linearly with the noise level (exhibiting a slope of 1 on a log-log scale). We demonstrate that this scaling is an inherent property of stable estimators in the variance-dominated regime, where the error is primarily driven by noise rather than model bias.


The analysis is structured as follows: we first define the problem setup, then derive the exact scaling for linear least-squares regression (where the result is analytical), and finally extend the analysis to nonlinear estimators (such as neural networks) using a Lipschitz stability argument.

\subsection{Problem setup}
We consider a scalar physical quantity \(y\) governed by a deterministic function \(f(x)\) subject to additive stochastic noise. The observed signal is modeled as:
\begin{equation}
    y(t) = f(x(t)) + \varepsilon(t) \;,
\end{equation}
where \(\varepsilon \sim \mathcal{N}(0, \sigma_{\mathrm{noise}}^2)\) is a Gaussian noise process with zero mean and standard deviation $\sigma_{\mathrm{noise}}$. 
In practice, we observe this process at a discrete set of \(N\) sampling points \(\{x_i\}_{i=1}^N\). The resulting observation vector \(\mathbf{y} \in \mathbb{R}^N\) is given by:
\begin{equation}
    \mathbf{y} = \mathbf{f}(\mathbf{x}) + \mathbf{\varepsilon} \,, 
\label{eq:app_function_noise}
\end{equation}
where $\mathbf{x}\in \mathbb{R}^N$ is the dynamical state vector, $\mathbf{f}: \mathbb{R}^N \to \mathbb{R}^N$ is a vector-valued function, and $\mathbf{\varepsilon}\in \mathbb{R}^N$  is an independent and identically distributed (i.i.d.) noise vector.
We define the relative noise level as \(\eta = \sigma_{\mathrm{noise}} / \sigma_{\mathrm{signal}}\), where \(\sigma_{\mathrm{signal}}\) is the standard deviation of the noiseless function values \(f(\mathbf{x})\).

Our goal is to analyze the dependence of the reconstruction error on the noise level \(\sigma_{\mathrm{noise}}\). We quantify this error using the RMSE between the estimated function \(\hat{f}\) and the true function \(f\) by taking the expectation over the noise distribution (denoted simply by \(\mathbb{E}\)) of the squared Euclidean distance, normalized by the number of time points:
\begin{equation}
    \mathrm{RMSE} = \sqrt{ \frac{1}{N} \mathbb{E} \left[ \| \hat{\mathbf{f}} - \mathbf{f} \|^2 \right] } \;.
    \label{eq:rmse_vector_def}
\end{equation}

\subsection{Fundamental limits: The Cramér-Rao bound}
To understand the scaling behavior, we first consider the lower bound given by the information theory for any unbiased parametric estimator. Let the function be approximated by a parametric model \(f(x; \boldsymbol{\theta})\) with parameters \(\boldsymbol{\theta} \in \mathbb{R}^p\).

For an estimator \(\hat{\boldsymbol{\theta}}\) that is unbiased (i.e., \(\mathbb{E}[\hat{\boldsymbol{\theta}}] = \boldsymbol{\theta}\)), the covariance of the parameter estimation error is bounded below by the inverse of the Fisher information matrix (FIM)\cite{Kay1993estimation,Rife1974,Richmond2006,Quirini2023,Jia2023,Mehmetcik2023} (denoted as \(\mathcal{I}(\boldsymbol{\theta})\)):
\begin{equation}
    \mathrm{Cov}(\hat{\boldsymbol{\theta}}) \succeq \mathcal{I}(\boldsymbol{\theta})^{-1} \;.
\end{equation}

Under the assumption of additive Gaussian noise with variance \(\sigma_{\mathrm{noise}}^2\), the log-likelihood of the observations is proportional to \(-\frac{1}{2\sigma_{\mathrm{noise}}^2} \sum (y_i - f(x_i; \boldsymbol{\theta}))^2\). The entries of the Fisher Information Matrix are given by:
\begin{equation}
    [\mathcal{I}(\boldsymbol{\theta})]_{jk} = - \mathbb{E} \left[ \frac{\partial^2 \ln p(\mathbf{y}|\boldsymbol{\theta})}{\partial \theta_j \partial \theta_k} \right] = \frac{1}{\sigma_{\mathrm{noise}}^2} \sum_{i=1}^N \frac{\partial f(x_i)}{\partial \theta_j} \frac{\partial f(x_i)}{\partial \theta_k} \;,
\end{equation}
or in matrix notation as:
\begin{equation}
\mathcal{I}(\boldsymbol{\theta}) =
\frac{1}{\sigma_{\mathrm{noise}}^2}
\left(\frac{\partial \mathbf{f}}{\partial \boldsymbol{\theta}}\right)^{\!\top}
\left(\frac{\partial \mathbf{f}}{\partial \boldsymbol{\theta}}\right)
\Bigg|_{\boldsymbol{\theta}=\boldsymbol{\theta}^*}.
\end{equation}

Crucially, the FIM can be factored as:
\begin{equation}
    \mathcal{I}(\boldsymbol{\theta}) = \frac{1}{\sigma_{\mathrm{noise}}^2} \mathbf{J}^T \mathbf{J} \;,
\end{equation}
where \(\mathbf{J}\) is the Jacobian matrix of sensitivities depending only on the model structure and inputs, not on the noise level. Consequently, the lower bound on the parameter covariance scales directly with the noise variance:
\begin{equation}
    \mathrm{Cov}(\hat{\boldsymbol{\theta}}) \succeq \sigma_{\mathrm{noise}}^2 (\mathbf{J}^T \mathbf{J})^{-1} \;.
    \label{eq:cov_scaling}
\end{equation}

To relate parameter error to the reconstruction error of the function vector \(\mathbf{f}\), we use a first-order Taylor expansion around the true parameters:
\begin{equation}
    \hat{\mathbf{f}} - \mathbf{f} \approx \mathbf{J} (\hat{\boldsymbol{\theta}} - \boldsymbol{\theta}) \;.
\end{equation}
The total mean squared error (MSE) (sum of variances across all points) is the trace of the covariance of \(\hat{\mathbf{f}}\):
\begin{equation}
    \mathbb{E} \left[ \|\hat{\mathbf{f}} - \mathbf{f}\|^2 \right] \approx \mathrm{Tr} \left( \mathrm{Cov}(\hat{\mathbf{f}}) \right) = \mathrm{Tr} \left( \mathbf{J} \, \mathrm{Cov}(\hat{\boldsymbol{\theta}}) \, \mathbf{J}^T \right) \;.
\end{equation}
Applying the inequality from Eq.~\eqref{eq:cov_scaling} and the property that \(\mathrm{Tr}(ABA^T) \ge \mathrm{Tr}(ACA^T)\) if \(B \succeq C\):
\begin{equation}
    \mathbb{E} \left[ \|\hat{\mathbf{f}} - \mathbf{f}\|^2 \right] \ge \sigma_{\mathrm{noise}}^2 \, \mathrm{Tr} \left( \mathbf{J} (\mathbf{J}^T \mathbf{J})^{-1} \mathbf{J}^T \right) \;.
\end{equation}
Notice that \(\mathbf{P} = \mathbf{J} (\mathbf{J}^T \mathbf{J})^{-1} \mathbf{J}^T\) is the projection matrix onto the column space of the Jacobian. Its trace is equal to the rank of the model (the number of effective parameters, \(p\)):
\begin{equation}
    \mathbb{E} \left[ \|\hat{\mathbf{f}} - \mathbf{f}\|^2 \right] \ge \sigma_{\mathrm{noise}}^2 \cdot p \;.
\end{equation}
Finally, substituting this back into the definition of RMSE (Eq.~\ref{eq:rmse_vector_def}):
\begin{equation}
    \mathrm{RMSE} \ge \sqrt{ \frac{1}{N} \sigma_{\mathrm{noise}}^2 p } = \sigma_{\mathrm{noise}} \sqrt{\frac{p}{N}} \;.
\end{equation}
This derivation proves that the minimum achievable reconstruction error is strictly proportional to \(\sigma_{\mathrm{noise}}\). Therefore, efficient estimators will exhibit a linear relationship between RMSE and noise level.

\subsection{Fixed-basis regression and approximation error}

We now consider a regression model in which \(f(x)\) is approximated using a fixed set of basis functions \(\{\phi_k(x)\}_{k=1}^M\). In the general case, the true function \(f(x)\) may not lie entirely within the span of these basis functions. Thus, we decompose the function as\cite{Ljung1999}:

\begin{equation}
f(x) = f_{\Phi}(x) + b(x),
\label{eq:linear_f_with_basis}
\end{equation}

where \(f_{\Phi}(x) = \sum_{k=1}^M \theta_k^* \phi_k(x)\) is the orthogonal projection of \(f\) onto the span of the basis, and \(b(x)\) is the approximation error (bias).

Let \(\boldsymbol{\theta}\in\mathbb{R}^M\) denote the coefficient vector, and let \(\Phi\in\mathbb{R}^{N\times M}\) be the design matrix evaluated at at \(N\) sampling points, then Eq.~\ref{eq:linear_f_with_basis} can be expressed in matrix form as:
\begin{equation}
\mathbf{f} = \Phi\,\boldsymbol{\theta}^* + \mathbf{b}\,.
\end{equation}
A crucial property of the orthogonal projection is that the bias vector \(\mathbf{b}\) is orthogonal to the columns of the design matrix \(\Phi\), meaning \(\Phi^\top \mathbf{b} = \mathbf{0}\). The observed noisy data of Eq.~\ref{eq:app_function_noise}, can be expressed as:
\begin{equation}
\mathbf{y} = \mathbf{f} + \boldsymbol{\varepsilon} = \Phi\, \boldsymbol{\theta}^* + \mathbf{b} + \boldsymbol{\varepsilon},
\end{equation}
where \(\boldsymbol{\varepsilon}\) is the noise vector. The ordinary least squares (OLS) estimator is given by:
\begin{equation}
\hat{\boldsymbol{\theta}} = (\Phi^\top \Phi)^{-1}\Phi^\top \mathbf{y}.
\end{equation}

To determine the error, we first substitute the expression for \(\mathbf{y}\) into the estimator equation:
\begin{align}
\hat{\boldsymbol{\theta}} &= (\Phi^\top \Phi)^{-1}\Phi^\top (\Phi \,\boldsymbol{\theta}^* + \mathbf{b} + \boldsymbol{\varepsilon}) \\
&= \underbrace{(\Phi^\top \Phi)^{-1}\Phi^\top \Phi}_{= \mathbf{I}} \boldsymbol{\theta}^* + \underbrace{(\Phi^\top \Phi)^{-1}\Phi^\top \mathbf{b}}_{= \mathbf{0}} + (\Phi^\top \Phi)^{-1}\Phi^\top \boldsymbol{\varepsilon} \\
&= \boldsymbol{\theta}^* + (\Phi^\top \Phi)^{-1}\Phi^\top \boldsymbol{\varepsilon}.
\end{align}
Note that the term involving \(\mathbf{b}\) vanishes because \(\Phi^\top \mathbf{b} = \mathbf{0}\). 
The reconstructed function values are \(\hat{\mathbf{f}} = \Phi \,\hat{\boldsymbol{\theta}}\). Substituting \(\hat{\boldsymbol{\theta}}\):
\begin{equation}
\hat{\mathbf{f}} = \Phi\, \boldsymbol{\theta}^* + \mathbf{P}\, \boldsymbol{\varepsilon},
\end{equation}
where \(\mathbf{P} = \Phi(\Phi^\top \Phi)^{-1}\Phi^\top\) is the projection matrix onto the basis space. The reconstruction error vector is:
\begin{equation}
\hat{\mathbf{f}} - \mathbf{f} = (\Phi\, \boldsymbol{\theta}^* + \mathbf{P}\, \boldsymbol{\varepsilon}) - (\Phi\, \boldsymbol{\theta}^* + \mathbf{b}) = \mathbf{P}\, \boldsymbol{\varepsilon} - \mathbf{b}.
\end{equation}

We now calculate the MSE. Using the squared Euclidean norm \(\|\mathbf{v}\|^2 = \mathbf{v}^\top \mathbf{v}\):
\begin{equation}
\|\hat{\mathbf{f}} - \mathbf{f}\|^2 = (\mathbf{P}\, \boldsymbol{\varepsilon} - \mathbf{b})^\top (\mathbf{P}\, \boldsymbol{\varepsilon} - \mathbf{b}) = \boldsymbol{\varepsilon}^\top \mathbf{P}^\top \mathbf{P}\, \boldsymbol{\varepsilon} - 2\mathbf{b}^\top \mathbf{P}\, \boldsymbol{\varepsilon} + \mathbf{b}^\top \mathbf{b}.
\end{equation}
Using the properties of projection matrices (\(\mathbf{P}^\top = \mathbf{P}\) and \(\mathbf{P}^2 = \mathbf{P}\)) and taking the expectation \(\mathbb{E}[\cdot]\):
\begin{equation}
\mathbb{E}\!\left[\|\hat{\mathbf{f}} - \mathbf{f}\|^2\right] = \mathbb{E}[\boldsymbol{\varepsilon}^\top \mathbf{P} \,\boldsymbol{\varepsilon}] - 2\mathbf{b}^\top \mathbf{P} \,\underbrace{\mathbb{E}[\boldsymbol{\varepsilon}]}_{=\mathbf{0}} + \|\mathbf{b}\|^2.
\end{equation}

The first term is the expected quadratic form of the noise. Using the trace trick \(\mathbb{E}[\boldsymbol{\varepsilon}^\top \mathbf{P} \boldsymbol{\varepsilon}] = \mathrm{Tr}(\mathbf{P} \,\mathbb{E}[\boldsymbol{\varepsilon}\boldsymbol{\varepsilon}^\top])\):
\begin{equation}
\mathbb{E}[\boldsymbol{\varepsilon}^\top \mathbf{P} \boldsymbol{\varepsilon}] = \mathrm{Tr}(\mathbf{P} \sigma_{\mathrm{noise}}^2 \mathbf{I}) = \sigma_{\mathrm{noise}}^2 \mathrm{Tr}(\mathbf{P}).
\end{equation}
The trace of the projection matrix \(\mathbf{P}\) is equal to the rank of \(\Phi\), which is \(M\) (the number of basis functions). Thus, the total MSE is:
\begin{equation}
\mathbb{E}\!\left[\|\hat{\mathbf{f}} - \mathbf{f}\|^2\right] = \|\mathbf{b}\|^2 + M \sigma_{\mathrm{noise}}^2.
\end{equation}

Finally, substituting this back into the definition of RMSE (Eq.~\ref{eq:rmse_vector_def}):
\begin{equation}
\mathrm{RMSE} = \sqrt{\frac{1}{N} \left( \|\mathbf{b}\|^2 + M \sigma_{\mathrm{noise}}^2 \right)}.
\label{eq:app:rmse_linear_with_bias}
\end{equation}

Consider the specific case where the chosen basis functions are sufficient to fully describe the target function (i.e., the physical model is correctly specified). Mathematically, this implies that the true function vector \(\mathbf{f}\) lies strictly within the column space of the design matrix \(\Phi\). Because the projection of a vector onto a subspace that already contains it is the vector itself, the approximation is exact (\(\mathbf{f}=\Phi\, \boldsymbol{\theta}^* \)) and the bias vector vanishes completely (\(\mathbf{b} = \mathbf{0}\)). Under this condition, the RMSE becomes purely a function of the noise variance: 
\begin{equation}
\mathrm{RMSE} \approx \sqrt{\frac{M}{N}} \, \sigma_{\mathrm{noise}} = \sigma_\text{signal}\sqrt{\frac{M}{N}} \, \eta \,.
\end{equation}

This result provides the theoretical justification for the empirical model of Eq.~\ref{eq:fitting_rmse_empirical}, yielding the linear relationship:
\begin{equation}
    \text{RMSE} [f_i(z)]=A\,\eta \,.
\end{equation}

Indeed, for linear-in-parameter models subject to additive Gaussian noise, the OLS estimator coincides with the maximum likelihood estimator (MLE). In the asymptotic regime (assuming sufficient data and correct model specification) the estimator is both unbiased and efficient, meaning it attains the CRB~\cite{Kay1993estimation}. 
The prefactor \(A=\sigma_{\text{signal}}\sqrt{M/N}\) encapsulates the conditioning of the regression problem, the model complexity (\(M\)), the sample size (\(N\)), and the signal standard deviation (\(\sigma_{\text{signal}}\)); these quantities remain constant regardless of variations in the noise level.

\subsection{Nonlinear estimators and Neural Networks}

For general nonlinear estimators (e.g.\ neural networks), we consider a reconstruction operator
\begin{equation}
\hat f(\cdot) = \mathcal{A}(\mathbf{x}, \mathbf{y}),
\end{equation}
where \(\mathbf{x}=(x_1,\dots,x_N)\) and \(\mathbf{y}=(y_1,\dots,y_N)\) are the measured input–output pairs generated by
\(
y_i = f(x_i) + \varepsilon_i
\). We denote the vector of predictions at the training inputs by $\hat{\mathbf{f}}=\hat{f}(\mathbf{x})$. 
The noiseless reconstruction is obtained by applying the same operator to the noise-free variables
\begin{equation}
\hat f_{\mathrm{clean}}(\cdot) = \mathcal{A}(\mathbf{x}, \mathbf{f}),
\end{equation}
where the noiseless vector is defined as \(\hat{\mathbf{f}}_\text{clean}=\hat{f}_\text{clean}(\mathbf{x})\). We also define the structural (approximation) error as 
\begin{equation}
\mathbf{e}_{\mathrm{approx}}
= \hat{\mathbf{f}}_{\mathrm{clean}} - \mathbf{f}\,,
\end{equation}
which captures the inability of the estimator (architecture, capacity, regularization) to represent the true function even with a dense database. For noisy observations $\mathbf{y}=\mathbf{f}+\mathbf{\varepsilon}$, the reconstruction error decomposes as
\begin{equation}
\hat{\mathbf{f}} - \mathbf{f}
=
\bigl[\mathcal{A}(\mathbf{x},\mathbf{y})
- \mathcal{A}(\mathbf{x},\mathbf{f})\bigr]
+ \mathbf{e}_{\mathrm{approx}} .
\end{equation}

Taking squared Euclidean norms and expanding gives 

\begin{align}
\|\hat{\mathbf{f}}-\mathbf{f}\|^2
&=
\|\mathcal{A}(\mathbf{x},\mathbf{y})-\mathcal{A}(\mathbf{x},\mathbf{f})\|^2
+\|\mathbf{e}_{\mathrm{approx}}\|^2  \nonumber\\
&\quad
+2\bigl(\mathcal{A}(\mathbf{x},\mathbf{y})-\mathcal{A}(\mathbf{x},\mathbf{f})\bigr)^{\!\top}
\mathbf{e}_{\mathrm{approx}} .
\end{align}

To control the noise-propagation term, we assume that the estimator is locally
stable in the sense that there exists a constant \(L>0\) such that
\begin{equation}
\|\mathcal{A}(\mathbf{x},\mathbf{y})-\mathcal{A}(\mathbf{x},\mathbf{f})\|
\le L\,\|\mathbf{y}-\mathbf{f}\|
= L\,\|\boldsymbol{\varepsilon}\|.
\end{equation}
This Lipschitz-type condition is a standard stability assumption for trained
neural networks and nonlinear regression mappings in a well-behaved operating
regime\cite{Fazlyab2019,Gouk2020}.

Taking expectations with respect to the noise and assuming that the
cross-term averages to zero (which holds exactly under a first-order
linearization with zero-mean noise and approximately for stable estimators),
we obtain the mean-squared error bound
\begin{equation}
\mathbb{E}\!\left[\|\hat{\mathbf{f}}-\mathbf{f}\|^2\right]
\le
\mathbb{E}\!\left[\|\mathcal{A}(\mathbf{x},\mathbf{y})-\mathcal{A}(\mathbf{x},\mathbf{f})\|^2\right]
+\|\mathbf{e}_{\mathrm{approx}}\|^2 .
\end{equation}
Applying the Lipschitz bound gives
\begin{equation}
\mathbb{E}\!\left[\|\mathcal{A}(\mathbf{x},\mathbf{y})-\mathcal{A}(\mathbf{x},\mathbf{f})\|^2\right]
\le
L^2\,\mathbb{E}[\|\boldsymbol{\varepsilon}\|^2]
=
L^2\,N\,\sigma_{\mathrm{noise}}^2,
\end{equation}
and therefore
\begin{equation}
\mathbb{E}\!\left[\|\hat{\mathbf{f}}-\mathbf{f}\|^2\right]
\le
\|\mathbf{e}_{\mathrm{approx}}\|^2 + L^2 N \sigma_{\mathrm{noise}}^2.
\end{equation}

Substituting this back into the definition of RMSE (Eq.~\ref{eq:rmse_vector_def}):
\begin{equation}
\mathrm{RMSE}
\le
\sqrt{\frac{1}{N}\|\mathbf{e}_{\mathrm{approx}}\|^2 + L^2 \sigma_{\mathrm{noise}}^2 }.
\end{equation}

This result identifies two regimes:
\begin{itemize}
\item[i)] \textit{Approximation-dominated regime (low \(\eta\)).}
When \(\sigma_{\mathrm{noise}}\) is small, the RMSE saturates at the irreducible
approximation error \(\|\mathbf{e}_{\mathrm{approx}}\|/\sqrt{N}\). Thus, the RMSE converges to a constant value representing the inability of the model to perfectly fit the noiseless curve.
\item[ii)] \textit{Noise-dominated regime (sufficiently large \(\eta\)).}
When \(L^2\sigma_{\mathrm{noise}}^2 \gg \|\mathbf{e}_{\mathrm{approx}}\|^2/N\),
the approximation term becomes negligible and
\begin{equation}
\mathrm{RMSE} \lesssim L\,\sigma_{\mathrm{noise}} = L\, \sigma_{\mathrm{signal}} \; \eta .
\end{equation}
\end{itemize}

Hence, for stable nonlinear estimators, including neural networks trained to
convergence, the linear scaling of RMSE with the noise level is recovered in the variance-dominated regime.

\subsection{Interpretation}

The theoretical derivation discussed in this appendix confirms that a slope of one on a log-log plot of RMSE vs.\ $\eta$ is the signature of a stable estimator in the noise-dominated regime:
\begin{equation}
\log(\mathrm{RMSE}) \approx \log(\eta) + \text{const}.
\end{equation}

On the other hand, at low noise values ($\eta\to 0$), the fixed approximation error dominates, causing the curve to flatten (slope tends to zero). This effect appears for all the methods because RMSE values should converge to the RMSE value at $\eta= 0$. However, this flattening can occur starting from relatively high $\eta$ if the function approximator is not flexible enough to represent the function.
This can be mathematically justified for the limit case of $\sigma_\text{noise}\rightarrow0$ in Eq.~\ref{eq:app:rmse_linear_with_bias}, which yields to a constant value of
    \begin{equation}
        \text{RMSE}\approx \sqrt{\frac{||\mathbf{b||^2}}{N}}
    \end{equation}

Deviations from these linear (noise- or variance-dominated) and constant (bias-dominated) regimes may correspond to:

\begin{itemize}    
    \item[i)] \textit{Structural instability:} Methods that do not propose a mathematical structure family a prior (e.g., SR and baseline SINDy), may introduce some spurious terms (for instance, involving terms such as $x\,\dot{x}$, that are not compatible with the system structure studied in this work). These terms may lead to divergences on the RMSE values above some noise threshold.
    \item[ii)] \textit{Staircase effects:} Methods that selects functional forms by a discrete search (e.g., SR) may present `staircase' effects. 
    Specifically, the algorithms switch between different mathematical expressions on the Pareto frontier, thus leading to jumps in the .
\end{itemize}

\section{Computational cost}\label{app:computational_cost}

Table \ref{tab:sim_times_hierarchical} details the computational time required for each identification technique using a dataset from the main manuscript generated with SNR$=20$ dB, with $N_\text{data}=1000$. NN-CC methods employ a NN architecture with three hidden layers, each comprising 100 neurons and rectified linear unit (ReLU) activation functions. Since hyperparameters were kept fixed for the two Duffing and stick-slip systems, the identification times for both systems were approximately the same. 
Regression and evolutionary methods were executed on an 11th Gen Intel Core i7-1165G7, with 8 processors (CPUs).
For the NN-CC method, we evaluated the training times on both the CPU and an Nvidia A10 GPU.

These results reveal that regression-based methods are orders of magnitude faster due to their reliance on efficient least-squares optimization, whereas the evolutionary and NN-CC approaches require significantly more time to explore broader functional spaces. Notably, the NN-CC$_{\text{+sym+post-SR}}$ with CPU incurs the highest cost ($\approx$ 3 minutes) primarily due to the optimization overhead required to enforce symmetry constraints. The use of GPU reduce this time up to 40\% (from $\approx$3 to $\approx$2 min).

\begin{table}[!htpb]
\centering
\caption{Comparison of simulation times per single training run. The NN-CC variants show times for both CPU and GPU implementations during the training phases. The final post-SR step is performed on the CPU.}
\label{tab:sim_times_hierarchical}
\begin{tabular}{@{}lrr@{}}
\toprule
Identification technique & Added cost & Total time \\ \midrule
Regression & & \\
\quad Parametric & --- & 0.1 ms \\
\quad SINDy & --- & 1.2 ms \\
\quad SINDy-CC & --- & 1.1 s \\
Evolutionary & & \\
\quad Symbolic regression (pySR) & --- & 28.9 s \\ \addlinespace
NN-CC variants on CPU & & \\
\quad NN-CC (CPU) & 76.4 s & 76.4 s \\
\quad NN-CC$_{\text{+sym}}$ (CPU) & + 100.5 s & 176.9 s \\
\quad NN-CC$_{\text{+sym+post-SR}}$ & + 6.7 s per CC & 190.3 s \\ \addlinespace
NN-CC variants on GPU  & & \\
\quad NN-CC (GPU) & 69.7 s & 69.7 s \\
\quad NN-CC$_{\text{+sym}}$ (GPU) & + 29.7 s & 99.4 s \\
\quad NN-CC$_{\text{+sym+post-SR}}$ & + 6.7 s per CC & 112.8 s \\ \bottomrule
\end{tabular}
\end{table}

\bibliographystyle{elsarticle-num} 
\biboptions{sort&compress}


\bibliography{References}


\end{document}